\documentclass[12pt]{article}
\usepackage{jheppub}

\usepackage{amsmath,amsopn,amssymb,amsfonts}
\usepackage{epsfig}
\usepackage{verbatim}
\usepackage{multirow}
\usepackage{arydshln}
\usepackage{amsthm}
\usepackage[margin=1cm]{caption}

\def\beq{\begin{equation}}
\def\eeq{\end{equation}}
\def\bea{\begin{eqnarray}}
\def\eea{\end{eqnarray}}
\def\half{\frac{1}{2}}

\newcommand{\CC}{{\mathbb C}}
\newcommand{\ZZ}{{\mathbb Z}}

\def\CD{{\cal D}}
\def\CE{{\cal E}}
\def\CH{{\cal H}}\def\CI{{\cal I}}
\def\CL{{\cal L}}
\def\CN{{\cal N}}\def\CO{{\cal O}}
\def\CQ{{\cal Q}}

\def\CW{{\cal W}}
\def\CC{{\cal C}}

\renewcommand{\Im}{{\rm Im}}
\renewcommand{\Re}{{\rm Re}}

\newcommand{\IC}{\mathbb{C}}
\newcommand{\IZ}{\mathbb{Z}}

\def\Z{\mathbb{Z}}

\newcommand{\ndt}{\noindent}

\def\e{\epsilon}

\def\bea{\begin{eqnarray}}
\def\eea{\end{eqnarray}}
\def\be{\begin{equation}}
\def\ee{\end{equation}}
\def\ba{\begin{align}}
\def\ea{\end{align}}
\def\bse{\begin{subequations}}
\def\ese{\end{subequations}}

\newcommand{\bem}{\begin{pmatrix}}
\newcommand{\eem}{\end{pmatrix}}

\def\={\;  = \;}
\def\+{\, + \,}

\def\wt{\widetilde}
\def\wh{\widehat}
\def\bar{\overline}

\def\rt2{\sqrt{2}}

\renewcommand{\Im}{\mbox{Im}}
\renewcommand{\Re}{\mbox{Re}}

\DeclareMathOperator*{\Res}{Res}
\DeclareMathOperator*{\JKres}{JK-Res}
\def\vth{\vartheta}
\def\ve{\varepsilon}
\def\v{\varphi}
\def\n{\nu}
\def\s{\sigma}
\def\g{\gamma}
\def\t{\tau}
\def\a{\alpha}
\def\b{\beta}
\def\d{\delta}
\def\k{\kappa}

\def\l{{\lambda}}

\def\w{{\omega}}

\def\G{\Gamma}
\def\D{\Delta}
\def\z{\zeta}\def\th{\theta}\def\vth{\vartheta}

\renewcommand{\th}{\theta}

\def\p{\partial}
\def\tbar{\bar \tau}

\def\ubar{\bar u}

\def\myth1{\vartheta_{1}}

\title{Elliptic genera of ALE and ALF manifolds from gauged linear sigma models}

\preprint{EFI-14-10}
\author[1]{Jeffrey A.~Harvey}
\author[2]{\hspace*{-0.2cm}, Sungjay Lee}
\author[3]{and Sameer Murthy}
\affiliation[1,2]{Enrico Fermi Institute, University of Chicago \\
5620 Ellis Av., Chicago Illinois 60637, USA}
\affiliation[3]{Department of Mathematics, King's College London \\
The Strand, London WC2R 2LS, U.K.}
\abstract{We compute the equivariant  elliptic genera of several classes of ALE and ALF manifolds using localization in gauged linear sigma models. In the sigma model computation the equivariant action corresponds to chemical potentials for $U(1)$ currents and the elliptic genera exhibit  interesting pole structure as a function of the chemical potentials. We use this to decompose the answers into polar terms that exhibit wall crossing and universal terms. We compare our results to previous results on the large radius limit of the Taub-NUT elliptic genus
and also discuss applications of our results to counting of BPS world-sheet spectrum of
monopole strings in the 5d $\CN=2$ super Yang-Mills theory and self-dual strings in the 6d
$\CN=(2,0)$ theories.
}
\keywords{GLSM, Elliptic genus, Non-compact CFT}

\begin{document}

\maketitle

\section{Introduction}

String theory on a compact Calabi-Yau manifold $X$ of complex dimension $d$ gives rise to a superconformal field theory (SCFT) with
$(2,2)$ supersymmetry. The two-variable elliptic genus $\CE(X,\t,z)$ can be defined in terms of the world-sheet superconformal field theory
as
\be
\CE_{ws}(X,\t,z)={\rm tr}_{{\cal H}_{RR}} \left( (-1)^{J_0 + \bar J_0} \, y^{J_0} \, q^{L_0-c/24} \, \bar q^{\bar L_0 - \bar c/24} \right)
\ee
where $c=\bar c=3d/2$ are the central charges of the left- and right-moving Virasoro algebras, ${\cal H}_{RR}$ is the space of states in the Ramond-Ramond sector of the SCFT and $(J_0,L_0)$  and $(\bar J_0, \bar L_0)$ are the
zero modes of the left-  and right-moving $U(1)$ and Virasoro generators of the $N=2$ superconformal algebra respectively. Here and in the following we set $q=e^{2 \pi i \tau}$ and $y=e^{2 \pi i z}$. For a SCFT with a discrete spectrum of states the elliptic genus is a holomorphic
function of $q$ because the contribution from states with $\bar L_0 \ne \bar c/24$ cancels between pairs of states with opposite values of
$(-1)^{ \bar J_0}$.

For compact Calabi-Yau spaces this definition agrees with a ``space-time" definition in terms of the Euler characteristic of the formal vector bundle
\be
E_{q,y}= y^{d/2} {\textstyle \bigwedge}_{-y^{-1} }T_X^* \bigotimes_{n \ge 1} {\textstyle  \bigwedge}_{-y^{-1} q^n} T_X^* \bigotimes_{n \ge 1} {\textstyle \bigwedge}_{-y q^n} T_X \bigotimes_{n \ge 0} S_{q^n}(T_X \oplus T^*_X)
\ee
where $T_X$ and $T^*_X$ are the holomorphic tangent bundle of $X$ and its dual and  ${\textstyle \bigwedge}_q V$ and $S_q V$ are shorthand notation for infinite sums of antisymmetrized and symmetrized powers of the vector bundle
$V$:
\be
{\textstyle \bigwedge}_q V = 1 + q V + q^2 {\textstyle \bigwedge}^2 V + \cdots \, , \qquad S_q V = 1 + q V + q^2 S^2 V + \cdots \, .
\ee
The elliptic genus is then
\be
\CE_{st}(X,\t,z) = \int_X ch(E_{q,y}) \, Td(X) \, .
\ee

The elliptic genus enjoys both modular and elliptic properties which are seen most clearly in terms of the world-sheet definition where the
modular properties follow from the definition as a path integral over the torus (or elliptic curve) labelled by the modular parameter $\tau$ and
the elliptic properties follow from spectral flow in the SCFT. These properties together imply that the elliptic genus is a Jacobi form of weight
zero and index~$d/2$.

The elliptic genus captures important topological information of $X$ and has played an important role in both mathematics and string theory.
Recently there has been much interest in extending the elliptic genus to include non-compact manifolds. In this case both the world-sheet and space-time definitions become problematic for closely related reasons. The world-sheet theory has a continuous spectrum of
conformal dimensions which can lead to divergences in the trace over states of the SCFT and the space-time index theoretic definition fails to make sense without some specification of boundary conditions at infinity.

Nonetheless it has been possible in certain simple models to make
sense of the elliptic genus for noncompact SCFT. The most studied example is the ``cigar'' conformal field theory analyzed in \cite{Eguchi:2004yi,Troost:2010ud,Eguchi:2010cb,Ashok:2011cy,Ashok:2013zka,Ashok:2013pya,Murthy:2013mya}.
One of the key new features is that, unlike the situation for compact superconformal field theories, the elliptic genus of the cigar model is not
holomorphic in~$\t$. It has an explicit dependence on the anti-holomorphic variable~$\bar \tau$, that reflects the fact that the density
of states of bosons and fermions in the continuum are not equal.

In special cases it is also possible to make sense of index theory on noncompact manifolds. One method was employed by Pope in \cite{Pope:1978zx,Pope:1981jx} to analyze the Dirac operator on Taub-NUT space. He coupled the Dirac operator to an additional self-dual electromagnetic field that
effectively localizes the solutions to the Dirac equation and introduced a boundary by putting a cutoff on the radial coordinate.
He was then able to compute the index of the Dirac operator using the
Atiyah-Patodi-Singer index theorem for manifolds with boundary
and took the radial cutoff to infinity at the end of the computation.   Another, related, approach can be applied when the manifold has $U(1)$ isometries
that allow one to compute an equivariant version of the elliptic genus. In conformal field
theory the latter approach corresponds to modifying the elliptic genus to be of the form
\be
\CE_{ws}(X,\t,z, \xi_j)={\rm tr}_{{\cal H}_{RR}} \left( (-1)^{J_0 + \bar J_0} \, y^{J_0} \,
e^{ 2 \pi i \sum_j \xi_j J_j} \, q^{L_0-c/24} \, \bar q^{\bar L_0 - \bar c/24} \right)
\ee
where the $\xi_j$ are chemical potentials for the world sheet currents $J_i$ which correspond to the $U(1)$ isometries of $X$.  This equivariant elliptic genus becomes singular as the $\xi_{j}$ all approach zero and in typical examples one finds poles of order one or two when some of
the $\xi_j$ vanish. In other words, the elliptic genus becomes a multi-variable Jacobi form that is meromorphic in the chemical potentials~$\xi_{j}$. This approach can naturally be extended to the space-time formula for the elliptic genus by computing equivariant indices. As a simple
example consider $\IC^2$ with coordinates $(z_1,z_2)$  with the $U(1) \times U(1)$ action $(z_1,z_2) \rightarrow (y_1 z_1, y_2 z_2)$ with
$y_i = e^{2 \pi i \xi_i}$. Application of the Atiyah-Bott fixed point formula gives the equivariant index of the Dirac operator on $\IC^2$ as
\be
{\rm ind} D_{Dirac}= \frac{y_1^{1/2} y_2^{1/2}}{(1-y_1)(1-y_2)}
\ee
which has poles at $y_i=1$.

Single variable meromorphic Jacobi forms $\psi(\tau,z)$ were analyzed in detail in \cite{Zwegers,Dabholkar:2012nd} where it was shown that they can be decomposed into a polar part and a free part
\be
\psi(\tau,z) = \psi^P(\tau,z) + \psi^F(\tau,z)
\ee
where the polar part contains all the poles in $z$ (and typically exhibits wall-crossing) and the free part has an expansion in terms of vector-valued mock modular forms $\{h_r \}$ of the form
\be
\psi^F (\t, z) = \sum_{r \in \ZZ/2m\ZZ}  h_r(\tau) \, \vartheta_{m,r}(\tau,z) \, .
\ee
It is therefore expected that there is a close relationship between the elliptic genera of non-compact manifolds and mock modular forms.

In this paper we explore  further the connection between elliptic genera of non-compact manifolds, meromorphic Jacobi forms and mock modular forms in the equivariant elliptic genera of a class of gravitational instantons that are given by hyper-Kahler four-manifolds of ALE and ALF type. The simplest of these spaces is the Taub-NUT manifold. More general constructions yield the  multi-center $A_k$ asymptotically locally Euclidean (ALE) and $A_k$ asymptotically locally flat (ALF) spaces of Gibbons and Hawking \cite{Gibbons:1979zt} and the $D_k$ and $E_6,E_7,E_8$ ALE spaces of Kronheimer \cite{kronheimer1989}.  It would also be interesting to extend our results to the $D_k$ ALF spaces discussed in \cite{Cherkis:1998xca,Cherkis:2003wk} but that is beyond the scope of this paper. We will focus exclusively on the world-sheet approach to the elliptic genus in this paper.
We find that the elliptic genera of the ALE spaces are meromorphic Jacobi forms. The elliptic genera of ALF spaces that we study
have poles in the chemical potentials~$\xi_{j}$ and, in addition, are non-holomorphic functions of~$\tau$. In this sense they combine the features of
the Dirac index on non-compact space as in~\cite{Pope:1978zx,Pope:1981jx}, as well as those of non-compact superconformal field theories as in the cigar model.

The outline of this paper is as follows. In the second section we discuss the first ingredient in our construction, a gauged linear sigma model (GLSM) that provides the ultraviolet definition of a two-dimensional field theory that flows in the infrared to a conformal field theory corresponding to string propagation on Taub-NUT space. We also give the extension of this to multi-center Taub-NUT. The third section defines the equivariant elliptic genus in these theories and carries out the computation of the elliptic genus for ALE and ALF spaces using the technique of localization.
In the fourth section we discuss the large radius limit of our result and compare it to previous results in the literature.
In the fifth section
we discuss the interpretation of our results in terms of BPS states and wall-crossing.
We also discuss the decomposition of the Taub-NUT elliptic genus into its holomorphic and non-holomorphic pieces.
Taub-NUT and its generalizations often occur as the moduli space of various supersymmetric brane and monopole configurations and our results have a number of interesting applications to the counting of BPS states in these systems which we discuss in Section 6.  In the final section we conclude.
The appendices contain some technical details of the computation of the Jeffrey-Kirwan residue
that appears in the localization computation and some technical details of the decomposition of
the elliptic genus into discrete and continuum contributions.

\section{Gauged Linear Sigma Models}

Gauged linear sigma models (GLSM) have been extensively used as a tool for understanding
various aspects of the non-linear sigma models (NLSM) following \cite{Witten:1993yc}.
Using the supersymmetric localization technique, the exact S$^2$ \cite{Benini:2012ui,Doroud:2012xw},
D$_2$ \cite{Sugishita:2013jca, Honda:2013uca, Hori:2013ika} and RP$^2$ \cite{Kim:2013ola}
partition functions of  GLSMs have been computed recently. From these new exact observables,
there has been remarkable progresses in understanding  Calabi-Yau spaces
\cite{Jockers:2012dk,Gomis:2012wy,Gerchkovitz:2014gta},
D-branes and Orientifolds therein. In this section we define GLSMs describing NLSMs on ALF and ALE spaces
in the infrared limit. We will use them to compute the equivariant elliptic genera in the next section.

\subsection{ALF spaces}

The  $\CN=(4,4)$ gauged linear sigma model (GLSM) which flows in the infrared to the Taub-NUT conformal
field theory was discussed in~\cite{Tong:2002rq,Harvey:2005ab} following  the work in~\cite{Hori:2001ax,Hori:2002cd}. Here we generalize
this to a GLSM that flows to the $A_{N-1}$ multi-center Taub-NUT metric with $N$ coincident centers.
The action is constructed from an~$\CN=(4,4)$ vector multiplet, a charged hypermultiplet, and a
neutral chiral multiplet.

The $\CN=(4,4)$ vector multiplet decomposes into an $\CN=(2,2)$ vector multiplet $V=(A_\mu,\s,\l_\pm)$
and a neutral chiral multiplet $\Phi=(\phi,\tilde \l_\pm)$. In terms of components,
\begin{align}
  \Phi & = \phi + \sqrt2 \th^+ \tilde \l_+ + \sqrt2 \th^- \tilde \l_- + \cdots
  \nonumber \\
  \Sigma & = \s - i\sqrt2 \th^+ \bar \l_+ - i \sqrt2 \bar\th^- \l_-
  + \sqrt2 \th^+ \bar \th^- \left( F_{12}+ iD \right)
  + \cdots\ .
\end{align}
where $\Sigma = \bar D_+ D_- V/\sqrt{2}$. Our superspace conventions are as in \cite{Tong:2002rq,Harvey:2005ab}.
The charged hypermultiplet decomposes into two chiral multiplets $Q=(q,\psi)$ and $\tilde Q=(\tilde q, \tilde \psi)$
with electric charges $+1$ and $-1$. The component field expansions of these superfields read as
\begin{align}
  Q & = q + \sqrt2 \th^+ \psi_+ +  \sqrt2 \th^- \psi_- + \cdots
  \nonumber \\
  \tilde Q & = \tilde q +  \sqrt2 \th^+ \tilde \psi_+ +  \sqrt2 \th^- \tilde \psi_- + \cdots\ .
\end{align}
Finally, the neutral hypermultiplet decomposes into a chiral multiplet $\Psi=(r_1,r_2,\chi_\pm)$
and a St\"uckelberg chiral multiplet $\G=(r_3,\gamma,\tilde \chi_\pm)$. Their component expansions are
\begin{align}
  \Psi & = \frac{r_1 + i r_2 }{\sqrt2} + \sqrt2 \th^+ \chi_+ +\sqrt2 \th^- \chi_- +\cdots\ ,
  \nonumber \\
  \Gamma & = - \frac{r_3 - i g^{-2} \g }{\sqrt2} + i \sqrt2 g^{-2} \th^+ \bar{\tilde \chi}_+
  + i \sqrt2 g^{-2} \th^- \bar{\tilde \chi}_- + \cdots\ .
\end{align}
Here the St\"uckelberg field $\g$ transforms under the $U(1)$ gauge rotation
$A_\mu \to A_\mu + \partial_\mu \alpha(x)$ as
\begin{align}
  \g \ \to \g - N \a(x)\ ,
\end{align}
and is periodic with $2\pi$ periodicity, i.e., $\g\simeq \g+2\pi$.

The Lagrangian for these fields takes the following form
\begin{align}
  2\pi \CL = \frac{1}{e^2} \CL_\text{vec} + \CL_{Q,\tilde Q} + \frac{1}{g^2 N^2} \CL_{\Psi}
  + g^2 \CL_\text{St} + \CL_{\CW}
\end{align}
with
\begin{align}
  \CL_\text{vec} & = \int d^4\th \ \left[ \bar \Sigma \Sigma + \bar \Phi \Phi \right]
  \nonumber \\
  \CL_{Q,\tilde Q} & = \int d^4\th \ \left[ \bar Q e^{-2 V} Q
  + \bar{\tilde Q} e^{+2V} \tilde Q \right]
  \nonumber \\
  \CL_{\Psi} & = \int d^4 \th \ \left[ \bar \Psi \Psi \right]
  \nonumber \\
  \CL_{St} & = \int d^4\th \ \frac12 \left[ \Gamma + \bar \Gamma + \sqrt2 N V \right]^2
  \nonumber \\
  \CL_\CW & = \int d^2\th \ \left[ \sqrt 2 \tilde Q \Phi  Q - \Phi \Psi \right] + \text{c.c.}\ .
  \label{mTNLag}
\end{align}
%

%
%
%
%

One can show that the above Lagrangian is invariant under the R-symmetry group
$SU(2)\times SO(4)\simeq SU(2)_1\times SU(2)_2 \times SU(2)_3$.
We summarize the representations of various fields and supercharges $\CQ_-^{a\dot \a},\CQ_+^{a\a}$
below:
\begin{align}
  (\CQ_-^{a\dot \a},\CQ_+^{a\a}) \ : & \ \left( {\bf 2},{\bf 1},{\bf 2}\right)_- \oplus
  \left( {\bf 2},{\bf 2},{\bf 1}\right)_+
  \nonumber \\
  (\sigma, \phi) \  : & \ \left( {\bf 1},{\bf 2},{\bf 2} \right)
  \nonumber \\
  (\l, \tilde \l) \ : & \ \left( {\bf 2},{\bf 2},{\bf 1} \right)_- \oplus
  \left( {\bf 2}, {\bf 1}, {\bf 2} \right)_+
  \nonumber \\
  (q,\tilde q) \ : & \ \left({\bf 2},{\bf 1},{\bf 1} \right)
  \nonumber \\
  (\psi, \tilde \psi) \ : & \ \left( {\bf 1},{\bf 1},{\bf 2} \right)_- \oplus
  \left( {\bf 1},{\bf 2},{\bf 1} \right)_+
  \nonumber \\
  (r_1,r_2,r_3) \ : & \ \left( {\bf 3},{\bf 1},{\bf 1} \right)
  \nonumber \\
  \g \ : & \ \left( {\bf 1}, {\bf 1},{\bf 1} \right)
  \nonumber \\
  (\chi,\tilde\chi) \ : & \ \left( {\bf 2},{\bf 1},{\bf 2} \right)_- \oplus
  \left( {\bf 2},{\bf 2},{\bf 1} \right)_+ \ ,
\end{align}
where the subscripts $\pm$ denote the helicities while $a,\a,\dot \a$ are for doublets
under $SU(2)_1 \times SU(2)_2 \times SU(2)_3$. The Lagrangian also enjoys
a $U(1)_f$ flavor symmetry under which $q$ and $\tilde q$ carry  charges $+1$ and $-1$ respectively
and all other fields are
neutral. We emphasize that the $U(1)_f$ flavor symmetry is distinct from the $U(1)$ gauge
symmetry. This is because the St\"uckelberg field $\g$ is neutral under $U(1)_f$.

By solving the triplet of D-term equations
%
%
and the Gauss constraint in the limit $e^2 \to \infty$,
one can show as in \cite{Tong:2002rq,Harvey:2005ab} that
the GLSM flows in the infrared to the $\CN=(4,4)$ Taub-NUT sigma model
whose bosonic Lagrangian becomes
\begin{align} \label{TNmetric}
  \CL_\text{b} = - \frac{1}{2N} \bigg[
  H\big(r\big) \partial_\mu \vec r \cdot \partial_\mu \vec r +
  H\big(r\big)^{-1} \left( \partial_\mu \k + \frac{N}{2} \vec w \cdot \partial_\mu \vec r \right)^2 \bigg]
\end{align}
with
\begin{align}
  H\big(r\big) = \frac{1}{g^2 N } + \frac{N}{2|\vec r|} \,  ,
\end{align}
$\vec w$ a vector field obeying $\vec \nabla \times \vec w = \vec \nabla (1/r)$
%
%
%
and  $\k = \g + \arg{q}$ denoting a gauge-invariant angle variable whose range is
$[0,2\pi]$. The Taub-NUT space has $SU(2)_{\mathbb{R}^3} \times U(1)_\kappa$
isometry group, which in the GLSM is identified as the $SU(2)_1$ R-symmetry and $U(1)_f$ flavor group.
The size of the asymptotic circle of the Taub-NUT space is parameterized by $g^2$.

Note that the kinetic Lagrangian for the St\"uckelberg superfield $\G$ contains the two-fermion terms
\begin{align}
  \CL_\text{St} = \cdots + \l_- \bar{\tilde \chi}_+ + \bar{\tilde \l}_- \bar \chi_+ -
  \bar \l_- \tilde \chi_+ - \tilde \l_- \chi_+ \ ,.
  \label{twof}
\end{align}
These will play an important role in the following discussion.


%
%
%

\subsection{ALE spaces}

If we consider the large size limit $g^2 \to \infty$ of the multi Taub-NUT
spaces with $N$ coincident centers we obtain an ALE space with an A$_{N-1}$ singularity.
There are two different GLSMs that can be used to describe these spaces,
one of which originates from the toric geometry of $\mathbb{C}^2/\mathbb{Z}_N$ and the other from Kronheimer's hyperK\"ahler quotient.
We denote the former by GLSM I and the latter by GLSM II.

\paragraph{GLSM I}
The GLSM I can be described by an  $\CN=(2,2)$ gauge theory in two dimensions that flows to
the non-linear sigma model on $A_{N-1}$ space in the infrared with the enhanced $\CN=(4,4)$
supersymmetry. The theory contains
$U(1)^{N-1}$ vector multiplets $V_a$ and $N+1$ chiral multiplets $\Phi^i=(X_1, Y_a, X_2)$ ($a=1,2,..,N-1$)
of charges $Q_a^i$
\begin{align}
  Q_a^i = \begin{pmatrix}
  1& -2 & 1 & & &
  \\
  & 1 & -2 & 1 & &
  \\
  & &  &  \ddots & &
  \\
  & & & 1 & -2 & 1
  \end{pmatrix}\ .
\end{align}
The Lagrangian of the model is
\begin{align}
  2\pi \CL = \int d^4 \th \ \Big[ \frac{1}{e^2} \bar \Sigma \Sigma + \sum_{i} \bar \Phi_i e^{-2 Q_a^i V^a}
  \Phi^i \Big]\ .
\end{align}
The above Lagrangian is invariant under $U(1)_1\times U(1)_2\times U(1)_R$ global symmetry where
the flavor group $U(1)_1\times U(1)_2$ is a subgroup of $U(1)\times SU(2)$ isometry group of the A$_{N-1}$
space. We summarize the global charges of chiral multiplets below.
\begin{center}
\setlength{\tabcolsep}{13pt}{
\begin{tabular}{c|ccc}
  \hline
  & $X_1$ & $Y_a$ & $X_2$
  \\ \hline \hline
  $U(1)_1$ & $1$ & $0$ & $0$
  \\
  $U(1)_2$ & $0$ & $0$ & $1$
  \\
  $U(1)_R$ & $0$ & $0$ & $0$
\end{tabular}
}
\end{center}

\paragraph{GLSM II}
\begin{table}[t]
\begin{center}
  \begin{tabular}{c|c| p{8cm} }
    \hline
    Orbifold & Hypersurface & \hspace*{2.6cm} Quiver Diagram \\
    \hline
    \hline
    $A_k$ & $x^2 + y^2 + z^{k+1} = 0$ & \parbox[c]{1em}{\includegraphics[width=8cm]{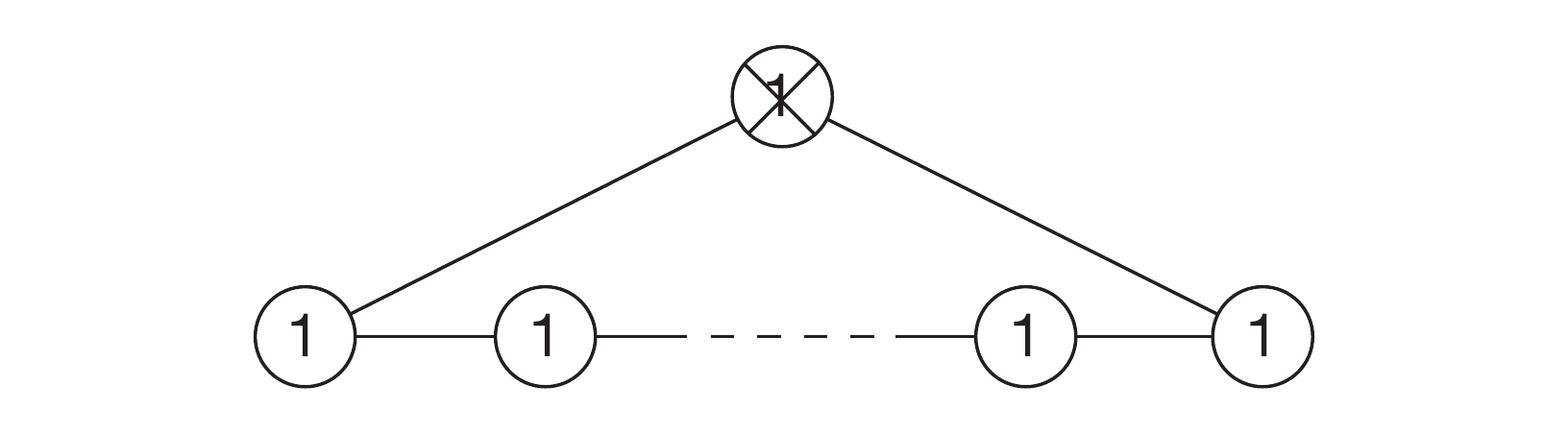}}
    \\
    $D_{k}$ & $x^2 + y^2 z+ z^{k-1} = 0$ & \parbox[c]{1em}{\includegraphics[width=8cm]{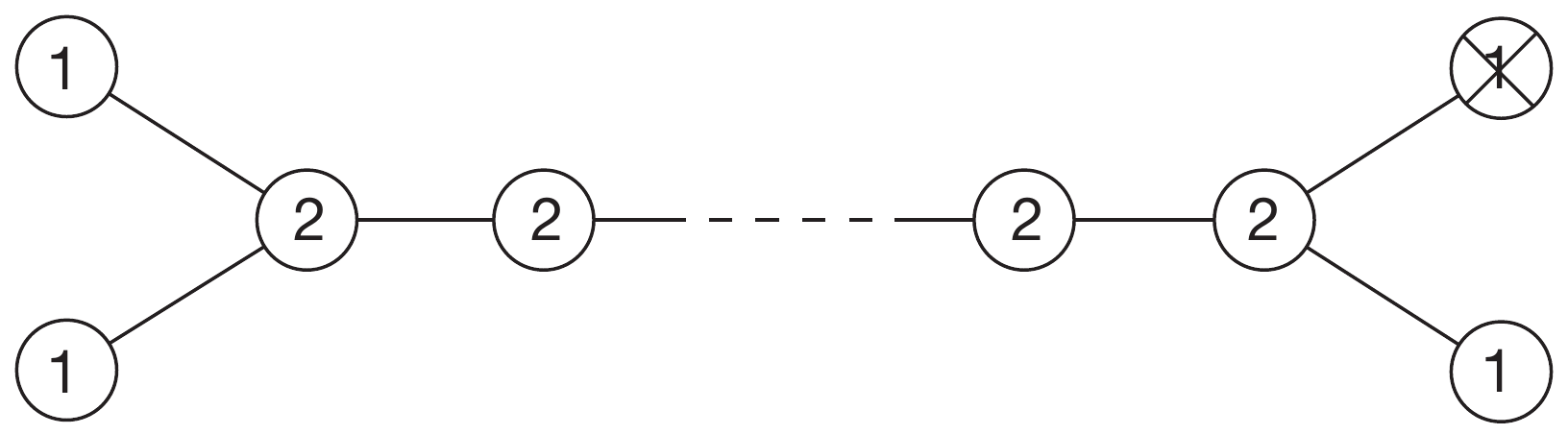}}
    \\
    $E_6$ & $x^2+y^3+z^4=0$ &  \parbox[c]{1em}{\includegraphics[width=8cm]{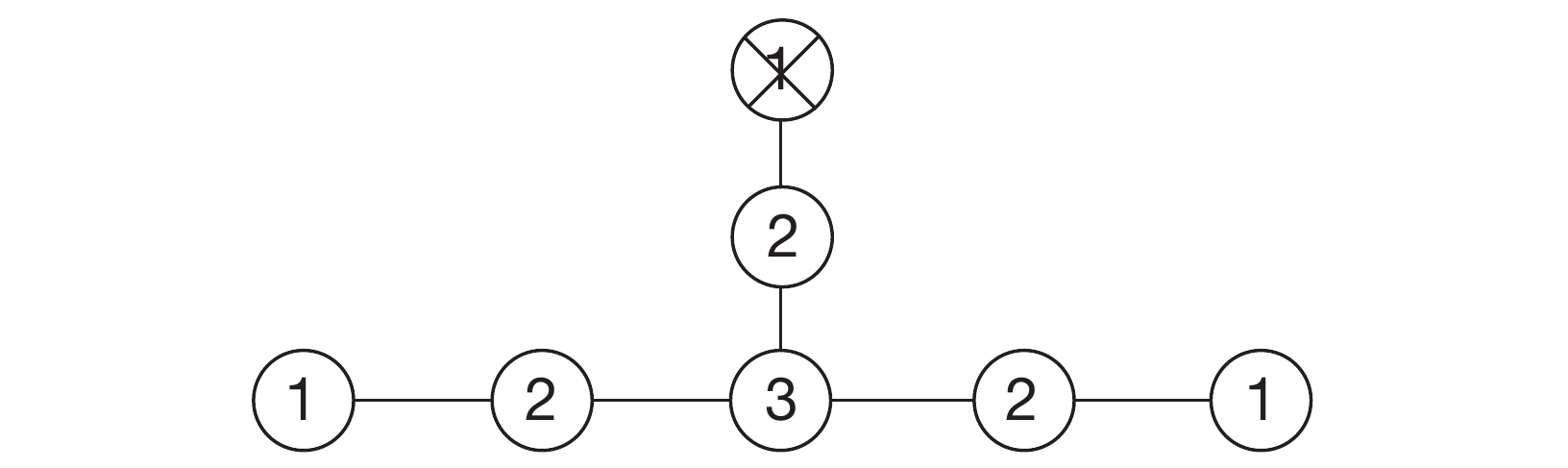}}
    \\
    $E_7$ & $x^2+y^3+ y z^3=0$ & \parbox[c]{1em}{\includegraphics[width=8cm]{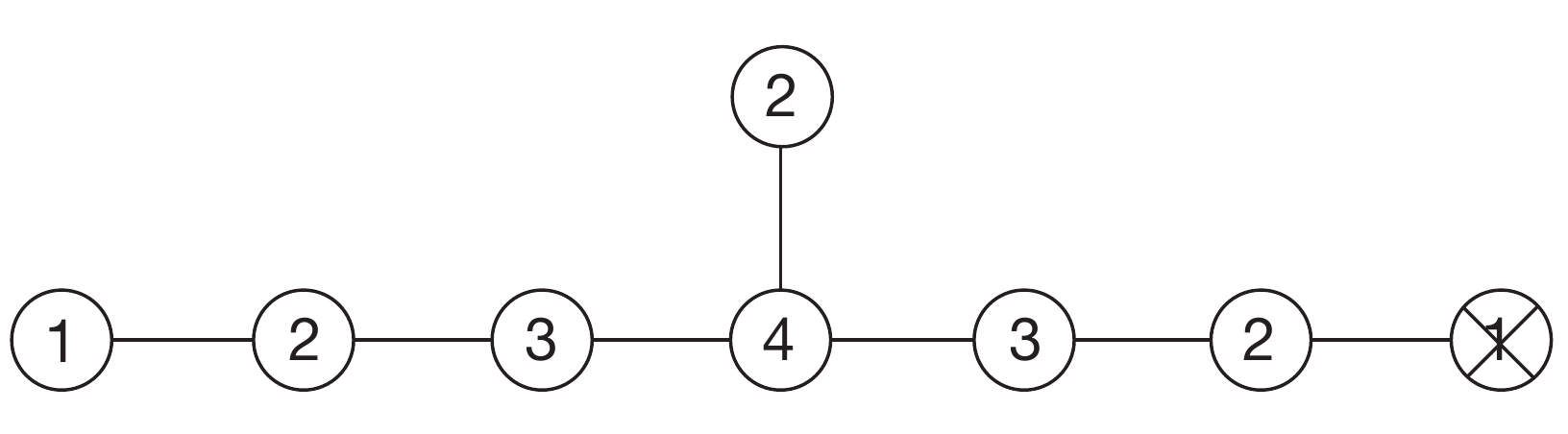}}
    \\
    $E_8$ & $x^2+y^3+ z^5=0$ & \parbox[c]{1em}{\includegraphics[width=8cm]{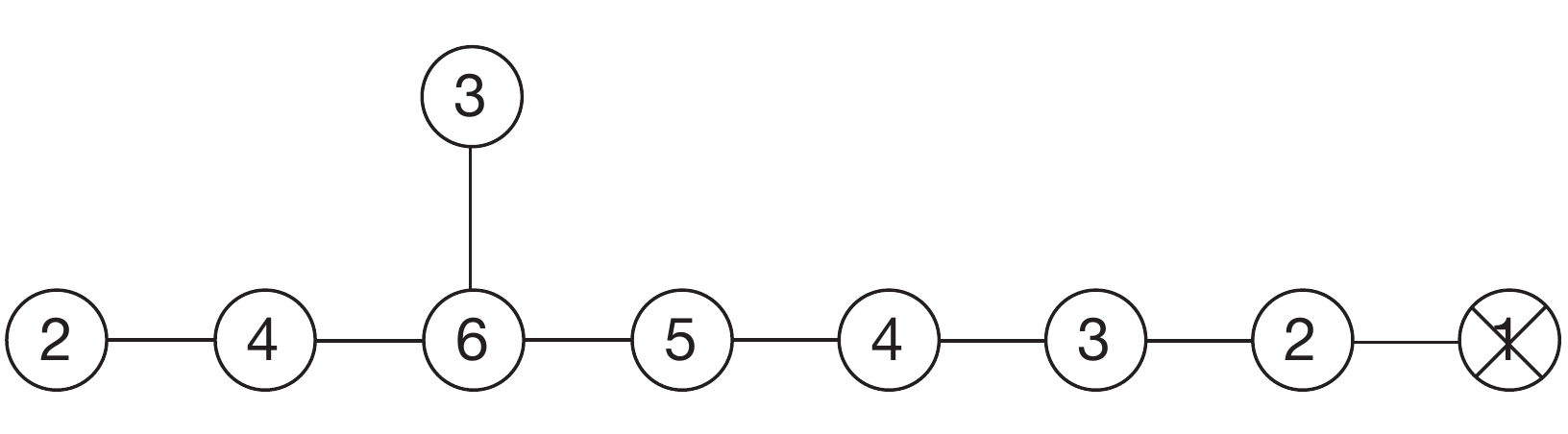}}
  \end{tabular}
\end{center}
\caption{The quiver diagrams of the ALE spaces with ADE singularities. Each circle with an
integer $n$ introduces a $U(n)$ vector multiplet and each line connecting two circles
represents a hypermultiplet in the bifundamental representation. For the symbol $\otimes$, we ungauge
the corresponding gauge group.}\label{(4,4)ALE}
\end{table}
Let us now in turn discuss the other GLSMs, denoted by GLSM II, that
describe the general ALE spaces $\mathbb{C}^2/\G_\text{ADE}$
via Kronheimer's hyperK\"ahler quotient \cite{Johnson:1996py}.
A GSLM II is in general a quiver gauge theory preserving the $\CN=(4,4)$ supersymmetry. Its gauge group
and matter content are read off from the quiver diagram summarized
in the table \ref{(4,4)ALE}. The GLSM II has both Coulomb and Higgs branch.

For instance, let us consider the quiver gauge theory for $\mathbb{C}^2/\mathbb{Z}_2$.
The GLSM consists of a two-dimensional $\CN=(4,4)$ $U(1)$ gauge theory, coupled to
two hypermultiplets $Q^i$ ($i=1,2$) of charge $+1$ and two hypermultiplets $\wt Q_i$ of charge~$-1$
\begin{align}
  2\pi \CL = \frac{1}{e^2} \CL_\text{vec} + \CL_{Q,\tilde Q}  + \CL_{\CW}
\end{align}
with
\begin{align}
  \CL_\text{vec} & = \int d^4\th \ \left[ \bar \Sigma \Sigma + \bar \Phi \Phi \right]
  \nonumber \\
  \CL_{Q,\tilde Q} & = \int d^4\th \ \sum_{i=1}^2 \left[  \bar Q_i e^{-2 V} Q^i
  + \bar{\tilde Q}^i e^{+2V} \tilde Q_i \right]
  \nonumber \\
  \CL_\CW & = \int d^2\th \ \left[ \sqrt 2 \tilde Q_i \Phi  Q^i  \right] + \text{c.c.}\ .
  \label{A1GLSMII}
\end{align}
The above Lagrangian has $SU(2)_1 \times SU(2)_2 \times SU(2)_3$
R-symmetry and $SU(2)_f$ flavor symmetry groups, under which the
vector multiplet scalars transform as $({\bf 1},{\bf 2},{\bf 2},{\bf 1})$ and
the hyper multiplet scalars transform as $({\bf 2},{\bf 1},{\bf 1},{\bf 2})$.
The four supercharges with the positive helicity transform in the $({\bf 2},{\bf 2},{\bf 1},{\bf 1})$,
while the other four supercharges with the negative helicity transform in the
$({\bf 2},{\bf 1},{\bf 2},{\bf 1})$.

One can show that the Higgs branch of the theory is the orbifold space $\mathbb{C}^2/\mathbb{Z}_2$.

Despite the lack of manifest $\CN=(4,4)$ SUSY in GLSM I and the worry about
the potential contribution from the extra Coulomb branch of both GLSM I and II,
we can show later in the present work
that the elliptic genera of two different GLSMs corresponding the A$_{N-1}$ space
perfectly agree. In fact one can verify that the contribution to the elliptic genus
from the Coulomb branch vanishes as in \cite{Harvey:2014zra}.

\section{Computation of Elliptic Genera}

In this section we compute the equivariant elliptic genus for the 2d GLSMs
described in the previous section. To define an elliptic genus,
we first choose two supercharges $\CQ_+$ and $\bar \CQ_+$ that generates
the $\CN=(0,2)$ supersymmetry,
\begin{align}
  \left\{ \CQ_+ , \bar \CQ_{+} \right\} = \bar L_0 \propto H + i P\ ,
\end{align}
where $+$ denotes the helicity. In the Hamiltonian formalism,
the elliptic genus is defined as
\begin{align}
  \CE(\t;z,\xi) = \text{Tr}_{\CH_\text{RR}} \Big[ (-1)^F \, q^{L_0} \, \bar q^{\bar L_0} \,
  e^{-2\pi i z Q_R} \, e^{-2\pi i  \xi \cdot  Q_f} \Big]\ ,
  \label{def0}
\end{align}
provided that the theory under study has a global symmetry
that commutes with the above particular choice of $\CN=(0,2)$ supersymmetry
inside the $\CN=(4,4)$ supersymmetry. We denote the left-moving
R-symmetry of $\CN=(4,4)$ superconformal symmetry  by~$Q_R$ and the rest
of the global symmetry by the vector~$Q_f$. The trace is performed over the Hilbert space in the
Ramond-Ramond sector.
The parameter $\t = \t_1 + i \t_2$ specifies the complex structure of a
torus $w\simeq w + 2\pi \simeq w + 2\pi \t$, and we have  $q = e^{2\pi i\t}$.

The elliptic genus (\ref{def}) can alternatively be viewed in terms of a path integral
on the torus  with complex coordinate $w$ subject to the identification $w\simeq w + 2\pi \simeq w+ 2\pi \t$. One can also describe the torus
as $w=\s_1 + \t \s_2$ where $\s_{1,2}$ are periodic with periodicity $2\pi$. We then need to specify
the boundary conditions for field variables in the ``space'' and ``time'' directions,
i.e., $\s_1$ and $\s_2$ directions. The Hilbert space in the Ramond sector implies
the untwisted boundary condition in the $\s_1$ direction.
The chemical potentials $\xi$ and $z$ leads to the boundary conditions in the time direction $\s_2$
twisted by the global charges. In summary, denoting the fields collectively by $\varphi(w,\bar w)$, the path-integral
is subject to the boundary conditions
\begin{align}
  \varphi(w+2\pi,\bar w + 2\pi) & = \varphi(w,\bar w) \ ,
  \nonumber \\
  \varphi(w+2\pi\t, \bar w + 2\pi \bar \t) & =
  e^{-2\pi i \xi \cdot Q_F[\Phi]}
  e^{-2\pi i z Q_R[\Phi]}
  \varphi(w,\bar w)\ .
  \label{twistedbd}
\end{align}

\subsection{ALF spaces}

Let us first focus on the Taub-NUT example (\ref{mTNLag}).
We choose two supercharges $\CQ^{++}_+$ and $\CQ^{--}_-$ generating the $\CN=(0,2)$ supersymmetry.
We write $\{Q_1,Q_2,Q_R\}$ for the Cartan generators of the R-symmetry group
$SU(2)_1\times SU(2)_2\times SU(2)_3$ and $Q_f$ for the $U(1)_f$ charge.
Note that the above $\CN=(0,2)$ supercharges are neutral under $Q_1-Q_2$, $Q_R$ and $Q_f$.
For later convenience,
we present the charges of various fields
in the table below.
\begin{table}[t]
\begin{center}
\setlength{\tabcolsep}{10pt}{
\begin{tabular}{c|ccc|c}
  \hline
  & $Q_1-Q_2$ & $Q_R$ & $Q_f$ & $\CN=(0,2)$ rep.
  \\ \hline \hline
  $q$ & $-1$ & $0$ & $+1$ & \multirow{2}{*}{chiral}
  \\
  $\psi_+$ & $-1$ & $0$ & $+1$ &
  \\ \hdashline
  $\tilde q$ & $-1$ & $0$ & $-1$& \multirow{2}{*}{chiral}
  \\
  $\tilde \psi_+$ & $-1$ & $0$ & $-1$ &
  \\ \hdashline
  $\psi_-$ & $0$ & $-1$ & $+1$ & fermi
  \\ \hdashline
  $\tilde \psi_-$ & $0$ & $-1$ & $-1$ & fermi
  \\ \hline
  $\bar \s$ & $-1$ & $+1$ & $0$ & \multirow{2}{*}{chiral}
  \\
  $\l_+$ & $-1$ & $+1$ & $0$ &
  \\ \hdashline
  $\phi$ & $+1$ & $+1$ & $0$ & \multirow{2}{*}{chiral}
  \\
  $\tilde \l_+$ & $+1$ & $+1$ & $0$ &
  \\ \hdashline
  $A_\mu$ & $0$ & $0$ & $0$ & \multirow{2}{*}{vector}
  \\
  $\l_-$ & $0$ & $0$ & $0$ &
  \\ \hdashline
  $\tilde \l_-$ & $+2$ & $0$ & $0$ & fermi
  \\ \hline
  $r_1,r_2$ & $-2$ & $0$ & $0$ & \multirow{2}{*}{chiral}
  \\
  $\chi_+$ & $-2$ & $0$ & $0$ &
  \\ \hdashline
  $r_3,\g$ & $0$ & $0$ & $0$ & \multirow{2}{*}{chiral}
  \\
  $\tilde \chi_+$ & $0$ & $0$ & $0$ &
  \\ \hdashline
  $\chi_-$ & $-1$ & $-1$ & $0$ & fermi
  \\ \hdashline
  $\tilde \chi_-$ & $-1$ & $+1$ & $0$ & fermi
  \\ \hline
\end{tabular}
}
\end{center}
\end{table}
The equivariant elliptic genus can be defined as follows
\begin{align}
  \CE(\t;\xi_1,\xi_2,z) = \text{Tr}_{\CH_\text{RR}} \Big[ (-1)^F q^{L_0} \bar q^{\bar L_0}
  e^{-2\pi i z Q_R}  e^{-2\pi i \xi_1 Q_f} e^{-2\pi i \xi_2 (Q_1-Q_2)}
   \Big]\ ,
  \label{def}\ .
\end{align}
It is noteworthy that two fermions $\l_-$ and $\tilde \chi_+$ are neutral
under $Q_f$, $Q_1-Q_2$ and $Q_R$, which leads to potentially dangerous fermionic zero modes.

\paragraph{Localization}

Using the localization method, we evaluate the path-integral on the torus exactly.
We start with the Taub-NUT example with a single center $N=1$.
Note first that the Lagrangian terms $\CL_\text{vec}$, $\CL_{Q,\tilde Q}$ and $\CL_{\Psi}$
are all $\CQ$-exact up to total derivatives. Therefore one can scale them up to infinity
leaving the final result unchanged. However, the Lagrangian term
for the St\"uckelberg field $\CL_\text{St}$ is not globally $\CQ$-exact \cite{Hori:2001ax}, and
we need to perform the path-integral in that sector exactly.


In other words, we compute the elliptic genus by taking a partial weak-coupling limit
of the theory, i.e.,
\begin{align}
  \CE(\t; \xi_1, \xi_2, z) = \int \CD \Phi \ e^{-t (S_\text{vec}+S_{Q,\tilde Q} + S_{\Psi}) - g^2 S_\text{St} - S_\CW}
  \label{defE1}
\end{align}
in the limit $t \to \infty$. The path-integral is subject to the twisted boundary conditions
(\ref{twistedbd}). Equivalently, one can also consider a path-integral with periodic boundary conditions
in both $\s_1$ and $\s_2$ directions by turning on suitable background gauge fields coupled to
$U(1)_f$ and R-symmetry currents
\begin{align}
  A^f & = \frac{\xi_1}{2i\t_2} dw - \frac{\xi_1}{2i\t_2} d\bar w\ ,
  \nonumber \\
  A^{Q_1-Q_2} & = \frac{\xi_2}{2i \t_2} dw - \frac{\xi_2}{2 i\t_2}d\bar w\ ,
  \nonumber \\
  A^{Q_R} & = \frac{z}{2i \t_2} dw - \frac{z}{2 i\t_2}d\bar w\ .
\end{align}
For instance, in the path-integral representation with periodic boundary conditions,
the covariant derivative  now takes the following form
\begin{align}
  D_w  = \partial_w  - i A_w -  \frac{\xi_1 Q_f + \xi_2 (Q_1-Q_2) + z Q_R}{2\pi \t_2} \ .
\end{align}
In what follows, we use the second representation of the index with periodic boundary conditions.

In the limit $t \to \infty$, the path-integral (\ref{defE1}) localizes to  supersymmetric
field configurations that minimize the positive definite Euclidean action terms
$S_\text{vec}+S_{Q,\tilde Q}+S_{\Psi}$.
One can show \cite{Benini:2013nda}  that the localization locus is characterized by the flat connections
on the torus
\begin{align}
  A = \frac{\bar u}{2i\tau_2}dw - \frac{u}{2i \t_2} d\bar w\ , \qquad
  u = \frac{1}{2\pi}\oint_{\a} A - \frac{\t}{2\pi}\oint_{\b} A\ ,
\end{align}
where $\a$ and $\b$ denote ``temporal'' ($\s_2$) and ``spatial'' ($\s_1$) cycles. All other fields
in the superfields $V$, $Q$, $\tilde Q$ and $\Psi$  vanish at the localization loci.
Due to the large gauge transformation, the Wilson lines around the torus should take values in $T^2$,
i.e, $u\in E(\t) = \mathbb{C}/(\mathbb{Z}+\t\mathbb{Z})$ ($u\simeq u+1 \simeq u+\t$).
We emphasize here that the path-integral over the St\"uckelberg field and its super partners
is not localized.

We then expand the Lagrangian terms $\CL_\text{vec}+\CL_{Q,\tilde Q}+\CL_{\Psi}$ around
the saddle points with care given to (fermionic) zero modes.
We have
\begin{align}
  t \CL_\text{vec} + t \CL_\text{ghost}  = &- \frac12 A'_\mu \partial^2 A'_\mu
  - 4 \bar \s D_{\bar w} D_w \s - 4 \bar \phi D_{\bar w} D_w \phi
  + 2i \bar \l_- \partial_{\bar w } \l_- - 2i \bar \l_+ D_w \l_+
  \nonumber \\
  & + 2i \bar{\tilde \l}_- D_{\bar w } \tilde \l_- - 2i \bar{\tilde \l}_+ D_w \tilde \l_+
  + \bar c' \partial^2 c'  + \frac12 b^2  + \CO(\frac{1}{\sqrt t})\ ,
  \\
  t \CL_{Q,\tilde Q}  = & - 4 \bar q  D_{\bar w} D_w q - 4 \bar{\tilde q} D_{\bar w}D_w \tilde q
  - 2 i \bar \psi_+ D_w \psi_+
  + 2 i \bar \psi_- D_{\bar w} \psi_-
  - 2i \bar{\tilde \psi}_+ D_w \tilde \psi_+
  \nonumber \\
  & + 2i \bar{\tilde \psi}_- D_{\bar w} \tilde \psi_-
  + i \sqrt2  \bar \psi_+ \bar \l^0_- q - i \sqrt2 \bar q \l^0_- \psi_+
  - i \sqrt2  \bar \psi_+ \bar \l^0_- \tilde q + i \sqrt2 \bar{\tilde q} \l^0_- \psi_+
  \nonumber \\
  & + \CO(\frac{1}{\sqrt t})\ ,
  \nonumber \\
  t \CL_\Psi  = & - 2 (r_1 - i r_2) D_{\bar w} D_w (r_1 + ir_2) - 2 i \bar \chi_+ D_w \chi_+
  + 2i \bar \chi_- D_{\bar w} \chi_- + \CO(\frac{1}{\sqrt t}) \ ,
  \nonumber
\end{align}
%
%
where $A_m'$ and $c'$ denote fluctuations with zero-modes
removed\footnote{The supersymmetric ghost terms $\CL_\text{ghost}$ contains various Lagrange multipliers
that remove the zero modes of the ghost and vector fields. For details see \cite{Murthy:2013mya,Ashok:2013pya}.}
while $\l_-^0$ and $\bar \l^0_-$ denote the zero-modes in gaugino fields.
The massive fluctuations are rescaled by $1/{\sqrt t}$ so that their
kinetic terms are canonically normalized
(e.g. $q \to \frac{1}{\sqrt t} q$)\footnote{One can show that the integration
measure is invariant under this rescaling.}. Near the saddle points,
the Lagrangian for the St\"uckelberg multiplet $\G$ can be expressed as
\begin{align}
  g^2 \CL_\text{St}  = & \frac{2}{g^2} \left| \partial_w r_3 \right|^2
  + 2g^2 \left| \partial_w \g + \frac{\bar u}{2i\t_2} \right|^2
  - \frac{2i}{g^2} \bar{\tilde \chi}_+ D_w \tilde \chi + \frac{2i}{g^2} \bar{\tilde \chi}_-
  D_{\bar w} \tilde\chi_-
  \nonumber \\ &
  + \l_-^0 \bar{\tilde \chi}_+^0 - \bar \l_-^0 \tilde \chi_+^0
  + i r_3^0  D^0 + \CO(1/{\sqrt t})\ .
\end{align}
After rescaling the massive fluctuations the F-terms in the action coming from $\CL_\CW$ are all suppressed by factors of $1/t$ and can be
dropped at large $t$.
Again $\tilde \chi_+^0$ and $\bar{\tilde \chi}_+^0$ denote potentially dangerous fermionic zero-modes, and
the zero-mode of the neutral field $r_3$ plays the role of Lagrangian multiplier that kills the
zero-mode of the auxiliary field $D$.

Collecting all the results, one can show that the path-integral reduces
to the following expression,
%
%
\begin{align} \label{elpip}
  \CE(\t;\xi_1,\xi_2,z) & = g^2\int_{E(\t)} \frac{dud\bar u}{\t_2} \int \CD\varphi_m
  \int \CD \tilde \chi_+^0 \CD \bar{\tilde \chi}_+^0 \CD \l^0_- \CD \bar \l^0_-
  \ e^{ - S_\text{quad}(\varphi_m) }
  \nonumber  \\ & \hspace*{1cm} \times e^{\int d^2 \sigma \left(-  \l_-^0 \bar{\tilde \chi}^0_+ +  \bar \l_-^0 \tilde \chi^0_+
  + i \sqrt2  \bar \psi_+ \bar \l^0_- q - i \sqrt2 \bar q \l^0_- \psi_+
  - i \sqrt2  \bar \psi_+ \bar \l^0_- \tilde q + i \sqrt2 \bar{\tilde q} \l^0_- \psi_+ + \CO(1/{\sqrt t}) \right) }\ ,
\end{align}
where $\varphi_m$ denotes massive fluctuation fields collectively. Note that the two-fermion terms
$\l_-^0 \bar{\tilde \chi}^0_+$ and $\bar \l_-^0 \tilde \chi^0_+$ saturate the fermionic zero-modes while the other two fermion terms
in (\ref{twof}) do not have zero modes and as mentioned earlier are suppressed at large $t$. The cubic terms in the exponent of
(\ref{elpip}) only involve the $\lambda^0_-, \bar \lambda_-^0$ zero mode and thus do not contribute after saturating the zero mode
integral with the quadratic zero mode terms.
We are thus finally left with computing the one-loop determinants of massive fluctuations.

To compute the one-loop determinant, it is useful to expand the fluctuation fields
$\varphi_m$ in terms of Fourier modes,
\begin{align}
  \varphi_m(w,\bar w) = \sum_{(m,n)\in\mathbb{Z}^2} c_{m,n} e^{i m \s_1-  i n\s_2}
  =  \sum_{(m,n)\in\mathbb{Z}^2} c_{m,n} e^{- \frac{n+\bar \t m}{2\t_2}w
  + \frac{n+\t m}{2\t_2}\bar w }\ ,
\end{align}
which satisfy the periodic boundary conditions in both $\s_1$ and $\s_2$ directions.
For the St\"uckelberg field $\g$\footnote{Unlike the GLSM flowing to the cigar CFT,
the R-symmetry currents in the present model are not anomalous. Therefore, the St\"uckelberg
field $\g$ carries no R-charges.}, one also has to consider momentum and winding modes
\begin{align}
  \g(w,\bar w) & = \sum_{(p,\w)\in \mathbb{Z}^2} \big( \w \s_1 - p \s_2 \big) + \sum_{(m,n)\neq (0,0)}
  c_{m,n} e^{i m \s_1-  i n\s_2}
  \nonumber \\ & = \sum_{(p,\w)\in \mathbb{Z}^2} i \big( \frac{p+\bar \t \w}{2\t_2}w
  - \frac{p+\t \w}{2\t_2}\bar w \big) + \sum_{(m,n)\neq (0,0)} c_{m,n} e^{- \frac{n+\bar \t m}{2\t_2}w
  + \frac{n+\t m}{2\t_2}\bar w }\ .
\end{align}
One can show \cite{Benini:2013nda,Benini:2013xpa,Gadde:2013dda}  that the one-loop determinants from various supermultiplets are given by
\begin{align}
  Z^\text{1-loop}_{V,\Phi} & = \prod_{(m,n)\in\mathbb{Z}^2}
  \frac{n + \t m + 2\xi_2}{(n + \t m - \xi_2+z) (n+\t m + \xi_2 + z)}
  \times \prod_{(m,n)\neq (0,0)} (n + m\t) \ ,
  \nonumber \\
  Z^\text{1-loop}_{Q,\tilde Q} & = \prod_{(m,n)\in\mathbb{Z}^2}
  \frac{(n + \t m + u + \xi_1 - z)(n + \t m - u - \xi_1 - z)}{(n + \t m + u +\xi_1 - \xi_2) (n+\t m - u - \xi_1 - \xi_2)}
  \\
  Z^\text{1-loop}_{\Psi,\G} & = \prod_{(m,n)\in\mathbb{Z}^2}
  \frac{(n+\t m -\xi_2-z)(n+\t m -\xi_2+z)}{n + \t m - 2 \xi_2} \times
  \prod_{(m,n)\neq (0,0)} \frac{n+m\bar \t}{|n + m\t|^2}
  \nonumber \\ &
  \times \sum_{(p,\w)\in\mathbb{Z}^2} e^{-\frac{g^2\pi}{\t_2}\big| u + p + \t\w\big|^2}\ .
  \nonumber
\end{align}
In total, up to an overall sign, one obtains
\begin{align}
  Z^\text{1-loop} & = \prod_{(m,n)\in\mathbb{Z}^2}
  \frac{(n + \t m + u +\xi_1 - z)(n + \t m - u -\xi_1 - z)}{(n + \t m + u +\xi_1 - \xi_2) (n+\t m - u -\xi_1- \xi_2)}
  \times \sum_{(p,\w)\in\mathbb{Z}^2} e^{-\frac{g^2\pi}{\t_2}\big| u + p + \t\w\big|^2}\ ,
  \nonumber \\ & = \
  \frac{\myth1(\t,u+\xi_1+z) \, \myth1(\t,u+\xi_1-z)}{\myth1(\t,u+\xi_1+\xi_2) \, \myth1(\t,u+\xi_1-\xi_2)}
  \sum_{(p,\w)\in\mathbb{Z}^2} e^{-\frac{g^2\pi}{\t_2}\big| u + p + \t\w\big|^2}\ ,
\end{align}
where, for the last equality, we used the regularization scheme in \cite{Benini:2013nda} that matches the
Hamiltonian computation. Note that, except the momentum and winding modes,
the contributions from the $\CN=(4,4)$ vector multiplet $V,\Phi$ and the hypermultiplet $\G,\Psi$
exactly cancel each other. This is consistent with the $\CN=(4,4)$ super Higgs mechanism.
The theta function $\vth_{1}(\t,z)$ appearing in the above expression is the odd Jacobi theta function,
we present its definition and some properties in Appendix~\S\ref{JacApp}.

To summarize, the elliptic genus of the Taub-NUT CFT is given by
\begin{align}
  \CE(\t;\xi_1,\xi_2,z) = g^2 \int_{E(\t)} \frac{dud\bar u}{\t_2}
  \  \frac{\myth1(\t,u+\xi_1+z) \, \myth1(\t,u+\xi_1-z)}{\myth1(\t,u+\xi_1+\xi_2) \, \myth1(\t,u+\xi_1-\xi_2)}
  \sum_{(p,\w)\in\mathbb{Z}^2}
e^{-\frac{g^2\pi}{\t_2}\big| u + p + \t\w\big|^2}\ .
  \label{result1}
\end{align}

It is straightforward to generalize the above computation to
the GLSM (\ref{mTNLag}) for the multi-center $A_{N-1}$ Taub-NUT space, which leads to
\begin{align}
  \CE_N(\t; \xi_1,\xi_2,z) & = \frac{g^2}{N} \int_{E(\t)} \frac{dud\bar u}{\t_2}
\  \frac{\myth1(\t,u+\xi_1+z)\, \myth1(\t,u+\xi_1-z)}{\myth1(\t,u+\xi_1+\xi_2)\, \myth1(\t,u+\xi_1-\xi_2)}
   \sum_{(p,\w)\in\mathbb{Z}^2} e^{-\frac{g^2\pi}{\t_2}\big|u + \frac{p + \t\w}{N}\big|^2}\, .
  \label{resultmulti}
\end{align}

\paragraph{Modularity, ellipticity, and holomorphy. \label{holprop}}
We now make some comments about the mathematical structure of the single-center Taub-NUT\footnote{Similar comments
hold for the multi-center $A_{N-1}$ Taub-NUT space.}
index~\eqref{result1},
which we write as:
\be \label{defE}
  \CE(\t, z, \xi_1, \xi_2)  \=  \int_{E(\t)} \frac{du \, d\bar u}{\t_2} \,
 \v(\t,z, u+\xi_{1},\xi_2) \, H_{g}(\t,u) \, ,
\ee
where
\be \label{defphi}
\v(\t,z, u,\xi_2) \= \frac{\myth1(\t,u+z)\, \myth1(\t,u-z)}{\myth1(\t,u+\xi_2) \,
  \myth1(\t,u-\xi_2)} \, ,
\ee
and
\be \label{defHg}
H_{g}(\t,u)  \= g^2 \sum_{n,w \in\IZ} e^{-\frac{g^2\pi}{\t_2}\big| u + n + \t w\big|^2} \, .
\ee
The pair~$(n,w)$ in the integrand of~\eqref{defHg} transforms into~$(n,w+1)$ under~$u \to u+\t$, and
into~$(n+1,w)$ under~$u \to u+1$. Since we are summing over all integer values of~$(n,w)$
the function~$H_{g}(\t,u)$ is invariant under the elliptic transformations~$u \to u+\IZ\t+ \IZ$.
The measure is clearly invariant under these transformations.
Using the elliptic property~\eqref{elliptic} of the Jacobi theta functions, we can check that the function~$\v$ is also invariant
under these transformations.
The integral over~$E(\t)$ is therefore independent of the coset representative.

Furthermore, we can exchange the sum over $(n,w) \in \IZ^{2}$ in the integrand with a sum over the
integration region, thus effectively unfolding it to the whole complex plane.
We thus obtain the equivalent representation:
\begin{align}
  \CE(\t;\xi_1,\xi_2,z) = g^2 \int_{\IC} \frac{dud\bar u}{\t_2}
  \  \frac{\myth1(\t,u+\xi_1+z)\, \myth1(\t,u+\xi_1-z)}{\myth1(\t,u+\xi_1+\xi_2)\, \myth1(\t,u+\xi_1-\xi_2)}
 \; e^{-\frac{g^2\pi}{\t_2}|u|^2}\ .
  \label{result1again}
\end{align}

To understand the modular properties of~$\CE$, we consider how the various pieces behave
under modular transformations:
\begin{align}
  \t \to \frac{a\t+b}{c\t +d}, \qquad z \to \frac{z}{c\t+d}, \qquad \xi_{1,2} \to \frac{\xi_{1,2}}{c\t +d}
\end{align}
with $a,b,c,d\in \mathbb{Z}$ and $ad-bc = 1$. We first make a change of variables~$u = u'/(c\tau+d)$.
The form of the measure is invariant under the combined transformations.
The integration region in~\eqref{result1again} does not change.
The exponential factor is also invariant under these combined transformations.
The function~$\v$ has weight 0 under these modular transformations, as can be checked using
the modular transformations of the Jacobi theta functions.
The elliptic genus therefore has weight 0 under modular transformations.

The function~$\CE(\t,z,\xi_1,\xi_2)$ has good transformation properties under independent translations of
$z, \xi_{1}, \xi_{2}$ by $\IZ \t + \IZ$ (the \emph{elliptic transformations}).
It has index~$m=1$ under shifts of $z$, index~$m=0$ for shifts of~$\xi_{1}$, and index~$m=-1$ under shifts
of~$\xi_{2}$, meaning that it obeys:
\be\label{elliptic}
\CE(\t, \xi+\l\tau+\mu)\= e^{-2\pi i m(\l^2 \t + 2 \l \xi)} \CE(\t, \xi)  \qquad \forall \quad \l,\,\mu \in \IZ \, ,
\ee
where~$\xi$ denotes any one of the three chemical potentials~$z,\xi_{1}, \xi_{2}$, and the above equation holds
for fixed values of the other two chemical potentials.

The function~$\CE$ is non-holomorphic in~$\tau$ in that
it explicitly depends on the anti-holomorphic variable~$\bar \tau$.
To extract the non-holomorphic dependence on~$\tau$, we use the same method as in~\cite{Harvey:2013mda}.
Writing~$u=a\t+b$ where~$a,b \in [0,1]$, the measure $\frac{du \, d\bar u}{\t_2} =2 da\,db$ is independent
of~$\tbar$. The function~$H_{g}(\t,u=a\t+b)$ obeys the heat
equation~$\p_{\tbar} H_{g}(\t,a\t+b) = \frac{i}{2\pi g^{2}} \, \p_{\ubar}^{2} H_{g}(\t,u) |_{u=a\t+b}$.
Using these expressions, we can express the~$\tbar$-derivative of the elliptic genus as a contour integral:
\bea
\p_{\tbar} \, \CE(\t,z,\xi_1,\xi_2)
  & = & \frac{i}{2\pi g^{2}} \, \int_{E^{\varepsilon}(\t)} \frac{du \, d\bar u}{\t_2} \,
 \v(\t,z, u+\xi_{1},\xi_2)\, \p_{\ubar}^{2} \, H_{g}(\t,u) \, \cr
  & = & \frac{i}{2\pi} \, \int_{E^{\varepsilon}(\t)} du \, d\bar u \, \p_{\ubar}
\Big( \frac{1}{\t_2}  \v(\t,z, u+\xi_{1},\xi_2) \, \p_{\ubar} \, H_{g}(\t,u) \Big) \,  \cr
  & = & \frac{i}{2\pi} \, \oint_{\p E^{\varepsilon}(\t)} du  \,
  \frac{1}{\t_2} \, \v(\t,z, u+\xi_{1},\xi_2)\, \p_{\ubar} \, H_{g}(\t,u)  \,  .
 \label{dtaubarE1}
\eea
Here we have used the meromorphicity of~$\v$ to obtain the second line,
and Stokes's theorem to obtain the third line. We then compute the contour integral
using the residue theorem:
\bea
\p_{\tbar} \, \CE(\t,z,\xi_1,\xi_2)
 & = & -\frac{1}{\t_2 \, g^{2}} \big( {\rm Res}_{u=\xi_{2}-\xi_{1}} + {\rm Res}_{u=-\xi_{2}-\xi_{1}} \big) \,
 \v(\t,z, u+\xi_{1},\xi_2)\, \p_{\ubar} \, H_{g}(\t,u)  \,
  \cr
\label{dtaubarE}
& = & -\frac{1}{\pi \t_2 \, g^{2}}
  \frac{\myth1(\t,\xi_{2}+z)\, \myth1(\t,\xi_{2}-z)}{\myth1(\t,2\xi_2) \,
  \eta(\t)^{3} }\, \p_{\ubar} \, H_{g}(\t,u) \mid_{u=\xi_{2}-\xi_{1}}  \,  \\
& = &
  \frac{\myth1(\t,z+\xi_{2})\, \myth1(\t,z-\xi_{2})}{\myth1(\t,2\xi_2) \,
  \eta(\t)^{3} }\,  \frac{g^{2}}{\t_2^{2}} \sum_{n,w \in\IZ} (n + \t w) \, e^{-\frac{g^2\pi}{\t_2}\big|
  \xi_{2} - \xi_{1} + n + \t w\big|^2} \, \nonumber .
\eea

We see from Equation~\eqref{dtaubarE} that in the large radius $g^{2} \to \infty$ limit, the right-hand side,
and consequently the anti-holomorphic dependence, vanishes for generic values of~$(\t,\xi_{1}, \xi_{2})$.
In \S\ref{Rtoinfty} we shall compute the large-radius limit of the elliptic genus by a saddle point method
and verify that it is holomorphic in~$\t$.
We shall also comment on the non-holomorphic dependence of the elliptic genus in the following sections.

The equation~\eqref{dtaubarE} is similar to that obeyed by a (mixed) mock Jacobi form~\cite{Dabholkar:2012nd}, but it is not
quite that. The anti-holomorphic
derivative of~$\CE$ is a multiple of a meromorphic function of~$(\t,u,z,\xi_{2})$ (the ratio of the theta functions in the last line above),
and a non-holomorphic function of~$(\t,\xi_{1},\xi_{2})$ (the sum over $(n,w)$).
This latter function~$\p_{\ubar} \, H_{g}(\t,u) \mid_{u=\xi_{2}-\xi_{1}}$ is not factorizable into a
sum of holomorphic and anti-holomorphic modular/Jacobi forms as in the most general definition of the (mixed)
mock modular form. When~$g^{2}$ hits an integer, then the expression does factorize.

Finally, we briefly comment on the nature of the dependence of the elliptic genus on the equivariant chemical
potential~$\xi_{2}$ that we introduced in~\eqref{def0}. We note that there is a divergence as $\xi_{2} = \ve \to 0 $:
\be
\p_{\tbar} \, \CE(\t,z,\xi_{1},\xi_{2} =\ve) \= \frac{1}{\ve} \, \frac{g^2}{4 \pi \t_2^{2}}
  \frac{\myth1(\t,z)^{2}}{\eta(\t)^{6} }\,
  \sum_{n,w \in\IZ} (n + \t w) \, e^{-\frac{g^2\pi}{\t_2}\big| -\xi_{1} + n + \t w\big|^2} \, .
\ee
This same divergence will be present even after integrating back $\tbar$.
This means that the Taub-NUT elliptic genus is ill-defined in the absence of the chemical potential~$\xi_{2}$.
We will discuss the physical reason for this in \S\ref{nonhol}.

\subsection{ALE spaces}

We now turn to an evaluation of the equivariant elliptic genus for ALE spaces where we can rely heavily on previous results in the literature.
Benini {\it et al.} computed the elliptic genus of a certain class of the
two-dimensional SUSY gauge theories preserving at least $\CN=(0,2)$ supersymmetry in \cite{Benini:2013nda,Benini:2013xpa} (see also \cite{Gadde:2013dda}).
For the theories they studied  one can show that all terms in the Lagrangian
are Q-exact. Taking the Q-exact kinetic terms for the vector and chiral multiplets as
the path-integral weight, one can evaluate the elliptic genus using the
SUSY localization technique. The computation is quite parallel to that presented in
section 3.1 except that the gaugino zero modes $\l_+$ and $\bar \l_+$ are saturated
via Yukawa couplings rather than two-fermion couplings (\ref{twof}).
At the end, they obtain the following formula
\begin{align}
  \CE(\t; z, \xi) = \frac{1}{|W|}\sum_{u_\ast \in \mathfrak{M}_\eta} \JKres_{u_\ast}(Q_\ast,\eta)
  Z^\text{1-loop}(u)
\end{align}
where `JK-Res' denotes the so-called Jeffery-Kirwan residue operation and is
explained briefly in Appendix \S\ref{AppJK}.
We refer the reader to \cite{Benini:2013xpa} for more details.
Here $|W|$ denotes the order of the Weyl group.

Note that the GLSMs for the ALE spaces fit into the above class of SUSY gauge theories.

\paragraph{GLSM I} Let us first consider the GLSM I which flows in the infrared to
the NLSM on the A$_{N-1}$ singularity.
Using the result in \cite{Benini:2013xpa}, one can obtain the
elliptic genus of the A$_{N-1}$ space as
\begin{align}
  \CE(\t, z, \xi_1,\xi_2) = \sum_{u_\ast \in \mathfrak{M}_\eta} \JKres_{u_\ast}(Q_\ast,\eta)
  Z^\text{1-loop}(u)\ ,
  \label{formula1}
\end{align}
where
\begin{align}
  Z^\text{1-loop}(u) & = \left( \frac{i\eta(q)^3}{\myth1(\t,-z)} \right)^{N-1}
  \frac{\myth1(\t, u_1 - \z_R  + \z_1 )}{\myth1(\t,u_1+\z_1)}
  \frac{\myth1(\t, u_{N-1} - \z_R + \z_2 )}{\myth1(\t,u_{N-1} + \z_2 )}
  \nonumber \\ & \times
  \prod_{a=1}^{N-1} \frac{\myth1(\t,-\z_R - C_{ab} u_b )}{\myth1(\t,-C_{ab}u_b)}\ .
\end{align}
The two parameters $\z_1$ and $\z_2$ are chemical potentials
for the flavor charges $U(1)_1\times U(1)_2$ while $\z_R$ is the chemical
potential for the left-moving U(1)$_R$ charge.
The matrix $C_{ab}$ is the Cartan matrix
of $SU(N)$.

As a warm-up exercise, let us consider first the simplest example $\mathbb{C}^2/\mathbb{Z}_2$.
Choosing the parameter $\eta$ required in the definition of the Jeffrey-Kirwan residue so that $\eta<0$, one can show that the elliptic genus of the A$_1$ space can be written as
\begin{align}
  \CE(\t;z, \z_1,\z_2) &  = -\sum_{a,b=0,1}  \Res_{u=\frac{a+b\t}{2}}
  Z^\text{1-loop}(u)
  \nonumber \\ & =
  \frac12 \sum_{a,b=0,1} \prod_{i=1}^{2}
  \frac{\myth1(\t, \frac{a+b\t}{2} -\z_R  + \z_i )}{\myth1(\t, \frac{a+b\t}{2} + \z_i )}
  e^{-2\pi i b \z_R}\ ,
  \nonumber \\ & =
  \frac12 \sum_{a,b=0,1}
  \frac{\myth1(\t, \frac{a+b\t}{2} -\z_R  + \z_1 )\myth1(\t, \frac{a+b\t}{2} + \z_R  - \z_2 )}
  {\myth1(\t, \frac{a+b\t}{2} + \z_1 )\myth1(\t, \frac{a+b\t}{2} - \z_2 )}
  \ ,
  \label{formula3}
\end{align}
where we used the properties~\eqref{propertytheta1}, \eqref{propertytheta2} of the theta function for the second and third equalities.
%
On the other hand we are also free to  choose $\eta>0$. We then get contributions from
the residues at $u+\z_1=0$ and $u+\z_2=0$. This gives the result
\begin{align}
  \CE(\t;z,\z_1,\z_2) & = -
  \frac{\myth1(\t,\z_2-\z_1-\z_R) \myth1(\t,2\z_1-\z_R)}
  {\myth1(\t,\z_2-\z_1) \myth1(\t,2\z_1)}
  \nonumber \\ & -
  \frac{\myth1(\t,\z_1-\z_2-\z_R) \myth1(\t,2\z_2-\z_R)}
  {\myth1(\t,\z_1-\z_2) \myth1(\t,2\z_2)} \ .
  \label{difrep}
\end{align}
From the fact that $Z^\text{1-loop}(u)$ is a meromorphic function
on the torus $E(\t)$, one can show as in \cite{Benini:2013nda} that
these two expressions (\ref{formula3}) and (\ref{difrep})
are different representations of the same elliptic genus of the A$_1$ singularity.

We now in turn compute the elliptic genus of  the$A_{N-1}$ ALE space. First
one can always choose a vector $\eta$ in $\mathbb{R}^{N-1}$ that allows only one ordered basis
\begin{align}
  \mathfrak{B} = \{ Q_{Y_1}, Q_{Y_2}, ..,Q_{Y_{N-1}} \}
\end{align}
with $\n(\mathfrak{B})=+1$ so that
\begin{align}
  Q_\eta = \{ Q_{Y_1} , Q_{Y_2} , .., Q_{Y_{N-1}}\}\ .
\end{align}
For this choice of $\eta$, the Jeffrey-Kirwan residue can be non-zero
at the singular points where $N-1$ hyperplanes $H_{Y_1},..,H_{Y_{N-1}}$ intersect, i.e.,
$Q_\ast = Q_\eta$ and
\begin{align}
  \sum_{b=1}^{N-1} C_{ab} u_b = m_a + \t n_a \ \text{ for } a=1,2,..,N-1
\end{align}
One can then show that there are
$N^2$ singular points $u_\ast=(u_1,u_2,..u_{N-1})$ in the torus $E(\t)$, parameterized by
\begin{align}
  u_1 & = \frac{(N-1) m_1+ (N-2)m_2 + .. + m_{N-1}}{N} + \t
  \frac{(N-1) n_1+ (N-2)n_2 + .. + n_{N-1}}{N} \ ,
  \nonumber \\ & \vdots
  \nonumber \\
  u_{N-1} & = \frac{m_1+ 2 m_2 + .. + (N-1) m_{N-1}}{N} +
  \t \frac{n_1+ 2 n_2 + .. + (N-1) n_{N-1}}{N} \ ,
  \label{singularA}
\end{align}
where non-negative integers $m_a$ and $n_a$ satisfy
\begin{align}
  \sum_{a=1}^{N-1} m_a  = 0 ,\text{or } 1  \ \text{ and } \
  \sum_{a=1}^{N-1} n_a  = 0 ,\text{or } 1\ .
  \label{conditionA}
\end{align}
Note that
\begin{align}
  u_1 + u_N = \sum_{a=1}^{N-1} m_a + \t \sum_{a=1}^{N-1} n_a  \ .
  \label{relationA}
\end{align}
Using the properties of theta function (\ref{propertytheta1}) and (\ref{propertytheta2}),
the elliptic genera of the $A_{N-1}$ singularities are
\begin{align}
  \CE(\t;\z_R,\z_1,\z_2) & = \frac{1}{N} \sum_{m_a,n_a}
  \frac{\myth1(\t, u_1 - \z_R  + \z_1 )}{\myth1(\t,u_1 +\z_1)}
  \frac{\myth1(\t, -u_{N-1} + \z_R - \z_2  )}{\myth1(\t,-u_{N-1} - \z_2 )}
  e^{-2\pi i \sum_{a=1}^{N-1} n_a \z_R}
  \nonumber \\
  & = \frac{1}{N} \sum_{m_a,n_a}
  \frac{\myth1(\t, u_1 -\z_R  + \z_1 )}{\myth1(\t,u_1 +\z_1)}
  \frac{\myth1(\t, u_1 + \z_R - \z_2  )}{\myth1(\t,u_1 - \z_2 )}\ ,
\end{align}
where $u_1$ and $u_{N-1}$ are given by (\ref{singularA}). Here non-negative intergers
$m_a$ and $n_a$ satisfy the condition (\ref{conditionA}). Note that we used for the second equality
the relation (\ref{relationA}). We can rewrite the above expression
simply as
\begin{align}
  \CE(\t;\z_R,\z_1,\z_2) = \frac1N \sum_{a=0}^{N-1} \sum_{b=0}^{N-1}
  \frac{\myth1(\t, \frac{a+\t b}{N} -\z_R  + \z_1 )}{\myth1(\t,\frac{a+\t b}{N} +\z_1)}
  \frac{\myth1(\t, \frac{a+\t b}{N} + \z_R - \z_2  )}{\myth1(\t,\frac{a+\t b}{N} - \z_2 )} \ .
  \label{formulaN}
\end{align}
One can easily identify the terms with $b\neq 0$ as the contribution from the twisted sector.

\paragraph{GLSM II}

Let us consider the simplest example (\ref{A1GLSMII}) for the A$_1$ singularities.
One can define the elliptic genus as in (\ref{def}) where $Q_f$ now denotes
the Cartan generator of $SU(2)_f$. The elliptic genus of the GLSM II is
\begin{align}
  \CE(\t; z,\xi) = \sum_{u_\ast \in \mathfrak{M}_\eta} \JKres_{u_\ast}(Q_\ast,\eta)
  Z^\text{1-loop}(u) \ ,
\end{align}
where
\begin{align}
  Z^\text{1-loop} = Z_{V,\Phi} \cdot Z_{Q^1,\tilde Q_1} \cdot Z_{Q^2,\tilde Q_2}
\end{align}
with
\begin{align}
  Z_{V,\Phi} & = \frac{i\eta(q)^3}{\myth1(\t,\xi_2-z)}
  \frac{\myth1(\t,2\xi_2)}{\myth1(\t,\xi_2+z)}\ ,
  \nonumber \\
  Z_{Q^1,\tilde Q_1} & = \frac{\myth1(\t,u+\xi_1-z)}{\myth1(\t,u+\xi_1-\xi_2)}
  \frac{\myth1(\t,-u-\xi_1-z)}{\myth1(\t,-u-\xi_1-\xi_2)} \ ,
  \nonumber \\
  Z_{Q^2,\tilde Q_2} & = \frac{\myth1(\t,u-\xi_1-z)}{\myth1(\t,u-\xi_1-\xi_2)}
  \frac{\myth1(\t,-u+\xi_1-z)}{\myth1(\t,-u+\xi_1-\xi_2)}  \ .
\end{align}
Choosing $\eta >0$ we get contribution from the residues at $u+\xi_1-\xi_2=0$ and
$u-\xi_1-\xi_2=0$, which results in
\begin{align}
  \CE(\t;z,\xi_1,\xi_2) & = - \frac{\myth1(\t,-2\xi_1+\xi_2-z)\myth1(\t,2 \xi_1-\xi_2-z)}
  {\myth1(\t,-2 \xi_1)\myth1(\t,2 \xi_1-2 \xi_2)}
  \nonumber \\ & -
  \frac{\myth1(\t,2 \xi_1+ \xi_2-z)\myth1(\t,-2 \xi_1-\xi_2-z)}
  {\myth1(\t,2 \xi_1)\myth1(\t,-2 \xi_1-2 \xi_2)}  \ .
\end{align}
It exactly agrees with (\ref{difrep}) once various parameters are identified as
\begin{align}
  \xi_1 - \xi_2 = \zeta_1, \qquad \xi_1+ \xi_2 = - \zeta_2\ , \qquad z-\xi_2 = \z_R \ .
  \label{parmatch2}
\end{align}

For the generic ADE singularities, the elliptic genus becomes
\begin{align}
  \CE(\t; z,\xi) = \frac{1}{|W|} \sum_{u_\ast \in \mathfrak{M}_\eta} \JKres_{u_\ast}(Q_\ast,\eta)
  Z^\text{1-loop}(u) \ ,
\end{align}
where
\begin{align}
  Z^\text{1-loop} = Z_{V,\Phi} \cdot Z_{Q,\tilde Q}
\end{align}
with
\begin{align}
  Z_{V,\Phi} & = \left( \frac{i\eta(q)^3}{\myth1(\t,\xi_2-z)}
  \frac{\myth1(\t,2\xi_2)}{\myth1(\t,\xi_2+z)}  \right)^{\text{rk}[G]}
  \nonumber \\ & \qquad \times
  \prod_{\a \in \D} \
  \frac{\myth1(\t,\a\cdot u + \xi_2-z)}{\myth1(\t,\a\cdot u + \xi_2-z)}
  \frac{\myth1(\t,\a\cdot u + 2\xi_2)}{\myth1(\t,\a\cdot u + \xi_2+z)} \ ,
  \nonumber \\
  Z_{Q,\tilde Q} & = \prod_{\rho \in \mathfrak{R} } \
  \frac{\myth1(\t,\rho \cdot u -z)}{\myth1(\t,\rho \cdot u -\xi_2)}
  \frac{\myth1(\t,-\rho \cdot u -z)}{\myth1(\t,-\rho \cdot u -\xi_2)} \ ,
\end{align}
where $\a$ denote the roots of the gauge group G while $\rho$ denote the weights
of the G-representation $\mathfrak{R}$ of hypermultiplets. Note that there is no
extra flavor symmetry in the GLSMs for D and E-type singularities.

The GLSM's we have studied for ALF spaces flow to metrics that have coincident center metrics.
It would be interesting to generalize this computation to the general multi-center metric.
Naively we expect the elliptic genus to be independent of the moduli that
determine the locations of the centers.

\section{Large Radius Limit and Properties of the Elliptic Genera\label{Rtoinfty}}

In this section we discuss some physical properties of the elliptic genera that we computed in the previous section.
When the radius of the ALF spaces becomes very large, they go over to the ALE spaces.
We shall verify, in accordance with this fact, the ALF elliptic genus equals the ALE elliptic genus in the large radius limit.
We also comment on the additional states that are present in the ALF spectrum, and make consistency checks with
known results in the literature.

\subsection{Large radius limit of Taub-NUT space}
We start with the Taub-NUT sigma model.
Let us consider a limit where the size of the Taub-NUT space becomes large, i.e., $g^2 \gg 1$.
Starting from the expression~\eqref{result1again}, and making a change of variable~$u \to u -\xi_{1}$,
we obtain the expression:
\begin{align} \label{Eagain}
  \CE(\t;\xi_1,\xi_2,z)= g^2 \int_\mathbb{C} \frac{du d\bar u}{\t_2} \  e^{-\frac{g^2\pi}{\t_2}|u-\xi_1|^2}
  \frac{\myth1(\t,u+z) \, \myth1(\t,u-z)}{\myth1(\t,u+\xi_2) \, \myth1(\t,u-\xi_2)}\, .
\end{align}
We see that the integral over the holonomy-plane $u,\bar u$ gets a dominant
contribution around $u=\xi_1$ in the limit $g^2 \to \infty$. Thus, one obtains
\begin{align}
  \lim_{g^2 \to \infty} \CE(\t;\xi_1,\xi_2,z) =
  \frac{\myth1(\t,\xi_1+z) \, \myth1(\t,\xi_1-z)}{\myth1(\t,\xi_1+\xi_2) \, \myth1(\t,\xi_1-\xi_2)}
  \ , \label{ell_largesize}
\end{align}
which  agrees perfectly with the result in \cite{Hohenegger:2013ala} obtained
from the topological vertex formalism.

Using the infinite product representation of the Jacobi theta function,
\begin{align}
  \myth1(\t,z) = - i q^{\frac18} y^{\frac12} \prod_{n=1}^\infty
  \left( 1- q^n \right)\left( 1- y q^n \right)\left( 1- y^{-1} q^{n-1} \right),
\end{align}
where $q=e^{2\pi i\t}$ and $y=e^{2\pi i z}$,
one can separate the elliptic genus into two pieces,
\begin{align}
  \lim_{g^2\to \infty} \CE(\t; \xi_1,\xi_2,z)
  = \CE_\text{zero}(\t;\xi_1,\xi_2,z) \times \CE_\text{osc}(\t;\xi_1,\xi_2,z)\ ,
\end{align}
one of which denotes the contribution from the oscillator-modes
\begin{align}
  \CE_\text{osc}(\t;\xi_1,\xi_2,z) & = \prod_{n=1}^\infty
  \frac{\left(1- e^{2\pi i (\xi_1+z)} q^n \right) \left(1- e^{-2\pi i (\xi_1+z)} q^n \right)}
  {\left(1- e^{2\pi i (\xi_1+\xi_2)} q^n \right) \left(1- e^{-2\pi i (\xi_1+\xi_2)} q^n \right)}
  \nonumber \\ & \ \ \ \ \times
  \frac{\left(1- e^{2\pi i (\xi_1-z)} q^n \right) \left(1- e^{-2\pi i (\xi_1-z)} q^n \right)}
  {\left(1- e^{2\pi i (\xi_1-\xi_2)} q^n \right) \left(1- e^{-2\pi i (\xi_1-\xi_2)} q^n \right)}
  \label{large1}
\end{align}
and the other represents the contribution from the zero-modes
\begin{align}
  \CE_\text{zero}(\t; \xi_1,\xi_2,z) & =
  \frac{\left( 1-e^{-2\pi i (\xi_1+z)} \right)\left( 1-e^{-2\pi i (\xi_1-z)} \right)}
  {\left( 1-e^{-2\pi i (\xi_1+\xi_2)} \right)\left( 1-e^{-2\pi i (\xi_1-\xi_2)} \right)}
  \label{large2}
\end{align}

Let us explain how to understand these results (\ref{large1}) and (\ref{large2}) along the lines of
discussion in \cite{David:2006yn}\footnote{The authors of \cite{David:2006yn} studied the $(0,4)$ superconformal field theory with
Taub-NUT target space, this is slightly different in detail from the $(4,4)$ model that we have discussed so far.
We shall also discuss the~$(0,4)$ model in section 6.2.}.
As far as the oscillator modes are concerned,
the Taub-NUT sigma model can reduce to a free theory on the flat $\mathbb{R}^4$
in the limit $g^2\to\infty$. Furthermore,
only the ground states of right movers can contribute to the elliptic genus.
Therefore, we just need to compute an index of
four left-moving free bosons and four left-moving free fermions
with $\bar L_0 = 0$,
\begin{align}
  \CE_\text{osc} (\t;\xi_1,\xi_2,z) & = \text{Tr}_\text{osc} \Big[ (-1)^F q^{L_0} \bar q^{\bar L_0}
  e^{-2\pi i \xi_1 Q_f} e^{-2\pi i \xi_2 (Q_1-Q_2)}
  e^{-2\pi i z Q_R} \Big]\ ,
\end{align}
where the trace is performed over oscillators.

We summarize the charges of four bosons and fermions under $Q_f$, $Q_1-Q_2$ and $Q_R$ below:
we identified earlier $Q_f$ and $Q_1$ as the Cartan of $U(1)\times SU(2)$ of the Taub-NUT space.
Since $U(1)\times SU(2) \subset SU(2)\times SU(2)\simeq SO(4)_{\mathbb{R}^4}$ near the origin,
four free bosons should carry
\begin{align}
  \big( +1, +1, 0 \big) \oplus \big( +1 , -1, 0 \big) \oplus \big( -1, +1 , 0\big)
  \oplus \big(-1,-1 , 0\big)\ ,
\end{align}
while four free left-moving fermions carry
\begin{align}
  \big( +1, 0, +1  \big) \oplus \big( +1, 0, +1 \big) \oplus \big(-1, 0 ,+1 \big)
  \oplus \big(-1,0,-1\big)\ .
\end{align}
Hence one can show that
\begin{align}
  \CE_\text{osc} (\t;\xi_1,\xi_2,z) & =
  \prod_{n=1}^\infty
  \frac{\left(1- e^{2\pi i (\xi_1+z)} q^n \right) \left(1- e^{-2\pi i (\xi_1+z)} q^n \right)}
  {\left(1- e^{2\pi i (\xi_1+\xi_2)} q^n \right) \left(1- e^{-2\pi i (\xi_1+\xi_2)} q^n \right)}
  \nonumber \\ & \ \ \ \ \times
  \frac{\left(1- e^{2\pi i (\xi_1-z)} q^n \right) \left(1- e^{-2\pi i (\xi_1-z)} q^n \right)}
  {\left(1- e^{2\pi i (\xi_1-\xi_2)} q^n \right) \left(1- e^{-2\pi i (\xi_1-\xi_2)} q^n \right)}
  \ ,
\end{align}
which exactly agrees with (\ref{large1}).

On the other hand, we have to carefully consider zero-modes of the Taub-NUT sigma model,
sensitive to the global geometry. The dynamics of these zero-modes should be governed by
SUSY quantum mechanics (under a certain potential)\footnote{In the monopole moduli space problem, this term arises when
the Higgs vacuum expectation value is misaligned. As explained in \cite{Gauntlett:1999vc},
the potential  generated by the tri-holomorphic Killing vector field
on  Taub-NUT space is the same as the potential employed  in \cite{Pope:1978zx} as an infrared regulator
when computing  the spectrum of the Dirac operator on Taub-NUT space.}
in the Taub-NUT space preserving complex four supercharge.

This quantum mechanical system has been studied as the low-energy dynamics of
two distinct $SU(3)$ monopoles
\cite{Gauntlett:1996cw,Lee:1996if,Lee:1998nv,Bak:1999da,Bak:1999ip,Gauntlett:1999vc}.
In particular, it is shown in \cite{Bak:1999ip} that
when $|p| \geq 2$ where $Q_f\doteq p$ is the momentum charge conjugate to the Taub-NUT circle,
the bound states come in four multiplets with angular momentum
$Q_1=\frac{|p|}{2}$, $\frac{|p|-1}{2}$,
$\frac{|p|-1}{2}$, and $\frac{|p|}{2}-1$. If $|p| = 1$,
there are three multiplets with $Q_1=\frac{|p|}{2}$, $\frac{|p|-1}{2}$ and
$\frac{|p|-1}{2}$. Finally, it is important that we have
the unique threshold bound state with $p=0$. A little more detail
on various aspects of the Taub-NUT quantum mechanics
will be discussed in section 4.

In order to see these quantum bound states and their degeneracies,
it is useful to expand $\CE_\text{zero}(\t;\xi_1,\xi_2,z)$ in powers of
$e^{-2\pi i \xi_1}<1$ as follows
\begin{align}
  \CE_\text{zero}(\t; \xi_1,\xi_2,z) & =
  1 + \Big( \left[ e^{2\pi i \xi_2} + e^{-2\pi i \xi_2}\right]
  - e^{2\pi i z}\big[1\big]- e^{-2\pi i z}\big[1\big] \Big) e^{-2\pi i \xi_1}
  \nonumber \\ & \hspace*{0.75cm}  +
  \Big( \left[ e^{4 \pi i \xi_2} + 1 + e^{-4\pi i \xi_2} \right]
  - e^{2\pi i z} \left[ e^{2\pi i \xi_2 } + e^{- 2\pi i \xi_2} \right]
  \\ & \hspace*{1.35cm}
  - e^{-2\pi i z} \left[ e^{2\pi i \xi_2 } + e^{- 2\pi i \xi_2} \right]
  + \big[ 1 \big] \Big) e^{-4\pi i \xi_1}
  \nonumber \\ & \hspace*{0.75cm}  +
  \Big( \left[ e^{6 \pi i \xi_2} + e^{2\pi i \xi_2} +  e^{-2\pi i \xi_2} +  e^{-6\pi i \xi_2}\right]
  - e^{2\pi i z} \left[ e^{4 \pi i \xi_2} + 1 +  e^{-4 \pi i \xi_2} \right]
  \nonumber \\ & \hspace*{1.35cm}
  - e^{-2\pi iz} \left[ e^{4 \pi i \xi_2}  + 1 +  e^{-4 \pi i \xi_2} \right]
  + \left[  e^{2\pi i \xi_2} +  e^{- 2\pi i \xi_2} \right]  \Big) e^{-6\pi i \xi_1}
  \nonumber \\ & \hspace*{0.75cm}
  + \cdots \ . \nonumber
  \label{large3}
\end{align}
Using the characters of $SU(2)_{\mathbb{R}^3}$ defined by
\begin{align}
  \chi_{\bf R} = \text{Tr}_{\bf R}\Big[ e^{4 \pi i \xi_2 Q_1}\Big]
\end{align}
where ${\bf R}$ denotes a representation of $SU(2)$ and
$j$ denotes the third component of $SU(2)$ angular momentum, the
expression of the index $\CE_\text{zero}$ can be simplified as
\begin{align}
  \CE_\text{zero}(\t; \xi_1,\xi_2,z) & =
  1 + \Big( \chi_{\bf \frac12} - e^{2\pi i z} \chi_{\bf 0} - e^{- 2\pi i z} \chi_{\bf 0}  \Big) e^{-2\pi i \xi_1}
  \nonumber \\ & \ \ \   +
  \Big( \chi_{\bf 1} - e^{2\pi i z} \chi_{\bf \frac12}-e^{-2\pi i z}\chi_{\bf \frac12} + \chi_{\bf 0}
  \Big) e^{-4\pi i \xi_1}
  \nonumber \\ & \ \ \   +
  \Big( \chi_{\bf \frac32} - e^{2\pi i z} \chi_{\bf 1}-e^{-2\pi i z}\chi_{\bf 1} + \chi_{\bf \frac12}
  \Big) e^{-6\pi i \xi_1} + \cdots\ .
  \label{large4}
\end{align}
From the expression (\ref{large4}), we can confirm the unique bound state with
$Q_f\doteq 0$.
Moreover one can see three multiplets with $Q_1=\frac12,0,0$ for $|Q_f| \doteq1$,
and four multiplets with $Q_1=\frac{|p|}{2},\frac{|p|-1}{2},\frac{|p|-1}{2},\frac{|p|-2}{2}$
for $Q_f \doteq p$ with $|p|\geq 2$.

Since $\CE_\text{zero}(\t;\xi_1,\xi_2,z)=
\CE_\text{zero}(\t;-\xi_1, \xi_2 ,z)$, we obtain the same result
if we expand $\CE_\text{zero}$ in powers of $e^{2\pi i \xi_1}<1$.

In summary, the elliptic genus (\ref{result1}) is consistent with the result in \cite{David:2006yn}
in the large size limit $g^2\to\infty$.

\subsection{Continuum states, and an associated non-holomorphicity \label{nonhol}}
Several comments are in order. The computation in \cite{David:2006yn}  assumed that the index is independent of
the size of the Taub-NUT space, but we see that this is not the case. The size dependence arises in
the non-holomorphic terms. It is known that scattering states can lead to non-holomorphic
contribution to the elliptic genus.
The contribution to the elliptic genus from scattering states however occurs only when
the non-compact target space have a finite circle $S^1$ at infinity.

For instance let us consider quiver gauge theories which flow to
the non-linear sigma model on the ALE spaces. As shown in section 3.2,
the elliptic genus of ALE space is holomorphic. This is because, although
the target space is non-compact, the chemical potentials for global charges
of quiver gauge theories essentially make the theories gapped.

As a simple toy model, let us consider a free theory of a single
massless chiral multiplet $(q,\psi)$.
This model preserve the $U(1)_f$ flavor symmetry, identified as $U(1)$ isometry of the
target space $\mathbb{C}$.
One can show that turning on the chemical potential $z$ for
the $U(1)_f$ flavor charge induces a scalar potential
\begin{align}
  V(|q|) \simeq z^2 |q|^2\ ,
\end{align}
so that the wavefunction is localized near the origin of $\mathbb{C}$. The elliptic genus
receives contribution from discrete states only and should be holomorphic.

However, one can show that, due to the St\"uckelberg field, turning on the chemical
potential $\xi_1$ for the $U(1)_f$ flavor charge (momentum charge of the circle of the Taub-NUT space)
in the present model leads to the following scalar potential
\begin{align} \label{VTN}
  V_\text{TN}(|q|,|\tilde q |) \simeq  \frac{{\xi_1}^2}{\frac{1}{g^2} + \frac{1}{|q|^2 + |\tilde q|^2}}
\end{align}
which is asymptotic to a finite value $(\xi_1 g)^2$. Thus, there should be scattering states with $E \geq (\xi_1 g)^2 $
which can contribute to the elliptic genus. This explains why the elliptic genus of the Taub-NUT space
can depend on the size parameter $g^2$. Moreover, in the large radius limit,
the given theory now becomes fully gapped,
\begin{align}
  V_\text{TN}(|q|,|\tilde q|) \xrightarrow{g^2\to\infty} (\xi_1)^2 \left( |q|^2 + |\tilde q|^2 \right)\ .
\end{align}
It explains why the non-holomorphic elliptic genus of the Taub-NUT space
can reduce to a holomorphic one in the large size limit.

\subsection{ALF to ALE space}
In the large radius limit $g^2\to\infty$,
the multi Taub-NUT space with $N$ coincident centers becomes an ALE space A$_{N-1}$.
It is important to see if the elliptic genus (\ref{resultmulti}) agrees with those of A$_{N-1}$
spaces (\ref{formulaN}). Indeed, one can show that this is the case.
In the decompactification limit $g^2\to\infty$, the elliptic genus (\ref{resultmulti}) becomes
\begin{align}
  \lim_{g^2\to\infty} \CE_{N}(\t,\vec z) & =
  \frac1N \sum_{a=0}^{N-1} \sum_{b=0}^{N-1}
  \frac{\myth1(\t,\frac{a+\t b}{N}+ z_1+z_3)\myth1(\t,\frac{a+\t b}{N}+z_1-z_3)}
  {\myth1(\t,\frac{a+\t b}{N}+ z_1+z_2)\myth1(\t,\frac{a+\t b}{N}+ z_1-z_2)}
\ ,
\end{align}
which is in perfect agreement with the elliptic genus of the A$_{N-1}$ space (\ref{formulaN})
\begin{align}
  \CE_{A_{N-1}}(\t,z,\xi_i) =
  \frac1N \sum_{a=0}^{N-1} \sum_{b=0}^{N-1}
  \frac{\myth1(\t, \frac{a+\t b}{N} -\zeta_R  + \zeta_1 )}{\myth1(\t,\frac{a+\t b}{N} +\zeta_1)}
  \frac{\myth1(\t, \frac{a+\t b}{N} +\zeta_R - \zeta_2  )}{\myth1(\t,\frac{a+\t b}{N} - \zeta_2 )} \ .
\end{align}
Various chemical potential in both expressions are identified as in (\ref{parmatch2})
\begin{align}
  \xi_1 - \xi_2 = \zeta_1, \qquad \xi_1+ \xi_2 = - \zeta_2\ , \qquad z-\xi_2 = \z_R \ .
\end{align}
%


\paragraph{Witten Index.}
When all the chemical potentials $\xi_1,\xi_2,z$ are set to zero, the elliptic genus can reduce
to the Witten index. The Witten index gives the Euler number of the target space.
One can indeed show that the Witten index of Taub-NUT space is given by
\begin{align}
  \CE(\t; \xi_1,\xi_2,z=0) & = g^2 \int_\mathbb{C} \frac{dud\bar u}{\t_2} \ e^{-\frac{g^2\pi}{\t_2} |u|^2}
  = 1 \ ,
\end{align}
which agrees with the Euler characteristic of the Taub-NUT
space computed in \cite{Eguchi:1980jx}. Note that, for closed smooth manifolds,
the Euler characteristic coincides with the Euler number.

Moreover, the Witten index of the multi Taub-NUT space with N coincident
centers is given by
\begin{align}
  \CE(\t;\xi_1,\xi_2,z=0) = \frac{g^2}{N} \sum_{a,b=0}^{N-1}
  \int_\mathbb{C} \frac{dud\bar u}{\t_2} \ e^{-\frac{g^2\pi}{\t_2} \left|u+ \frac{a+\t b}{N} \right|^2}
  = N\ ,
\end{align}
which is also in perfect agreement with the Euler number of the
multi-center Taub-NUT space.

\section{The BPS Spectrum of Taub-NUT Space}

In this section we analyze the spectrum of states contained in the Taub-NUT elliptic genus $\CE(\t;\xi_1,\xi_2,z)$.
The elliptic genus receives contributions from discrete bound states
that are localized in the interior of Taub-NUT space as well as from states in the continuum.
We first explain how the elliptic genus can be additively decomposed
into a piece containing the discrete states and the rest:~$\CE = \CE_{\rm disc}+\CE_{\rm rest}$.
The discrete piece~$\CE_\text{disc}$ is a meromorphic Jacobi form with poles as a function
of the chemical potentials~$(\xi_{i})$. These poles and the corresponding residues capture the
jumps in the discrete spectrum across walls of marginal stability. We discuss the structure of these walls.
We further decompose~$\CE_\text{disc}(\t;\xi_1,\xi_2,z)$ into a polar part that captures all the wall-crossings and a finite
part that is stable across the space of chemical potentials, each having interesting modular properties.
Using this decomposition, we compute explicit Fourier expansions of the various pieces at generic values of the chemical potentials.

\subsection{Separation of the discrete states \label{sepdisc}}

The elliptic genus of Taub-NUT space was computed in Equation~(\ref{result1}). We redisplay it here for convenience:
\begin{align}
  \CE(\t;\xi_1,\xi_2,z) = g^2 \int_{E(\t)} \frac{dud\bar u}{\t_2}
  \  \frac{\myth1(\t,u+\xi_1+z) \, \myth1(\t,u+\xi_1-z)}{\myth1(\t,u+\xi_1+\xi_2) \, \myth1(\t,u+\xi_1-\xi_2)}
  \sum_{(p,\w)\in\mathbb{Z}^2}
e^{-\frac{g^2\pi}{\t_2}\big| u + p + \t\w\big|^2}\ .
  \label{result1againrep}
\end{align}
In order to have a Hamiltonian interpretation of this result, we use the method of \cite{Ashok:2013zka}
to first separate the elliptic genus into two pieces, one containing the right-moving ground states
and the other containing the scattering states.
The computation, which is fairly detailed, are presented in appendix~\S\ref{decompApp}.

The result of the calculation is that  one can separate the elliptic genus
into a discrete part and a continuum part as follows:
\begin{align} \label{Edecomp}
  \CE = \CE_\text{disc} + \CE_\text{rest}\ ,
\end{align}
where
\begin{align}
   \CE_\text{disc} & =   \frac{1}{\pi \eta(q)^6}\oint_{\cal C}du \
   \sum_{{\hat r}_i,{\hat s}_i,p,w}'
   \frac{\left(q{\bar q} \right)^{\frac{(p+g^2w)^2 }{4g^2} +\frac{u^2}{g^2}}}{2iu + (p+g^2w)}
   \left(q{\bar q} \right)^{iu+\frac{p+g^2w}{2}} q^{-pw} q^{\frac{\sum_i({\hat s}_i-1/2)^2}{2}}
   \nonumber \\ & \qquad \bigg. \times(-1)^{{\hat s}_1+{\hat s}_2}
   S_{{\hat r}_1}(q)S_{{\hat r}_2}(q)
   e^{2\pi i \xi_1 p} e^{2\pi i \xi_2 ({\hat r}_1-{\hat r}_2)} e^{-2\pi i z({\hat s}_1-{\hat s}_2)}
   \nonumber \\ & = \frac{1}{\eta(q)^6} \sum_{{\hat r}_i,{\hat s}_i,p}
  \sum_{w\in I_p}
  \d_{{\hat r}_1+{\hat r}_2 - {\hat s}_1 -{\hat s}_2 +2 - p}
  \ q^{- pw} q^{\frac{\sum_i({\hat s}_i-1/2)^2}{2}}
  \nonumber \\ & \qquad \bigg. \times(-1)^{{\hat s}_1+{\hat s}_2}
  S_{{\hat r}_1}(q)S_{{\hat r}_2}(q)
  e^{2\pi i \xi_1 p} e^{2\pi i \xi_2 ({\hat r}_1-{\hat r}_2)} e^{-2\pi i z({\hat s}_1-{\hat s}_2)}  \
  \label{holell}
\end{align}
with
\begin{align}
  S_r(q) = \sum_{n=0}^\infty (-1)^n q^{\frac{n(n+2r+1)}{2}} \nonumber\ .
\end{align}
Here we take the chemical potentials $z$ and $\xi_{1,2}$ to be real.
In the first line above, the symbol $\sum'$ denotes a constrained summation
\begin{align}
  \sum_{p,\w,r_i,s_i}' = \sum_{p,\w,r_i,s_i} \d_{r_1+r_2-s_1-s_2+2-p} \, .
\end{align}

The contour $\cal C$ above encloses the poles at $u = \frac i2 \left( p + g^2 w\right)$
where $(p,q)$ satisfy the condition
\begin{align}
  \e -g^2  < p+g^2 w < \e\ .
\end{align}
Equivalently, one can say that
\begin{align}
 w \in
  I_p \equiv \big\{ w \in \mathbb{Z} \big|  \e - 1 -\frac{p}{g^2} < w < \e -\frac{p}{g^2}\big\}
  \text{ for each }  p
  \label{range}\ .
\end{align}

At the end, the discrete part of the elliptic genus becomes
\begin{align}
  \CE_\text{disc} & = \frac{1}{\eta(q)^6} \sum_{{\hat r}_i,{\hat s}_i,p}
  \sum_{w\in I_p}
  \d_{{\hat r}_1+{\hat r}_2 - {\hat s}_1 -{\hat s}_2 +2 - p}
  \ q^{- pw} q^{\frac{\sum_i({\hat s}_i-1/2)^2}{2}}
  \nonumber \\ & \qquad \bigg. \times(-1)^{{\hat s}_1+{\hat s}_2}
  S_{{\hat r}_1}(q)S_{{\hat r}_2}(q)
  e^{2\pi i \xi_1 p} e^{2\pi i \xi_2 ({\hat r}_1-{\hat r}_2)} e^{-2\pi i z({\hat s}_1-{\hat s}_2)}
  \ .
  \label{holell}
\end{align}
It is useful to rewrite the above expression into
the following form
\begin{align}
  \CE_\text{disc} =
  \left(\frac{\myth1(\t,\xi_2-z)}{\eta(q)^3} \right)^2
  \sum_{{\check r}_i,p} \sum_{\w\in I_p} \d_{{\check r_1}+{\check r}_2 - p} \
  q^{-p\w}
  \frac{e^{2\pi i (\xi_1 + \xi_2) {\check r}_1}}{1-e^{2\pi i (z - \xi_2)} q^{{\check r}_1}}
  \frac{e^{2\pi i (\xi_1-\xi_2){\check r}_2}}{1-e^{2\pi i (\xi_2 - z)} q^{{\check r}_2}}\ ,
  \label{useful1}
\end{align}
where we used the formula \cite{Ashok:2013zka}
\begin{align}
  \frac{i\myth1(\t, z)}{1-e^{2\pi i z} q^p}
  = \sum_{m\in\mathbb{Z}} (-1)^m q^{\frac{(m-1/2)^2}{2}}
  \left(e^{2\pi i z}\right)^{m-\frac12} S_{-m+p}(q)\ ,
\end{align}
and made another change of variables
\begin{align}
  {\check s}_i \equiv - {\hat s}_i + 1\ , \qquad
  {\check r}_i \equiv {\hat r}_i - {\hat s}_i + 1 \ .
\end{align}
Note from the above expression (\ref{useful1}) that $\CE_\text{disc}$
respects the charge conjugation symmtry.
\begin{align}
  \CE_\text{disc}(\t;\pm \xi_1,\pm \xi_2,\pm z) = \CE_\text{disc}(\t;\xi_1,\xi_2,z)\ .
\end{align}

It is natural to decompose the discrete part $\CE_\text{disc}$ in a Fourier series with each term labelled by
the momentum charge $p$ as:
\begin{align}
  \CE_\text{disc} = \sum_{p\in \mathbb{Z}} e^{2\pi i p \xi_1} \CE^p_\text{disc}(\t,\xi_2,z)\ .
\end{align}
Using the expression (\ref{useful1}), one can show that
\begin{align}
  \CE^{p=0}_\text{disc}(q,\xi_2,z) & = 1
  + q \left[ \chi_{\bf 1}
  - 2 \tilde \chi_{\bf \frac12}\chi_{\bf \frac12}
  + \tilde\chi_{\bf 1} + 2 \right]
  \nonumber \\ &
  + q^2 \left[ \chi_{\bf 2} - 2 \tilde \chi_{\bf \frac12}\chi_{\bf \frac32}
  + \left( \tilde \chi_{\bf 1} + 6 \right) \chi_{\bf 1}
  -  8 \tilde \chi_{\bf \frac12} \chi_{\bf \frac12} + 3\tilde \chi_{\bf 1}+7
  \right]
  \nonumber\\ &
  + q^3 \left[ \chi_{\bf 3} - 2 \tilde \chi_{\bf \frac12}  \chi_{\bf \frac 52}
  + \left( \tilde \chi_{\bf 1} + 6 \right) \chi_{\bf 2}
  - 10 \tilde \chi_{\bf \frac12} \chi_{\bf \frac 32}
  + \left( 7 \tilde \chi_{\bf 1} + 20 \right) \chi_{\bf 1} \right.
  \nonumber \\ & \qquad \left.
  - \big( 2\tilde \chi_{\bf \frac32} + 26 \tilde\chi_{\bf \frac12}\big)\chi_{\bf \frac12}
  + 10 \tilde \chi_{\bf 1} + 19  \right] + \CO(q^4) \ ,
  \label{holp0}
\end{align}
where $\chi_{\bf R}(\xi_2)$ and $\tilde \chi_{\bf \tilde R}(z)$ denote
the characters of $SU(2)_{\mathbb{R}^3}$ and $SU(2)_3$ R-symmetry,
\begin{align}
  \chi_{\bf R}(\xi_2) & = \text{Tr}_\mathbf{R}\left[ e^{4\pi i \xi_2 j}\right] \
  \nonumber \\
  \tilde \chi_{\bf \tilde R}(\xi_2) & = \text{Tr}_\mathbf{\tilde R}
  \left[ e^{4\pi i z \tilde j}\right]\ .
\end{align}
%
The quantity $\CE_\text{disc}^{p=0}$ is universal in that it does not depend on the size of asymptotic circle $g^2$.
In section~\S\ref{AppGen} we will discuss a physical origin of this
universal factor in two ways -- as the unique threshold bound state of $SU(3)$ two distinct
monopoles in 4d $\CN=4$ super Yang-Mills theory, and
the BPS spectrum of $SU(3)$ self-dual string in 6d $\CN=(2,0)$ superconformal theory,
both of which are indeed independent of $g^2$.

For positive momentum charge $p>0$\footnote{When $|p|=1$ we define $\chi_{\bf \frac{|p|}{2} -1}=0$.}, we have:
\begin{align}
  \CE_\text{disc}^{p>0}  =  q^{|pw|} \Bigg[ \left( \chi_{\bf \frac p2}
  - \tilde \chi_{\bf \frac12} \chi_{\bf \frac{p-1}{2}}
  + \chi_{\bf \frac p2 -1}\right) + \CO(q) \Bigg] \text{ with }
  -g^2 <p + g^2 w < 0\ ,
  \label{holpp}
\end{align}
while, for negative charge $p<0$,
\begin{align}
  \CE_\text{disc}^{p<0} & = q^{|p(w+1)|} \Bigg[ \left( \chi_{\bf \frac{|p|}{2}}
  - \tilde \chi_{\bf \frac12} \chi_{\bf \frac{|p|-1}{2}}
  + \chi_{\bf \frac{|p|}{2} -1}\right) + \CO(q) \Bigg] \text{ with }
  0 < p + g^2 (w+1) < g^2  \ .
  \label{holnp}
\end{align}
These terms do depend on the size parameter $g^2$. We will show in section~\S\ref{AppGen} that
the first term of each $\CE_\text{disc}^{p\neq0}$ can be identified as
a contribution from supermultiplets of $\frac14$-BPS states in 4d
$\CN=4$ super Yang-Mills theory with $G=SU(3)$ that exhibit wall-crossing
as the parameter $g^2$ varies. This wall-crossing behavior explains why
the holomorphic part of the elliptic genus can depend on the continuous parameter $g^2$.

More generally, one can rewrite the full expression (\ref{useful1}) in terms of a contour integral as follows:
\begin{align}
  \CE_\text{disc} & =  \left(\frac{\myth1(\t,\xi_2-z)}{\eta(q)^3} \right)^2
  \sum_{{\check r}_i,p} \sum_{\w\in I_p} \d_{{\check r_1}+{\check r}_2 - p} \ q^{-pw}
  \frac{ e^{2\pi i (\xi_1 + \xi_2)\check r_1}}{1-e^{2\pi i (z - \xi_2)} q^{{\check r}_1}}
  \frac{ e^{2\pi i (\xi_1-\xi_2)\check r_2}}{1-e^{2\pi i (\xi_2 - z)} q^{{\check r}_2}}\ ,
  \nonumber \\ & =
  \left( \frac{\myth1(\t,\xi_2-z)}{\eta(q)^3}\right)^2 \frac{1}{2\pi i}
  \oint_{\CC^{(0)}} \frac{dx}{x} \sum_{p}\sum_{\w \in I_p} \   q^{-pw} x^{-p}
  \nonumber \\ & \qquad\qquad\qquad \times  \bigg.
  \sum_{{\check r}_1} \frac{\left( x e^{2\pi i (\xi_1+\xi_2)} \right)^{\check r_1}}{1-e^{2\pi i (z-\xi_2)}q^{{\check r}_1}}
  \sum_{{\check r}_2} \frac{\left( x e^{2\pi i (\xi_1-\xi_2)} \right)^{\check r_2}}{1-e^{2\pi i (\xi_2-z)}q^{{\check r}_2}} \, ,
\end{align}
where $x=e^{2\pi i \g}$ and the contour $C^{(0)}$ is chosen as $|x|=x_0$ where $|q|<x_0<1$.
Using the following identity \cite{Kac:1994kn},
\begin{align}
  \sum_{n\in\mathbb{Z}} \frac{x^n}{1-zq^n} =  i\eta(q)^3
  \frac{\myth1(\t,\g+\a) }{\myth1(\t,\g)\, \myth1(\t,\a)}\
\end{align}
where $x=e^{2\pi i \g}$ and $z=e^{2\pi i \a}$, one obtains a contour integral representation
of the discrete part of the elliptic genus
\begin{align}
  \CE_\text{disc}  & =
  \sum_{p}\sum_{w \in I_p} q^{-pw}\oint_{\CC^{(0)}} \frac{dx}{x} \ x^{-p}
  \frac{\myth1(\t,\xi_1+z+\g)\, \myth1(\t,\xi_1-z+\g)}
  {\myth1(\t,\xi_1+\xi_2+\g)\, \myth1(\t,\xi_1-\xi_2+\g)}\ ,
  \nonumber \\ & =
  \sum_{p}\sum_{w \in I_p} e^{2\pi i p \xi_1 }\oint_{\CC^{(w)}} \frac{dx}{x} \ x^{-p}
  \frac{\myth1(\t,\g+z-w\t)\, \myth1(\t,\g-z-w\t)}
  {\myth1(\t,\g+\xi_2-w\t)\, \myth1(\t,\g-\xi_2-w\t)}\ ,
  \nonumber \\ & =
  \sum_{p}\sum_{w \in I_p} e^{2\pi i p \xi_1 }\oint_{\CC^{(w)}} \frac{dx}{x} \ x^{-p}
  \frac{\myth1(\t,\g+z)\, \myth1(\t,\g-z)}
  {\myth1(\t,\g+\xi_2)\, \myth1(\t,\g-\xi_2)}\ .
  \label{useful2}
\end{align}
Here we used  the property of the theta function (\ref{propertytheta1}) to obtain the third equality,
and the contour $\CC^{(w)}$ is defined as $|x|=|q|^w x_0$. Here we assume that $|e^{2\pi i \xi_1}|=1$.
It is worth noting that, for each $p$, we have only one winding mode $w$ satisfying
(\ref{range}).

The expression in the last line of (\ref{useful2}) looks like the Fourier expansion
of the elliptic genus of the Taub-NUT space in the large size limit, and it is tempting to write:
\begin{align} \label{Ediscsim}
  \CE_\text{disc} \text{ ``$\simeq$'' }
   \frac{\myth1(\t,\xi_1+z) \, \myth1(\t,\xi_1-z)}
  {\myth1(\t,\xi_1+\xi_2) \, \myth1(\t,\xi_1-\xi_2)} \equiv \v(\t,z, \xi_{1},\xi_2) \, .
\end{align}
However, as we have indicated in the equation, this identification is not yet completely precise since
the last line of (\ref{useful2}) involves a summation
over~$p$ and $w$ (which in turn depends on~$p$), and a choice of contour. These complications are not
completely unexpected, as the integrand in (\ref{useful2}) is a meromorphic Jacobi form, for which the
Fourier expansion is a subtle problem related to the wall-crossing behavior of its Fourier
coefficients~\cite{Dabholkar:2012nd}.

In the rest of this section we address these issues. In \S\ref{walls} we discuss a physical model
for the wall-crossing in terms of winding strings and an associated potential in the quantum mechanics
of the center of mass of the string. In \S\ref{secdecomp} we present a
precise treatment of the Fourier decomposition of the discrete spectrum, which yields a simple
formula for the Fourier coefficients including the effects of wall-crossing.

%

\subsection{Wall-crossing \label{walls}}

To understand various terms in the discrete part of the elliptic genus,
it is useful to consider a dimensional reduction of the present model down to the quantum
mechanics in the sector of winding number $w$, i.e., in the variables of~\eqref{TNmetric},
\begin{align}
  \partial_1 \vec r = 0 \ , \qquad \partial_1 \k = w_\text{QM}\ .
\end{align}
Due to the above twisted dimensional reduction, known as Scherk-Schwarz reduction,
one can end up with a scalar potential in the QM on the Taub-NUT space
\begin{align}
  \CL_\text{QM}^b = \frac{H(r)}{2} \frac{d \vec r}{dt} \cdot \frac{d \vec r}{dt}
  + \frac{H(r)^{-1}}{2} \left( \frac{d \k}{dt} + \frac12 \cos\th \frac{d\th }{dt}\right)^2
  - \frac{1}{2} H(r)^{-1}w_\text{QM}^2\ .
  \label{TNQM1}
\end{align}
This quantum mechanical model describes
the center of mass motion of a string winding a circle and
the above attractive potential accounts for the tension of the string which prefer to
slip off at the origin.
However one can show below that there are ``stable'' BPS configurations rotating around the circle where the above attractive potential can be balanced by the angular momentum barrier.
These BPS configurations can explain various features of the holomorphic
part of the elliptic genus, as we will discuss shortly.

This system arises from the low-energy dynamics of a pair of distinct monopoles in
$\CN=4$ $SU(3)$ SYM in four dimensions. At the generic point
on the Coulomb branch, it is known that there can be $1/4$ BPS states
carrying nonparallel magnetic and electric charge. These quarter BPS
dyons can be identified as three-pronged strings ending on
misaligned three D3-branes. As shown in \cite{Bak:1999da}, one can study
various quantum aspects of $1/4$ BPS dyons from the $1/2$ BPS states of the
$\CN=4$ QM mechanics whose bosonic part is given exactly by (\ref{TNQM1}).
The asymptotic value of the attractive potential is proportional to misalignment of
the vev of scalar fields, denoted by a parameter $a$, i.e.
\begin{align}
  g w_\text{QM} \propto a\ .
\end{align}
Note that the momentum $p$
conjugate to the angle $\kappa$ of the Taub-NUT space can be identified as the electric charge of a
$1/4$ BPS particle.

\begin{figure}
\begin{center}
  \includegraphics[width=12cm]{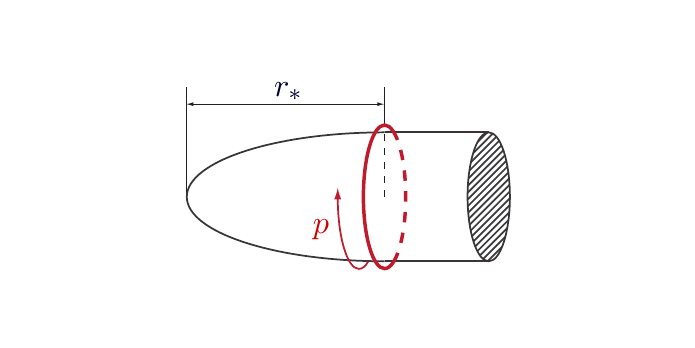}
  \caption{The tension of a string that winds the Taub-NUT circle can be balanced by the angular momentum
  barrier, which leads to a ``stable'' BPS state contributing to the elliptic genus. However, as the size
  of the circle varies, this BPS state can become unstable.}\label{spinning}
\end{center}
\end{figure}

For instance, it is shown in \cite{Lee:1998nv} that if $|p| \geq 2$ the bound states are in four
multiplets with angular momentum $j=\frac{|p|}{2}$, $\frac{|p|-1}{2}$,
$\frac{|p|-1}{2}$, and $\frac{|p|}{2}-1$. Total degeneracy is
\begin{align}
  \left( 2|p| + 1 \right) + 2 \times \left( 2|p| \right) + \left(2|p|-1\right) = 8 |p|\ .
\end{align}
When $|p| = 1$, the corresponding supermultiplet consists of three
angular multiplets with $j=\frac{1}{2}$, $0$ and
$0$. Finally, it is important that there is the unique
threshold bound state carrying no electric charge $p=0$. These bound states
describe stable spinning-string configurations that wind the Taub-NUT circle,
as depicted in the figure \ref{spinning}.

However these $1/2$ BPS states in the SQM (or $1/4$ BPS states in the 4d gauge
theory) can become unstable as the parameter $a$ varies, which is known as the wall-crossing
phenomenon. Let us discuss when the wall-crossing happens.
For a half-BPS state of the momentum charge $p$, the effective potential becomes
\begin{align}
  V_\text{eff}(r) = \frac12 H(r) p^2 + \frac{1}{2} H(r)^{-1}w_\text{QM}^2\ ,
\end{align}
from which one can show that the bound state distance $r_\ast$ is given by
\begin{align}
  2r_\ast = \frac{g^2 |p|}{g^2 |w_\text{QM}| - |p|} \ ,
\end{align}
and the energy $E$ of such a $\frac12$-BPS state, or equivalently the absolute value of the central charge $Z$,
is given by
\begin{align}
  E = |Z| = V_\text{eff}(r_\ast) = |pw_\text{QM}|\ .
\end{align}
It implies that there exist BPS states only when
\begin{align}
  g^2 |w_\text{QM}| - |p| > 0 \ ,
\end{align}
and the wall of marginal stability is located at
\begin{align}
  g^2 |w_\text{QM}| - |p| = 0\ ,
  \label{wcf}
\end{align}
where the bound state distance $r_\ast$ becomes infinity.

Upon the Scherk-Schwarz compaction, the elliptic genus defined in (\ref{def}) can reduce to a
supersymmetric index
\begin{align}
  \CI_{w_\text{QM}}(\t;\xi_1,\xi_2,z) = \text{Tr}\Big[ (-1)^F q^{L_0} \bar q^{\bar L_0}
  e^{-2\pi i \xi_1 Q_f} e^{-2\pi i \xi_2 (j_1-j_2)}
  e^{-2\pi i z j_3} \Big]\ ,
  \label{def2}
\end{align}
which counts the $\frac12$-BPS states with $\bar L_0 = \frac12 (E+Z) \doteq 0$ satisfying
\begin{align}
  L_0 \doteq |pw_\text{QM}| \ , \qquad p \cdot w_\text{QM} \leq 0 \ .
\end{align}
%
Putting the above facts altogether, the index should become
\begin{align}
  \CI_{w_\text{QM}>0}(\t;\xi_1,\xi_2,0) = \hspace*{-0.5cm}\sum_{ - g^2 w_\text{QM}<p<0}\hspace*{-0.5cm}
  q^{|pw_\text{QM}|} e^{2\pi i p\xi_1}
  \left( \chi_{\bf \frac{|p|}{2}} - 2 \chi_{\bf \frac{|p|-1}{2}}
  + \chi_{\bf \frac{|p|}{2}-1} \right) + \cdots\ ,
  \label{resultI1}
\end{align}
where the contribution from the bound states are presented only. Similarly,
\begin{align}
  \CI_{w_\text{QM}<0}(\t;\xi_1,\xi_2,0) = \hspace*{-0.5cm}\sum_{ 0<p < - g^2 w_\text{QM}}\hspace*{-0.5cm}
  q^{|pw_\text{QM}|}  e^{2\pi i p\xi_1}
  \left( \chi_{\bf \frac{|p|}{2}} - 2 \chi_{\bf \frac{|p|-1}{2}}
  + \chi_{\bf \frac{|p|}{2}-1} \right) + \cdots\ .
  \label{resultI2}
\end{align}
The above two indices for $w_\text{QM}>0$ and $w_\text{QM}<0$ are not completely independent of
the parameter $g^2$, but piecewise constant.
Finally, one obtains
\begin{align}
  \CI_{w_\text{QM}=0}(\t;\xi_1,\xi_2,0) = 1 + \cdots\ ,
  \label{resultI3}
\end{align}
which is independent of $g^2$.

From (\ref{resultI1}), (\ref{resultI2}), and (\ref{resultI3}), one can thus expect that
the discrete part of the elliptic genus $\CE_\text{disc}$ (with $z=0$) should contain the following terms
\begin{align}
  \CE_\text{disc}(\t;\xi_1,\xi_2,0) & = 1 + \sum_{p>0}\sum_{p+g^2w_\text{QM}<0}
  q^{|pw_\text{QM}|}  e^{2\pi i p\xi_1}
  \left( \chi_{\bf \frac{|p|}{2}} - 2 \chi_{\bf \frac{|p|-1}{2}}
  + \chi_{\bf \frac{|p|}{2}-1} \right)
  \nonumber \\ & +
  \sum_{p<0}\sum_{p+g^2w_\text{QM}>0}
  q^{|pw_\text{QM}|}  e^{2\pi i p\xi_1}
  \left( \chi_{\bf \frac{|p|}{2}} - 2 \chi_{\bf \frac{|p|-1}{2}}
  + \chi_{\bf \frac{|p|}{2}-1} \right) + \cdots\ .
\end{align}
However, unlike the index in SQM, it is not easy to see clearly
the contribution to $\CE_\text{disc}$ from $|w_\text{QM}|>|w_\text{QM}^{min}|$ for each $p$ where
\begin{align}
  0<|p+g^2w_\text{QM}^{min}|<g^2 \ .
\end{align}
This is because these contributions are mixed up with the contributions from the world-sheet oscillator modes. Therefore,
one can expect to see the following terms clearly in the
discrete part of the elliptic genus
\begin{align}
  \CE_\text{disc}(\t;\xi_1,\xi_2,0) & = 1 + \sum_{p>0}\sum_{-g^2<p+g^2w_\text{QM}<0}
  \hspace*{-0.7cm}
  q^{|pw_\text{QM}|}  e^{2\pi i p\xi_1}
  \left( \chi_{\bf \frac{|p|}{2}} - 2 \chi_{\bf \frac{|p|-1}{2}}
  + \chi_{\bf \frac{|p|}{2}-1} \right)
  \nonumber \\ & +
  \sum_{p<0}\sum_{0<p+g^2w_\text{QM}<g^2}
  \hspace*{-0.7cm}
  q^{|pw_\text{QM}|}  e^{2\pi i p\xi_1}
  \left( \chi_{\bf \frac{|p|}{2}} - 2 \chi_{\bf \frac{|p|-1}{2}}
  + \chi_{\bf \frac{|p|}{2}-1} \right) + \cdots\ ,
  \label{result4}
\end{align}
which is in perfect agreement with what we obtain in (\ref{holpp}), (\ref{holnp}), and (\ref{holp0}).

\vspace{0.2cm}

We can now make the equation~\eqref{Ediscsim} for the generating function of the discrete states
more precise. The dependence of the discrete spectrum on~$g^{2}$ was given in Equation~\eqref{useful2},
which was derived for real values of~$\xi_{1}$.
Since for each value of~$p$ we have only one value
of~$w \in I_p = \big\{ w \in \mathbb{Z} \big|  \e - 1 -\frac{p}{g^2} < w < \e -\frac{p}{g^2}\big\}$,
and since~$w$ only enters the equation via the contour, we can rewrite~\eqref{useful2} as:
\be \label{FexpEdisc}
 \CE^{(g)}_\text{disc} (\t,z,\xi_{1},\xi_2) \=
  \sum_{p} e^{2\pi i p \xi_1 }\int_{u_{0}(p,g)}^{u_{0}(p,g)+1} \, du \ e^{-2 \pi i p u} \, \v(\t,z, u,\xi_2) \, ,
\ee
where~$\Im(u_{0}(p,g))$ is determined by the condition $-w \, \Im(\t) < \Im(u_{0}) < (-w+1) \Im (\t)$.
The poles of~$\v(\t,z,u,\xi_{2})$ lie at~$u + \xi_{2} \in \IZ \t + \IZ $,
so that for real values of~$\xi_{2}$, this choice of contour determine a chamber
in which there is an unambiguous Fourier expansion of~$\v$.
Since~$u$ is a dummy variable on the right hand side of~\eqref{FexpEdisc}, we can choose~$u=\xi_{1}$,
and thus identify the partition function of discrete states with the function~$\v$, with a choice of
chamber for the imaginary part of~$\xi_{1}$.
\be
\CE^{(g)}_\text{disc} (\t,z,\xi_{1},\xi_2) \= \v(\t,z,\xi_{1},\xi_2) \, , \quad \text{with}
\ee
\be \label{defcontour}
\quad -w \, \Im(\t) < \Im(\xi_{1}) < (-w+1) \Im (\t) \, , \quad  w \in I_{p} \, .
\ee
Thus, while~$\Re(\xi_{1})$ is the chemical potential for the momentum~$p$, the value of
$\Im(\xi_{1})$ (more precisely, the chamber in which it lives) is determined by the winding number~$w$.
This picture is reminiscent of the spectrum of fundamental strings in~$AdS_{3}$ \cite{Maldacena:2000kv}.

\subsection{Fourier expansion of the discrete spectrum \label{secdecomp}}

In this section we study the function,
\be \label{defF}
\v(\t,z, u,\xi_2) \= \frac{\myth1(\t,u+z)\, \myth1(\t,u-z)}{\myth1(\t,u+\xi_2) \,
  \myth1(\t,u-\xi_2)} \, ,
\ee
and its Fourier coefficients. We assume below that~$z,u,\xi_{2}$ are complex variables.
There are three related reasons that we study this function.
Firstly, the  function~$\v$ with~$u\to\xi_{1}$ is the large radius
limit $g^{2} \to \infty$ of the Taub-NUT elliptic genus~\eqref{ell_largesize}. Secondly, the same function
with $u \to u+\xi_1$ appears in the integrand of the expression~\eqref{result1} for
the full elliptic genus~$\CE$ of Taub-NUT space. Thirdly, the same function, written with~$u \to \gamma$ is the generating
function for the discrete part of the elliptic genus~$\CE_\text{disc}$ \eqref{useful2}.

In all these applications, it is interesting to understand the Fourier coefficients of~$\v$ as a function of the variable~$u$.
With this in mind, we shall sometimes use~$\v(u) = \v(\t,z,u,\xi_{2})$.
The function~$\v$ is a periodic function in~$u$ with period~$\IZ \t + \IZ$:
\be
\v(u+m \t +n) \= \v(u) \, , \qquad m, n \in \IZ \, ,
\ee
i.e.~it has index 0 under elliptic transformations\footnote{Further, $\v$ has index~$1$ under elliptic
transformations of $z$ and index~$-1$ under those of~$\xi_{2}$. See Appendix \S\ref{JacApp} for the relevant
definitions.} of~$u$. There are two poles in any fundamental domain of~$u$, each corresponding to a certain
$\IZ \t + \IZ$ translate of~$u = \pm \xi_2 $. The residues at these poles are given by:
\be \label{resphi}
R(\t, z, \xi_{2}) \equiv \pm 2 \pi i \, {\rm Res}_{u = \pm \xi_2+\IZ \t + \IZ} \, \v(u) \=
- i \frac{\myth1(\t,\xi_2+z)\, \myth1(\t,\xi_2-z)}{\myth1(\t,2\xi_2) \, \eta(\t)^{3}} \, .
\ee
Note that the residues are the same at all the points by the periodicity of~$\v$. This implies that
the sum of the residues of~$\v$ inside any fundamental domain vanishes.

The presence of these poles in~$\v$ implies that its Fourier expansion is ambiguous, and that the Fourier
coefficient with respect to~$u$ depends on the location of~$u$. The general treatment of such a problem has
been presented in~\cite{Dabholkar:2012nd}, which we now follow.
We will decompose the meromorphic function~$\v$ into two pieces, one (the ``finite piece'')
having a Fourier expansion that is independent of the location of~$u$, and the other (the ``polar piece'')
containing all the wall-crossing information in~$u$.

To construct the polar part, we define the sum:
\be \label{T1}
T_{1}(\t,u,\xi_2) \= \sum_{\l \in \IZ} \Big( \frac{1}{1 - q^{\l} \, e^{2 \pi i (u+\xi_2)}} - \frac{1}{1 - q^{\l} \, e^{2 \pi i (u-\xi_2)}} \Big) \, .
\ee
The function~$T_{1}$ is a difference of two terms, each of which is an Appell-Lerch sum of index 0, i.e. an average of the
function~$\frac{1}{1-e^{2 \pi i (u+\xi_{2})}}$ over the lattice~$\IZ\t + \IZ$. One of the subtleties that often appears in the
treatment of such sums is the precise choice of rational function -- for example, the
function~$\half \frac{1+e^{2 \pi i (u+\xi_{2})}}{1-e^{2 \pi i (u+\xi_{2})}}$ has the same limiting behavior as $\frac{1}{1-e^{2 \pi i (u+\xi_{2})}}$ as~$(u+\xi_{2}) \to 0$ but the Fourier expansion is different
(see the discussion in~\S8.2 of \cite{Dabholkar:2012nd}).
In our case, this discussion is moot because we can also write the function~$T_{1}$ as:
\be
T_{1}(\t,u,\xi_2) \= \sum_{\l \in \IZ} \Big( \frac{1+q^{\l} \, e^{2 \pi i (u+\xi_2)}}{1 - q^{\l} \, e^{2 \pi i (u+\xi_2)}} -
\frac{1+q^{\l} \, e^{2 \pi i (u-\xi_2)}}{1 - q^{\l} \, e^{2 \pi i (u-\xi_2)}} \Big) \, .
\ee
We now write down the decomposition of the function~$F$ into a finite and a polar part.

\vspace{0.2cm}

\ndt {\bf Decomposition of the function~$\bf \v$.}
We have:
\be \label{decomp1}
\v \= \v^{F} + \v^{P} \, ,
\ee
with
\be \label{defphiF}
\v^{F}(\t,z,\xi_{2}) \= -R(\t, z, \xi_{2})\, T_{1}(\t,z,\xi_2) \, ,
\ee
\be\label{defphiP}
\v^{P}(\t,z,u,\xi_{2}) \= R(\t, z, \xi_{2}) \, T_{1}(\t,u,\xi_2) \, ,
\ee
where the function~$R$ is defined in~\eqref{resphi}, and the function~$T_{1}$ is defined in~\eqref{T1}.

\begin{proof}
Consider the function
\bea \label{defG}
G(u) \= G(\t, z, u, \xi_2) &\=& \frac{\v(\t,z,u,\xi_{2})}{R(\t,z,\xi_{2})} \cr
& \= & \frac{\myth1(\t,u+z)\, \myth1(\t,u-z)}{\myth1(\t,u+\xi_2) \, \myth1(\t,u-\xi_2)} \,
  \frac{i \, \myth1(\t,2\xi_2) \, \eta(\t)^{3}}{\myth1(\t,\xi_2+z)\, \myth1(\t,\xi_2-z)} \,.
\eea
This function~$G(u)$ has a periodicity $G(u+m \t +n) \= G(u)$,
with $m, n \in \IZ$ that descends from the periodicity of~$\v$.
It is a meromorphic function of~$u$ with poles at~$u=\pm \xi_2$ (and its translates by $(\IZ \t + \IZ)$),
with residues~$\pm 1$.  The function~$T_{1}(\t,u,\xi_2)$ is a periodic function of~$u$ with the same poles and
residues as~$G$. Thus the function~$G(u)-T_{1}(\t,u,\xi_2)$ is a periodic function of~$u$ that is holomorphic
in~$u$, implying that it is independent of~$u$. The last implication holds because of Liouville's theorem --
extending the function to the complex plane by periodicity, we get a bounded analytic function on the
complex plane which is therefore constant.

The function~$G$ is also periodic in~$z$ with period~$\IZ\t+\IZ$, and it is meromorphic in~$z$ with
poles at~$z = \pm \xi_2+\IZ\t+\IZ$ with residues~$\mp 1$.
Since~$T_{1}(\t,u,\xi_2)$ is independent of~$z$, the $z$-dependence
of~$G(\t,z,u, \xi_2) -T_{1}(\t,u,\xi_2)$ is the same as that of~$G$. By construction,
$-T_{1}(\t,z,\xi_2)$ has the same poles in~$z$ and the same residues at those poles
as~$G(\t,z,u, \xi_2) -T_{1}(\t,u,\xi_2)$ which implies that
$G(\t,z,u, \xi_2)-T_{1}(\t,u, \xi_2)+T_{1}(\t,z, \xi_2)$ is periodic and holomorphic in~$z$, and
therefore independent of~$z$.
Since it is independent of~$u$ and~$z$, the function~$G(\t,z,u, \xi_2)-T_{1}(\t,u, \xi_2)+T_{1}(\t,z, \xi_2)$
is equal to its value at any choice of~$u$ and~$z$.
A convenient choice is~$u=z$ at which $G(\t,z,u, \xi_2)-T_{1}(\t,u, \xi_2)+T_{1}(\t,z, \xi_2)$
vanishes, thus proving that it identically vanishes.
\end{proof}

\ndt {\bf Comments}:
\begin{enumerate}
\item \emph{The function~$\v^{F}$ is independent of~$u$}. This is consistent with the
analysis in \S\ref{sepdisc} where we found that the BPS states that are independent of~$g^{2}$ are those with
with zero momentum~$p=0$ around the Taub-NUT circle (recall that the charge conjugate to~$u$ is precisely this momentum~$p$).
The contour~$u=u_{0}$ defined for the finite part is, in general,~$2m u_{0}=p\t$ where~$m$ is the index of the
meromorphic Jacobi form. When $m=0$, as in our case, the coefficient of~$p=0$ does not depend on the contour~$u_{0}$.
This is borne out explicitly in our analysis by the fact that the sum of the residues of~$\v$ inside any fundamental
domain vanishes, as mentioned below~\eqref{resphi}.

\item \emph{Elliptic properties of~$\v^{F}$ and~$\v^{P}$.}
The functions~$\v^{F}$ and~$\v^{P}$ are products of the residue function~$R$~\eqref{resphi} and the Appell-Lerch
sum~$T_{1}$ \eqref{T1}.
The function~$R$ inherits its elliptic transformation properties from those of the Jacobi theta function (see Appendix~\S\ref{JacApp}).
The Appell-Lerch sum~$T_{1}(\t,u,\xi_{2})$, by construction, is an average over the lattice~$\IZ\t+\IZ$ and consequently
is invariant under elliptic transformations of~$u$ and~$\xi_{2}$. Putting together the various pieces, we deduce that the
functions~$\v^{P}$ and~$\v^{F}$ have the same elliptic transformation properties as those of the function~$\v$, i.e.
they both have index~$1$ in the variable $z$, index 0 in~$u$, and index~$-1$ in~$\xi_{2}$.

\item \emph{Modular properties of~$\v^{F}$ and~$\v^{P}$.}
The modular transformations of the Jacobi theta and Dedekind eta functions are well-understood, and
the new element is the modular transformation of the function~$T_{1}(\t,u,\xi_2)$. This can be derived by expressing~$T_{1}$
in terms of the Jacobi~$\vth_{1}$ function. We have:
\be \label{T1log}
T_{1}(\t,u,\xi_2) \= \frac{1}{2 \pi i } \frac{d}{du} \log \frac{\vth_{1}(\t,u+\xi_2)}{\vth_{1}(\t,u-\xi_2)} \, ,
\ee
which can be verified using the expansion (with $' \equiv \frac{1}{2 \pi i } \frac{d}{dz}$)
\be \label{logth}
\frac{\vth'_{1}(\t,z)}{\vth_{1}(\t,z)} \= \frac{1}{2 \pi i } \frac{d}{dz} \log \vth_{1} (\t,z) \=
\sum_{\l \in \IZ} \Big( \frac{1}{1 - q^{\l} \, e^{2 \pi i z}} - {\rm sgn}(\l) \Big) \, .
\ee
The function~$\vth'/\vth$ is invariant under~$\t \to \t+1$. Its modular property can be derived from that
of the function~$\vth_{1}$. We have:
\be \label{thlogmod}
\frac{\vth'_{1}(-1/\t,z/\t)}{\vth_{1}(-1/\t,z/\t)} = z + \frac{\vth'_{1}(\t,z)}{\vth_{1}(\t,z)} \, .
\ee
From the expressions \eqref{T1log}, \eqref{logth}, \eqref{thlogmod}, we can write down the modular properties of the function~$T_{1}$:
\be
T_{1}(\frac{a\t+b}{c\t+d}, \frac{u}{c\t+d},  \frac{\xi_2}{c\t+d}) \= 2\xi_2 + T_{1}(\t,u,\xi_2) \, .
\ee

\item We can combining the expressions~\eqref{decomp1} and~\eqref{T1log} to rewrite the function~$\v$ as:
\bea
\v(\t,z,u,\xi_2) & \= & R(\t,z,\xi_{2}) \, \frac{1}{2 \pi i} \,\frac{d}{du} \, \v(\t,z,u,\xi_{2}) \\
& \= & - \frac{\myth1(\t,\xi_2+z)\, \myth1(\t,\xi_2-z)}{2 \pi \, \myth1(\t,2\xi_2) \, \eta(\t)^{3}}
\,\frac{d}{du}
\log \frac{\vth_{1}(\t,u+\xi_2) \, \vth_{1}(\t,z-\xi_2)}{\vth_{1}(\t,u-\xi_2) \, \vth_{1}(\t,z+\xi_2)} \, . \nonumber
\eea

\end{enumerate}

\vspace{0.2cm}

\ndt {\bf Wall-crossing}

We can now easily write Fourier expansions in all regimes of all the variables.
As we saw above, the finite piece~$\v^{F}$  is independent of~$u$ and is holomorphic in~$z$.
It does not experience wall-crossing as we vary~$u$. It does, however, have poles in~$\xi_{2}$ which also causes its
own wall-crossing -- this can also be worked out using the methods of~\cite{Dabholkar:2012nd}, but we shall not do so here.

The wall-crossing in~$u$ is governed by the polar part~$\v^{P}$, and it happens each time~$\Im(u + \l \t) \pm \Im(\xi_2)$ changes sign.
Let's denote these values of~$\l$
by~$\l_{\pm}$. Then we have the following Fourier expansion for each term in the sum in~\eqref{T1}
\be
\frac{1}{1 - e^{2 \pi i (u\pm \xi_2 + \l_{\pm} \t)}} \=
\begin{cases} \label{FExpJumps}
\sum_{n \ge 0} \, e^{2 \pi i n u} \, e^{2 \pi i n (\xi_2\pm \l_{\pm} \t)} \, , \qquad \Im(u + \l \t) \pm \Im(\xi_2)> 0 \, ,  \\
- \sum_{n < 0} \, e^{2 \pi i n u} \, e^{2 \pi i n (\xi_2\pm \l_{\pm} \t)} \, , \qquad \Im(u + \l \t) \pm \Im(\xi_2) < 0 \, .
\end{cases}
\ee
From~\eqref{decomp1}, the Fourier coefficient of~$e^{2 \pi i p u}$ changes
(on crossing both the walls in an upward direction) by:
\be
\frac{\myth1(\t,\xi_2+z)\, \myth1(\t,\xi_2-z)}{\myth1(\t,2\xi_2) \, \eta(\t)^{3}} \,
\big(e^{2 \pi i p (\l_{+} \t + \xi_2)} - e^{2 \pi i p (\l_{-} \t -\xi_2)} \big)\, .
\ee

We thus obtain the following picture of wall-crossing:
As~$g^{2} \to \infty$, only the $w=0$ mode contributes, and the Fourier coefficients are given by an
expansion of~$\v$ with~$\xi_{1}$ having a
small positive imaginary part:~$0 < \Im(\xi_{1}) < \Im(\t)$. As we reduce~$g^{2}$, the Fourier coefficient
of~$e^{2 \pi i p}$ for each value of~$p\neq 0$ jumps
when~$p/g^{2}$ crosses an integer. The function~$\v^{P}$ encodes the
jumps in the spectrum in a simple manner. In particular, each wall corresponds to exactly one term
in the function~$T_{1}$ \eqref{T1} becoming singular, and its Fourier expansion on either side of each wall
has the simple expression given in~\eqref{FExpJumps}. The $p=0$ coefficient is
independent of~$g^{2}$, and is captured by the function~$\v^{F}$.

\subsection{Double pole case}
As we have discussed in~\S\ref{holprop} and \S\ref{nonhol},
the finite~$g^{2}$ elliptic genus is divergent as~$\xi_{2} \to 0$. The large radius limit~\eqref{ell_largesize}, on the other hand,
has no such issues and we can also study the Fourier coefficients of the generating function~$\v$ defined in \eqref{defF}
with~$\xi_{2}=0$.
In this case, the function~$\v$ has a double pole, and the mathematical properties are somewhat different compared
to the single pole case studied in the previous section. We proceed to study this case below.
We begin by defining the new generating function:
\be
\wt \v(\t,z,u) \, \equiv \, \v(\t,z,u,\xi_2=0)  \= \frac{\myth1(\t,u+z)\, \myth1(\t,u-z)}{\myth1(\t,u)^{2}} \, ,
\ee
which has double poles at $u \in \IZ \t + \IZ$, with a behavior near~$u =0$:
\be
\wt \v(\t,z,u=\ve) \= -\frac{1}{(2 \pi i \ve)^{2}} \, \v_{-2,1}(\t,z) \+ O(1)\, ,
\ee
where
\be
 \v_{-2,1}(\t,z)\, := \, \frac{\myth1(\t,z)^{2}}{\eta(\t)^{6}} \, .
\ee

In this case as well, we can write a decomposition of~$\wt \v$ into finite and polar parts.
Define the function
\be \label{T2}
T_{2}(\t,u) \= - \sum_{\l \in \IZ}  \frac{q^{\l} \, e^{2 \pi i u}}{(1 - q^{\l} \, e^{2 \pi i u})^{2}}  \, .
\ee

\ndt {\bf Decomposition of the function~$\bf \wt \v$.}
The main proposition in this case is:
\be \label{decomp2}
\wt \v(\t,z,u) \=  \v^{F}(\t,z) + \v^{P}(\t,z,u)
\ee
where
\be
\wt \v^{F}(\t,z) \=  -  \v_{-2,1}(\t,z) \, T_{2}(\t,z) \, ,
\ee
\be
\wt \v^{P}(\t,z,u) \=   \v_{-2,1}(\t,z) \, T_{2}(\t,u) \, .
\ee
The proof of this proposition is similar to the proof of the decomposition of the function~$\v$ in \S\ref{secdecomp} --
we compare the poles and residues of each side of the equation~\eqref{decomp2}, and fix constants by computing a finite number
of coefficients. We shall not spell out the details.

The function~$T_{2}$ is related to the Weierstrass~$\wp$ function as follows~(see \cite{Dabholkar:2012nd}, \S 8.5):
\be \label{T2wp}
12\,T_{2}(\t,u) \= \frac{3}{\pi^{2}} \wp(\t,u) + E_{2}(\t) \, ,
\ee
from which we can deduce the modular and elliptic behavior of ~$\v^{F}$ and~$\v^{P}$.
The Weierstrass~$\wp$ function is a (meromorphic) Jacobi form of weight~$k=2$ and index~$0$.
The only lack of modularity comes from the transformation of the function~$E_{2}$, which is
a quasi-modular form, i.e.~$\wh E_{2} (\t)= E_{2}(\t) - \frac{3}{\pi \t_{2}}$ transforms as a modular form of weight~2.

The Fourier coefficients of the finite part~$\wt \v^{F}$ can be easily read off from the expansion~\eqref{T2}.
As a check, we compute the coefficients at~$z=-1/2$:
\bea
\wt \v^{F}(\t,z=-1/2) = &&  1 + 16\, q + 96\, q^2 + 448\, q^3 + 1728\, q^4 + 5856\, q^5 + 18048\, q^6 + \cr
&& \quad 51584\, q^7 + 138624\, q^8 + 353872\, q^9 + O(q^{10}) \, ,
\eea
 which agrees with the numbers computed in \cite{Kim:2011mv}.

The Weierstrass~$\wp$ function obeys the relation:
\be
- \frac{3}{\pi^{2}} \wp(\t,z) \= \frac{\v_{0,1}(\t,z)}{\v_{-2,1}(\t,z)} \, ,
\ee
where~$2 \v_{0,1}(\t,z)$ is the elliptic genus of the $K3$ surface.
Using the relations~\eqref{decomp2}, \eqref{T2wp}, we obtain an alternate expression for~$\wt \v$:
\bea \label{exp2again}
\wt \v(\t,z,u) &\= & \frac{3}{\pi^{2}} \v_{-2,1}(\t,z) \, (-\wp(\t,z)+\wp(\t,u)) \, , \cr
 &\= & \v_{0,1}(\t,z) +\frac{3}{\pi^{2}} \, \v_{-2,1}(\t,z) \, \wp(\t,u) \, .
\eea

\section{Application and Generalization \label{AppGen}}

\subsection{1/4-BPS State Counting in 5d $\CN=2$ SYM}
Recently the maximally supersymmetric Yang-Mills theories in five dimensions
have been studied to understand various aspects of $\CN=(2,0)$ superconformal
theories in six dimensions. See for instance \cite{Kallen:2012va,Kim:2012ava,Kim:2012qf}.
In particular, the Witten
index counting the $1/4$-BPS states in the Coulomb phase of the
5d $\CN=2$ $U(N)$ SYM is computed in \cite{Kim:2011mv}, from which one can read
off the BPS spectrum of self-dual strings in 6d $\CN=(2,0)$ A$_{N-1}$ theories.
The BPS spectrum of self-dual strings have been also studied in
\cite{Haghighat:2013gba,Haghighat:2013tka}
from the dual IIA picture and in \cite{Hosomichi:2014rqa} from the ABJM model on the boundary.
In this section we will show that the elliptic genus of the Taub-NUT space
can capture certain $1/4$ BPS states in the 5d gauge theory\footnote{
In \cite{Bak:2014xwa}, the authors addressed a similar question and proposed an elliptic
genus formula of the Taub-NUT space. However their elliptic genus is a Jacobi form of weight zero
on $\G(2)$ rather than on the full modular group and thus does not agree with our result even in the large radius limit.}.

Let us first describe such $1/4$ BPS states in
type IIA string theory or M-theory.
\begin{figure}[t]
  \begin{center}
    \hspace*{1cm}\includegraphics[width=13cm]{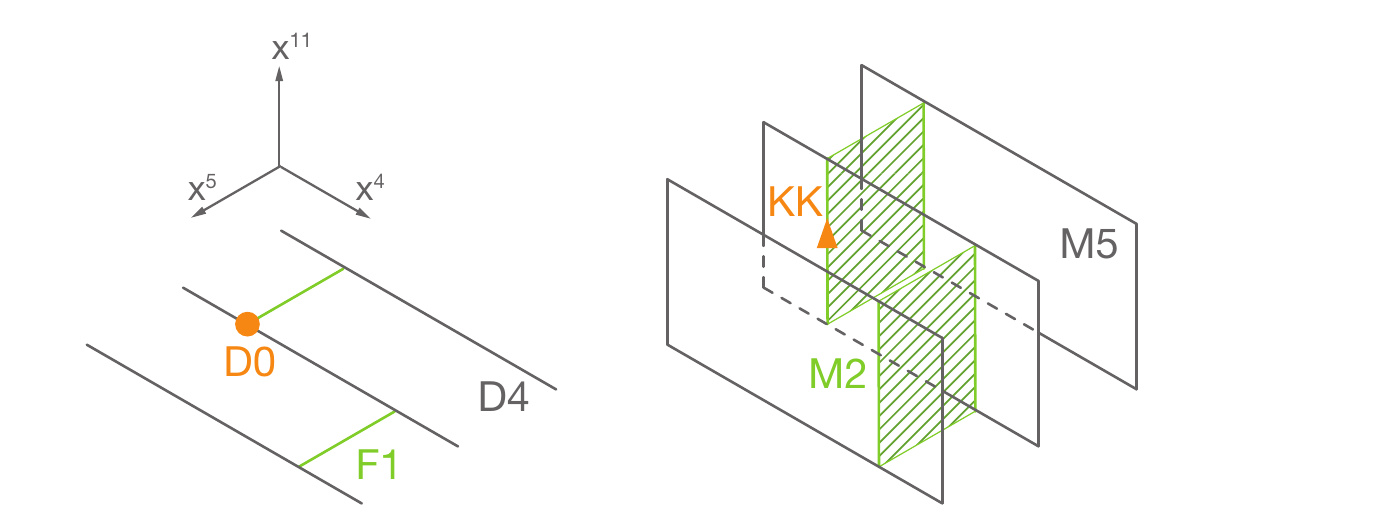}
    \caption{The IIA brane configuration of a $1/4$ BPS state carrying
    both instanton and electric charges in 5d $\CN=2$ super Yang-Mills theory.
    These $1/4$ BPS states uplift to the BPS spectrum
    of M2 self-dual strings.}\label{m2-m5}
  \end{center}
\end{figure}
The brane realization consists of $N$ parallel D4-branes and $k$ D0-branes
together with several fundamental strings. They can be uplifted to the non-zero KK momenta on
M2-M5 brane systems. See Fig. \ref{m2-m5} for the brane configuration.
We also summarize the world-volume directions of the branes below
\begin{center}
\setlength{\tabcolsep}{0.25cm}
\begin{tabular}{c|c|cccc|c|cccc|c}
  \hline
  & 0 & 1 & 2 & 3 & 4 & 5 & 6 & 7 & 8 & 9 & 11 \\
  \hline
  \hline
  M5 & $\times$ & $\times$ & $\times$ & $\times$ & $\times$ & & & & & & $\times$ \\
  M2 & $\times$ &  &  &  &  & $\times$ & & & & &  $\times$ \\
\end{tabular}\ .
\end{center}

The $1/4$ BPS states carry both $U(1)^N$ electric and instanton charges.
A supersymmetric index that counts both multi- and single-$1/4$ BPS particles in
the 5d $\CN=2$ SYM theory is defined in \cite{Kim:2011mv} as
\begin{align}
  \CI(q,\g,\mu) = \sum_{k=0}^\infty q^k \CI_k
\end{align}
with the fugacity $q$ of the instanton charge (or KK momentum charge along $x^{11}$)
and
\begin{align}
  \CI_k = \text{Tr}_k \Big[ (-1)^F e^{-\vec\mu \cdot \vec \Pi} e^{-\g_L (2J_L)}
  e^{-{\tilde \g}_L (2{\tilde J}_R)} e^{-i\g_R(J_R+{\tilde J}_R)} \Big]\ ,
\end{align}
where the trace is performed over states in the gauged quantum mechanics of
$k$ D0-branes bounded to $N$ parallel D4-branes.
Here $\g_L$ and
$\tilde \g_L$ denote the chemical potentials for the Cartan sub algebra of
$SU(2)_L \subset SO(4)_{1234}$ and $SU(2)_L \subset SO(4)_{6789}$ respectively
while $\g_R$ is the chemical potential for the Cartan sub-algebra of the diagonal subgroup of
$SU(2)_R\subset SO(4)_{1234}$ and $SU(2)_R \subset SO(4)_{6789}$.
The parameter $\vec \mu$ ($\mu_1>\mu_2>..>\mu_N$)
denotes the chemical potential for $U(1)^N$ electric charges of the
5d $U(N)$ gauge theory in the Coulomb phase. One can read off the single-particle index $i_{sp}$
by taking the Plethystic logarithm of the multi-particle index $\CI$, i.e.,
\begin{align}
  \CI(q,\g,\mu) = \text{Exp}\Big[ \sum_{n=1}^\infty \frac{1}{n} i_{sp}(q^n,n \g, n\mu) \Big]\ .
\end{align}
\begin{figure}[t]
  \begin{center}
    \includegraphics[width=13cm]{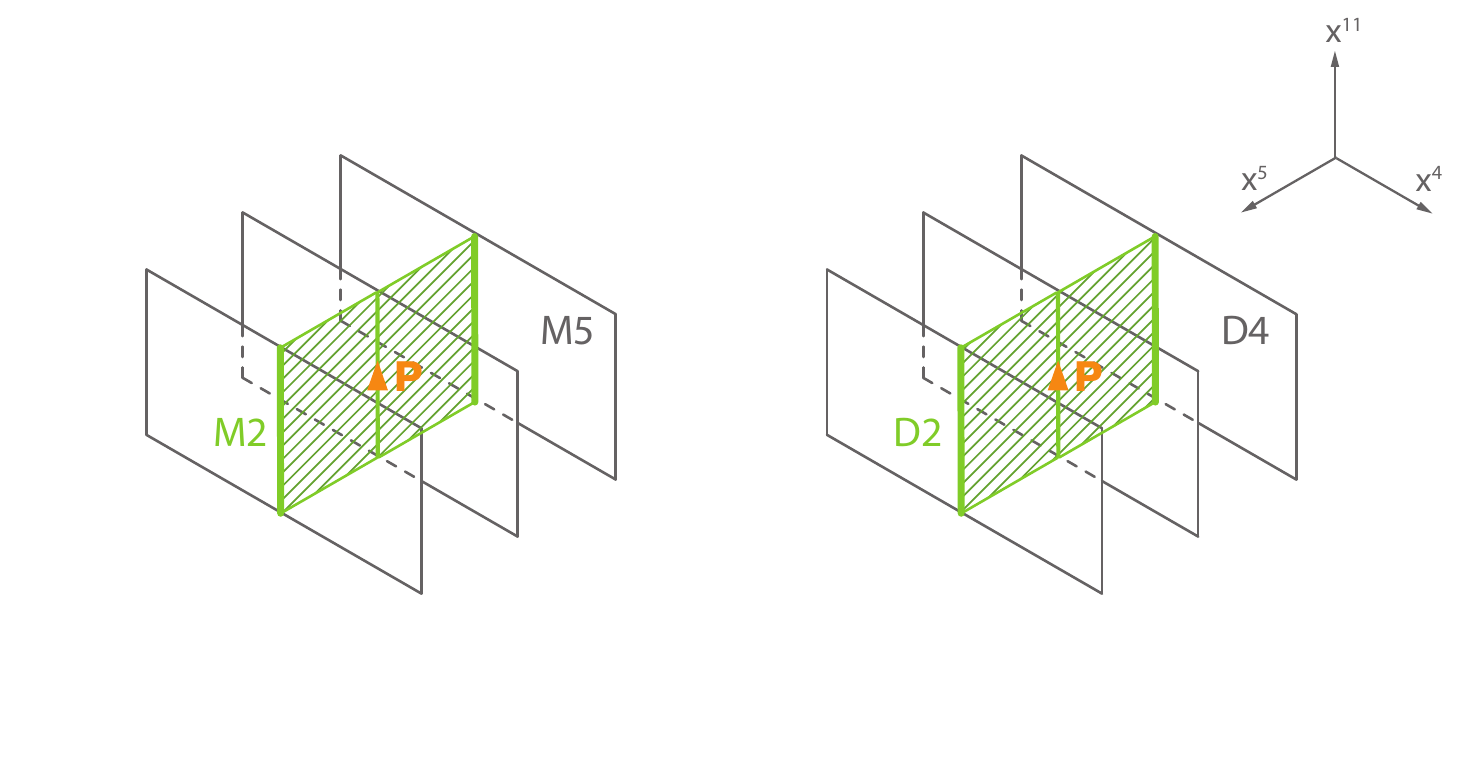}
   \caption{$SU(3)$ self-dual string and a pair of distinct monopole strings. }
    \label{SU(N)}
  \end{center}
\end{figure}
From now on, let us focus on particular $1/4$ BPS single-particle states of
electric charge $\Pi=(1,0,-1)$ and arbitrary instanton charges $k$ in
the 5d $\CN=2$ $U(3)$ gauge theory. These $1/4$ BPS single-particle states describe the BPS states
of the so-called $SU(3)$ self-dual string. The $SU(3)$ self-dual string can
be identified as a M2-brane stretched between
the first and the last M5-branes as depicted in Fig. \ref{SU(N)}.
Note that these states are weighted by the factor $e^{-(\mu_1-\mu_2)}$ in the
single-particle index $i_{sp}^{U(3)}(q,\g,\mu)$ of the $U(3)$ theory.
Expanding the index $i_{sp}^{U(3)}$ in powers of
the chemical potential $e^{-(\mu_i-\mu_j)}$ with $i>j$ where $i$ and $j$
run over $1,2,3$,
\begin{align}
  i_{sp}^{U(3)}(q,\g,\mu) = \sum_{i>j} \sum_{n=0}^\infty i_{ij}^{U(3)}(n,q,\g) e^{-n(\mu_i - \mu_j)}
\end{align}
one can therefore read off the contribution from the above $1/4$ BPS states of electric charge
$\Pi=(1,0,-1)$ from the the coefficient of $e^{-(\mu_1-\mu_3)}$, i.e., $i_{13}^{U(3)}(0,q,\g)$.

Factoring out the contribution from the c.o.m of the above self-dual string $i_{com}$,
\begin{align}
  i^{U(3)}_{13}(q,\g) = i_{com}(q,\g) \times i_{rel}(q,\g) \ ,
\end{align}
it is shown in \cite{Kim:2011mv} that the index $i_{rel}$ is given by\footnote{For the simplicity, the index $i^{U(3)}_{13}$ only with $\g_L =\g_R= 0$ and $\tilde \g_L = \pi$ is presented in \cite{Kim:2011mv}.}
\begin{align}
  i_{rel} & =
  \oint \frac{dx}{x} \prod_{n=1}^\infty \ \Bigg[
  \frac{\big( 1 - q^n  e^{i {\tilde \g}_L}x \big) \big(1-q^{n-1}e^{-i{\tilde \g}_L}x^{-1}\big)}
  {\big( 1 - q^n  e^{i \g_L } x \big) \big(1-q^{n-1} e^{- i\g_L} x^{-1} \big)}
  \frac{\big( 1 - q^n  e^{i {\tilde \g}_L}x \big) \big(1-q^{n-1}e^{- i {\tilde \g}_L}x^{-1}\big)}
  {\big( 1 - q^n  e^{i \g_R}x \big) \big(1-q^{n-1} e^{-i\g_R} x^{-1}\big)} \Bigg]\ ,
  \nonumber \\ & \Bigg. \xrightarrow[\g_L,\g_R=0]{\tilde{\g}_L=\pi}
  1+ 16 q +  96q^2 + 448q^3 + 1728q^4 + 5856q^5 +\cdots \ \Bigg. .
  \label{index2}
\end{align}

Let us compactify the M2-M5 system in Fig. \ref{SU(N)} further on a circle $x^1$, which breaks
the global symmetry $SO(4)_{1234}$ down to $SO(3)_{234}$. We thus need to set
$\g_L =\g_R$. If we take the type IIA reduction along the $x^1$ direction rather than the $x^{11}$ direction, one obtains the D2-D4 system and the BPS states of the $SU(3)$ self-dual string
are mapped to those of a pair of distinct monopole strings in 5d $U(3)$ gauge theory.
More precisely we expect that the discrete part of the elliptic genus of the Taub-NUT space (\ref{useful2}) in the zero momentum sector $p=0$ can capture the BPS states counted by
$i_{rel}(q,\g_L,\tilde \g_L)$. This is because the relative moduli space of the pair of $SU(3)$ distinct monopole strings is the Taub-NUT space.

Indeed, the expression (\ref{index2}) exactly agrees with the discrete part of the elliptic
genus of the Taub-NUT spaces in the sector of $p=0$,
\begin{align}
  \CE_\text{disc}^{p=0} & =
  \int_0^{1} d\g  \
  \frac{\myth1(\t,\g+z)\myth1(\t,\g-z)}
  {\myth1(\t,\g+\xi_2)\myth1(\t,\g-\xi_2)}\ ,
   \\ & =
  \oint \frac{dx}{x} \prod_{n=1}^\infty \ \Bigg[
  \frac{\big( 1 - q^n  e^{2\pi i z}x \big) \big(1-q^{n-1}e^{-2\pi iz}x^{-1}\big)}
  {\big( 1 - q^n  e^{2\pi i \xi_2 } x \big) \big(1-q^{n-1} e^{-2\pi i\xi_2} x^{-1} \big)}
  \frac{\big( 1 - q^n  e^{-2\pi i z}x \big) \big(1-q^{n-1}e^{2\pi i z}x^{-1}\big)}
  {\big( 1 - q^n  e^{-2\pi i \xi_2}x \big) \big(1-q^{n-1} e^{2\pi i \xi_2} x^{-1}\big)} \Bigg]\ ,
  \nonumber
\end{align}
where we have to set
\begin{align}
  \g_L = 2\pi \xi_2\ , \qquad \tilde \g_L = 2\pi z
\end{align}
to match the notation used in \cite{Kim:2011mv}.

\subsection{$\CN=(0,4)$ Elliptic Genera }

The $\CN=(0,4)$ elliptic genus of the Taub-NUT space
plays a key role in understanding the physical interpretation of the Igusa form $\Phi_{10}^{-1}$, which counts the supersymmetric index of quarter BPS dyons in $\CN=4$ string theory \cite{David:2006yn} using the 4d-5d lift \cite{Gaiotto:2005gf} and
which is directly related to the degeneracies of quarter-BPS black holes~\cite{Dabholkar:2010rm}.
It is thus important to generalize our result to the $\CN=(0,4)$ NLSM on the Taub-NUT space.

Let us first start with the $\CN=(0,4)$ GLSM which flows to the $\CN=(0,4)$ NLSM on the multi Taub-NUT space with $N$ coincident centers,
which is nothing but a $\CN=(0,4)$ truncation of the $\CN=(4,4)$ model (\ref{mTNLag}).
The $\CN=(0,4)$ GLSM involves $\CN=(0,4)$ $U(1)$ vector multiplet $(A_\mu,\l_-,\tilde \l_-)$,
coupled to a charged $\CN=(0,4)$ hyper multiplet $(q,\tilde q, \psi_+, \tilde \psi_+)$ and a
St\"uckelberg $\CN=(0,4)$ hyper multiplet $(\vec r,\g, \chi_+, \tilde \chi_+)$.
Note that one can decompose various $\CN=(0,4)$ multiplets into $\CN=(0,2)$ multiplets as follows,
%
\begin{center}
\setlength{\tabcolsep}{15pt}{
\begin{tabular}{c|c}
  \hline
  $\CN=(0,4)$ & $\CN=(0,2)$
  \\ \hline \hline
  \multirow{2}{*}{vector} &  vector  $V=(A_\mu,\l_-)$
  \\
  & fermi  $\Lambda=(\tilde \lambda_-) $
  \\ \hline
  \multirow{2}{*}{charged hyper} & chiral $Q=(q,\psi_+)$
  \\
  & chiral $\tilde Q=(\tilde q, \tilde \psi_+)$
  \\ \hline
  \multirow{2}{*}{St\"uckelberg hyper} & chiral $\Psi=(r_1,r_2,\chi_+)$
  \\
  & chiral $\G=(r_3,\g,\tilde \chi_+)$
  \\ \hline
\end{tabular}
}
\end{center}
One might worry that the above theory suffers from the gauge anomaly. As discussed in \cite{Hori:2001ax},
one can actually construct a modified $U(1)$ gauge current
that is conserved and gauge-invariant due to the compensator field $\g$.

The above $\CN=(0,4)$ GLSM is invariant under $SU(2)_1\times SU(2)_2$ R-symmetry.
Various fields in the model and supercharges $\CQ^{\a\dot \a}_+$
transform under the R-symmetry group as follows:
\begin{align}
  \CQ^{\a\dot\a}_+ \ : & \
  \left( {\bf 2},{\bf 2}\right)
  \nonumber \\
  (\l_-, \tilde \l_-) \ : & \ \left( {\bf 2},{\bf 2}\right)
  \nonumber \\
  (q,\tilde q) \ : & \ \left({\bf 2},{\bf 1} \right)
  \nonumber \\
  (\psi_+, \tilde \psi_+) \ : & \
  \left( {\bf 1},{\bf 2} \right)
  \nonumber \\
  (r_1,r_2,r_3) \ : & \ \left( {\bf 3},{\bf 1} \right)
  \nonumber \\
  \g \ : & \ \left( {\bf 1}, {\bf 1} \right)
  \nonumber \\
  (\chi_+,\tilde\chi_+) \ : & \ \left( {\bf 2},{\bf 2} \right) \ ,
\end{align}
where the indices $\a, \dot a$ are for doublets under $SU(2)_1, SU(2)_2$.

The $\CN=(0,4)$ equivariant elliptic genus is defined by
\begin{align}
  \CE^{(0,4)}(\t; \xi_1,\xi_2) = \text{Tr}_{\CH_\text{RR}} \Big[ (-1)^F q^{L_0} \bar q^{\bar L_0}
  e^{-2\pi i \xi_1 Q_f} e^{-2\pi i \xi_2 (Q_1-Q_2)}
  \Big]\ ,
  \label{(0,4)def}
\end{align}
where $Q_1$ and $Q_2$ are the Cartan generators of $SU(2)_1$ and $SU(2)_2$ respectively while
$Q_f$ is the $U(1)_f$ charge.
The computations are quite parallel to those in section 3. At the end, one can show that
the elliptic genus of the multi-center $A_{N-1}$ Taub-NUT space is
\begin{align}
  \CE_N^{(0,4)}(\t;\xi_1,\xi_2) = \frac{g^2}{N} \int_{\mathbb{C}} \frac{dud\bar u}{\t_2}
  \frac{\eta(q)^2}{\myth1(\t,u+\xi_1+\xi_2)\myth1(\t,u+\xi_1-\xi_2)}
  \sum_{a,b=0}^{N-1} e^{-\frac{g^2}{\t_2}\left|u + \frac{a+\t b}{N} \right|^2}\ .
  \label{(0,4)TN}
\end{align}

\paragraph{Large radius limit}
In the large radius limit, one can obtain
\begin{align}
  \lim_{g^2\to\infty}\CE_N^{(0,4)}(\t;\xi_1,\xi_2)  = \sum_{a,b=0}^{N-1} \frac{\eta(q)^2}{\myth1(\t,\frac{a+b \t}{N}-\xi_1+\xi_2)\myth1(\t,\frac{a+b \t}{N}-\xi_1-\xi_2)}\ .
\end{align}
As discussed in section 4.1, one expects that the above result is the same as the $\CN=(0,4)$ elliptic genus of the A$_{N-1}$ ALE space.
A closely related but slightly different counting problem that is directly relevant to black hole state counting has
been discussed in~\cite{Dabholkar:2008zy}.

Combining the result $\CE_{N=1}^{(0,4)}$ and the additional contribution from four free left-moving fermions, one can show that
\begin{align}
  \lim_{g^2\to\infty}\CE_{N=1}^{(0,4)}(\t,\xi_1,\xi_2=0)  \times \CE_\text{free fermions}
  = \frac{4\eta(q)^6}{\myth1(\t,\xi_1)^2} = \frac{4}{\varphi_{-2,1}(\t,\xi_1)}\ ,
\end{align}
which agrees with the result of David and Sen.

\paragraph{Discrete part} Note that the result of David and Sen
is valid only in the large radius limit. For finite values of $g^2$, the $\CN=(0,4)$
elliptic genus of the Taub-NUT space is non-holomorphic.  As in section 5.1, one can
separate the elliptic genus into a discrete part and a continuum part
\begin{align}
  \CE^{(0,4)}_{N=1} = \CE^{(0,4)}_{N=1,\text{disc}} + \CE^{(0,4)}_{N=1,\text{rest}} \ ,
\end{align}
with
\begin{align}
  \CE_{N=1,\text{disc}}^{(0,4)} = - \frac{1}{\eta(q)^4} \sum_{p,r_1,r_2} \sum_{w\in I_p}
  \d_{r_1+r_2+1+2w-p} \
  q^{w^2-pw} e^{2\pi i p \xi_1 } e^{2\pi i \xi_2 (r_1-r_2)} S_{r_1}(q) S_{r_2}(q)\ ,
  \label{(0,4)TN3}
\end{align}
where $I_p$ is defined as in (\ref{range}). The computational details are presented
in appendix B. Note that $\CE_\text{disc}$ satisfies the charge conjugation
symmetry
\begin{align}
  \CE_{N=1,\text{disc}}^{(0,4)}(\t;\xi_1,\xi_2) =
  \CE_{N=1,\text{disc}}^{(0,4)}(\t;\pm \xi_1,\pm \xi_2) \ .
\end{align}

One can rewrite the discrete part of the elliptic genus (\ref{(0,4)TN3})
as a Fourier expansion of $\lim_{g^2\to\infty} \CE^{(0,4)}_{N=1}$
with a certain choice of contour: first massage
the discrete part of the elliptic genus into the form below
\begin{align}
  \CE_{N=1,\text{disc}}^{(0,4)} & = - \frac{1}{\eta(q)^4}
  \oint_{C^{(0)}} \frac{dx}{2\pi i x}\  \sum_p e^{2\pi i p \xi_1}
  \sum_{w\in I_p}  q^{w^2-pw} x^{2w -p}
  \sum_{r_1} \left( x e^{2\pi i \xi_2} \right)^{r_1+\frac12} S_{r_1}(q)
  \nonumber \\ & \times
  \sum_{r_2} \left( x e^{-2\pi i \xi_2} \right)^{r_2 + \frac12} S_{r_2}(q)
  \nonumber \\ & = \sum_p e^{2\pi i p \xi_1}
  \oint_{C^{(0)}} \frac{dx}{2\pi i x}\  \sum_{w\in I_p}  q^{w^2-pw} x^{2w -p}
  \frac{\eta(q)^2}{\myth1(\t, \g+\xi_2)\myth1(\t, \g-\xi_2) }\ ,
\end{align}
where $x=e^{2\pi i \g}$ and the contour $C^{(0)}$ is chosen as $|x|=x_0$ where $|q|<x_0<1$.
Using the properties of the theta functions (\ref{propertytheta1}),
one can then express the discrete part of the elliptic genus as follows
\begin{align}
  \CE_{N=1,\text{disc}}^{(0,4)} & = \sum_p e^{2\pi i p \xi_1} \sum_{w\in I_p} \oint_{C^{(w)}}
  \frac{dx}{2\pi i x} \ x^{2w - p}  q^{-w^2} \frac{\eta(q)^2}
  {\myth1(\t, \g - \t w + \xi_2)\myth1(\t, \g-\t w -\xi_2) }
  \nonumber \\ & =
  \sum_p e^{2\pi i p \xi_1} \sum_{w\in I_p} \oint_{C^{(w)}}
  \frac{dx}{2\pi i x} \ x^{- p} \frac{\eta(q)^2}
  {\myth1(\t, \g  + \xi_2)\myth1(\t, \g -\xi_2) }\ ,
\end{align}
where the contour $C^{(w)}$ is chosen as $|x| = x_0 q^w$.

An analysis similar to that in~\S\ref{secdecomp} can be performed for the~$(0,4)$ model.
The discrete part of the elliptic genus is now a meromorphic Jacobi form of negative index.
The mathematical analysis of the decomposition of such Jacobi forms into finite and polar pieces
has recently been done in~\cite{Bringmann:2014nba}.

\section{Conclusions and Discussion}

In this paper we computed the (equivariant) elliptic genera of a simple class of non-compact Calabi-Yau spaces using localization
in a gauged non-linear sigma model that flows in the infrared to a two-complex dimensional ALE or ALF space. By studying the equivariant elliptic genus we were able to compute a mathematically well defined quantity and to isolate the divergences which appear in the non-compact elliptic genus when we remove the equivariant action. These divergences appear as poles in the variables $y_1,y_2$ that determine the equivariant action and imply that the equivariant elliptic genus is a
meromorphic, multi-variable Jacobi form. The Fourier expansion of meromorphic Jacobi forms has an ambiguity involving a choice
of contour. For instance,  there are a number of discussions on how to chose a contour in the dyon counting formula in $\CN=4$ string theory related to the attractor mechanism for black
holes~\cite{Dabholkar:2007vk, Cheng:2007ch, Banerjee:2008yu}
The present work seems to be the first example where the contour is chosen from the world-sheet point of view rather than relying on the space-time point of view.
This approach may shed new light on the contour ambiguity in other counting formulae for black hole and BPS states in string theory.
We also gave both a mathematical treatment of wall crossing in the equivariant elliptic genus and a physical quantum mechanical model which partially captures and explains this phenomenon.

There is clearly much more that can be done. It would be interesting to extend these results to a larger class of non-compact Calabi-Yau spaces and we also anticipate further applications of these results to BPS state counting problems.  For example it would be interesting to use our results to study some aspects of the 5d $\CN=2$ super Yang-Mills theories
in the presence of surface operators that can be understood as an external monopole string. This is because
the multi-centered Taub-NUT spaces arise in the moduli space of framed BPS states in the presence of the monopole (string) defects, see e.g. \cite{Cherkis:1997aa,Moore:2014gua}. The well known T-duality between fivebranes and ALE spaces \cite{Ooguri:1995wj, Kutasov:1995te, Ooguri:1997ih, Anselmi:1993sm, Gregory:1997te} implies that our results also have implications for the elliptic genus of fivebrane backgrounds in string theory and the counting of BPS states in these theories.  It would be interesting to see if the techniques developed here can be used to further extend the program of
connecting the ADE classification of the fivebrane conformal field theory to the ADE classification of Umbral Moonshine \cite{Cheng:2013wca} initiated in \cite{Harvey:2013mda} and studied further in \cite{Cheng:2014zpa}.

\section*{Acknowledgments}

We thank Francesco Benini, Nima Doroud, Jaume Gomis, Kazuo Hosomichi, Sheldon Katz, Seok Kim, David Kutasov,
Kimyeong Lee, Jan Troost, and Piljin Yi  for helpful conversations. JH acknowledges the support of NSF grant 1214409.
The work of SL is supported by the Ernest Rutherford fellowship of
the Science \& Technology Facilities Council ST/J003549/1.

\newpage
\centerline{\Large \bf Appendix}
\appendix

\section{Jacobi forms and theta functions \label{JacApp}}

In this appendix, we briefly review the basic facts about Jacobi forms and Jacobi theta functions that we
used in the main text. A Jacobi form of weight~$k$ and index~$m$ is a holomorphic
function\footnote{These definitions can be extended in a simple manner to multiple elliptic variables.}~$\v(\tau, u)$
from~$\mathbb{H} \times\IC$ to $\IC$ which is ``modular in $\tau$ and elliptic in $u $''
in the sense that it transforms under the modular group as
  \be\label{modtransform}  \v\Bigl(\frac{a\t+b}{c\t+d},\frac{u}{c\t+d}\Bigr) \=
   (c\t+d)^k\,e^{\frac{2\pi imc u^2}{c\t+d}}\,\v(\t,u)  \qquad \forall \quad
   \Bigl(\begin{array}{cc} a&b\\ c&d \end{array} \Bigr) \in SL(2; \Z) \ee
and under the translations of $u$ by $\mathbb{Z} \tau + \mathbb{Z}$ as
  \be\label{elliptic}  \v(\t, u+\lambda\tau+\mu)\= e^{-2\pi i m(\lambda^2 \t + 2 \lambda u)} \v(\t, u)
  \qquad \forall \quad \lambda,\,\mu \in \Z \, . \ee
We consider $k,m \in \half \IZ$.

The transformation laws~\eqref{modtransform}, \eqref{elliptic} include the periodicities $\v(\t+1,z) = \v(\t,z)$ and $\v(\t,z+1) = \v(\t,z)$, so $\v$ has a Fourier expansion
\be\label{fourierjacobi} \v(\t,z) \= \sum_{n, r} c(n, r)\,q^n\,e^{2 \pi i r z} \; , \qquad\qquad
   (q :=e^{2\pi i \t}) \ .
 \ee

The Dedekind eta function:
\be\label{defeta}
  \eta(\tau)\;:=\; q^{1/24} \prod_{n=1}^{\infty} (1-q^{n}) \,
\ee
is a modular form of weight~$1/2$.
The odd Jacobi theta function:
\be
\vartheta_1(\tau,z) = -iq^{1/8} \zeta^{1/2} \prod_{n=1}^\infty (1-q^n) (1-\zeta q^n) (1-\zeta^{-1} q^{n-1})
= i \sum_{m \in \ZZ} e^{\pi i (m+\half)}  \, q^{(m+1/2)^2/2} \, \zeta^{m+\half} \,
\ee
is a Jacobi form of weight~$1/2$ and index~$1/2$, and obeys the relations
\begin{align}
  \myth1(\t,z) & = - \myth1(\t,-z)\ ,
  \nonumber \\
  \myth1(\t, a+b\t + z ) & = ( - 1)^{a+b} e^{-2\pi i b z - i \pi b^2\t}
  \myth1(\t, z )
  \label{propertytheta1}
\end{align}
for $a,b\in \mathbb{Z}$, and
\be\label{propertytheta2}
 \left. \frac{1}{2\pi i} \, \frac{d}{dz} \vth_{1}(\t,z) \right|_{z=0} \= -i \, \eta(\t)^{3} \, .
\ee

\section{Jeffrey-Kirwan Residue \label{AppJK}}

Let us denote by $Q_\a$ the charge vectors in $\mathbb{R}^r$ where $r$ is the rank of a given
gauge group G. We consider a hyperplane in the $u$-plane for each of charge vectors
where $Z_\text{1-loop}$ becomes singular, i.e.,
\begin{align}
  H_{Q_\a} = \big\{ u \big| Q_\a(u) + z_\a =0 \big\} \ ,
\end{align}
where $z_\a$ denote chemical potential collectively. The elliptic genus can be
obtained from the Jeffrey-Kirwan residue sum over singular points where
$n$ hyperplanes intersect. Here $n\geq r$.
Let us denote by $\mathfrak{M}$ a set of
such singular points. More precisely, the Jeffrey-Kirwan
residue at a singular point $u_\ast$ is defined by
\begin{align}
  \JKres_{u_\ast \in \mathfrak{M}_\eta}(Q_\ast,\eta)
  = \sum_{\mathfrak{B}_a} \n(\mathfrak{B}_a) \Res_{a}
\label{def1}
\end{align}
where $\Res_a$ is the so-called {\it iterative residue}. We will explain the definition of  various symbols in the formula (\ref{def1}) below.

\begin{enumerate}

  \item Let us suppose that $n$ hyperplanes intersect at $u=u_\ast$.
  We denote by $Q_\ast$ an ordered set of charge vectors that define such $n$ hyperplanes
  \begin{align}
    Q_\ast = \{ Q_1,...,Q_n \}\ .
  \end{align}

  \item Choose an arbitrary vector $\eta$ in $\mathbb{R}^r$.
  We denote by $\mathfrak{M}_\eta$ a subset of $\mathfrak{M}$ satisfying the following condition:
  $u\in \mathfrak{M}_\eta$ if
  \begin{align}
    \eta \in \text{Cone}(Q_{i_1}, Q_{i_2}, .., Q_{i_r})
  \end{align}
  for any $r$ linearly independent charge vectors in $Q_\ast$.

  \item Find all the possible ordered basis $\mathfrak{B}_a$ in $\mathbb{R}^r$
  \begin{align}
    \mathfrak{B}_a = \{ Q_{a_1}, Q_{a_2}, ..., Q_{a_r} \} \ ,
  \end{align}
  where $Q_{a_j} \in Q_\ast$ ($j=1,2,..,r$) satisfying the following condition
  \begin{align}
    \eta \in \text{Cone}\big( \k^a_1, ..,\k^a_r \big)\ .
  \end{align}
  Here we introduce vectors $\k^a_j$ that are sums of charge vectors $Q_i \in Q_\ast$:
  \begin{align}
    \k^a_j = \sum_{Q_i \in F_j } Q_i \ \text{ for } j=1,2,..,r
  \end{align}
  where $F_j = \text{span}\{Q_{a_1},..,Q_{a_j}\}$.

  For each of the above ordered basis, one can define a number $\n(\mathfrak{B}_a)$
  \begin{align}
    \n(\mathfrak{B}_a) = \text{sign}\det \big[ \k^a_1 , \k^a_2 , .., \k^a_r \big]
  \end{align}

    \item The Jeffrey-Kirwan residue is defined by
  \begin{align}
    \JKres_{u_\ast \in \mathfrak{M}_\eta}(Q_\ast,\eta) = \sum_{\mathfrak{B}_a} \n(\mathfrak{B}_a) \Res_{a}
  \end{align}
  where $\Res_a$ is the so-called {\it iterative residue}:
  let $\tilde u_j=Q_{a_j} {u}+ z_{a_j}$ and $w = w_{1..r} d\tilde u_1 \wedge
  ..\wedge d\tilde u_r$, then the iterative residue is given by
  \begin{align}
    \Res_{a} w = \Res_{\tilde u_r=0}  \cdots  \Res_{\tilde u_1=0} w_{1..r} \ .
  \end{align}
  In other words, the JK-residue of $Z_\text{1-loop}$ at $u_\ast\in \mathfrak{M}_\eta$
  is
  \begin{align}
    \JKres_{u_\ast \in \mathfrak{M}_\eta}(Q_\ast,\eta) Z_\text{1-loop}
    = \sum_{\mathfrak{B}_a} \n(\mathfrak{B}_a)
    \Res_{\tilde u_r=0}  \cdots
    \Res_{\tilde u_1=0} \frac{Z_\text{1-loop}}{\det\big[Q_{a_1},..,Q_{a_r}\big]}\ .
  \end{align}

  \item If $n=r$, one can show that the Jeffrey-Kirwan residue becomes
  \begin{align}
    \JKres_{u_\ast \in \mathfrak{M}_\eta} \frac{d^r u}{Q_1(u) Q_2(u) .. Q_r(u) } =
    \frac1{|\text{det}(Q_1\cdots Q_r)|}\ .
  \end{align}
  %

%
%
%
%
%
%
%
%

\end{enumerate}

\section{Decomposition of Elliptic Genus into Discrete and Continuum Parts \label{decompApp}}

\paragraph{$\CN=(4,4)$ elliptic genus}
In this appendix we begin with the expression~\eqref{result1againrep} and show the details of the steps that lead up
to~\eqref{Edecomp}. Using the Poisson resummation formula, one can rewrite the elliptic genus as
\begin{align}
  \CE(\t;\xi_1,\xi_2,z)  = \sqrt{g^2\t_2} \int_0^1 du_1 du_2 \
  \Theta(\t;u_1+\t u_2+\xi_1,\xi_2,z)
  \sum_{(p,w)\in \mathbb{Z}^2} q^{l_0} {\bar q}^{\bar l_0} e^{-2i \pi u_1 p} \ ,
  \label{6.1}
\end{align}
where
\begin{align}
  \Theta(\t;\xi_1,\xi_2,z) \equiv \frac{\myth1(\t,\xi_1+z)\myth1(\t,\xi_1-z)}
  {\myth1(\t,\xi_1+\xi_2)\myth1(\t,\xi_1-\xi_2)}\ ,
\end{align}
and
\begin{align}
  l_0 = \frac{1}{4g^2} \big( p- g^2 (w+u_2) \big)^2 \ ,
  \qquad
  \bar l_0 = \frac{1}{4g^2} \big( p + g^2 (w+u_2) \big)^2\ .
\end{align}
Let us assume that all the chemical potentials $z$, $\xi_1$ and $\xi_2$
are real. Using the formulas\footnote{This formula is valid when $|q|<|x|<1$},
\begin{align}
  \frac{1}{i\myth1(\t,z)} = \frac{1}{\eta(q)^3} \sum_{r\in \mathbb{Z}} x^{r+\frac12} S_r(q)\ ,
  \qquad
  S_r(q) = \sum_{n=0}^\infty (-1)^n q^{\frac{n(n+2r+1)}{2}}\ ,
  \label{identity1}
\end{align}
and
\begin{align}
  \myth1(\t,z) = i \sum_{s\in \mathbb{Z}} (-1)^s q^{\frac{(s-\frac12)^2}{2}} y^{-s+\frac12 }\ ,
  \label{identity2}
\end{align}
where $q=e^{2\pi i\t}$ and $x=e^{2\pi i z}$, one can expand the elliptic genus
as follows\footnote{Note that the above expansion is valid only when
\begin{align}
  \left| q \right| < \left|x_i \right| < 1\ \rightarrow \
  |q| < |q^{u_2}| < 1\ ,
\end{align}
which is the case for $u_2 \in (0,1)$.}
\begin{align}
  \CE(\t;\xi_1,\xi_2,z) & = \sqrt{g^2\t_2} \int_0^1 \prod_i du_i
  \sum_{p,\w,r_i,s_i} \frac{ q^{l_0} {\bar q}^{\tilde l_0}}{\eta(q)^6}
  e^{-2i \pi u_1 p} \prod_{i=1}^2 (-1)^{s_i}
  q^{\frac{(s_i-1/2)^2}{2}}  S_{r_i}(q) x_i^{r_i+\frac12}y_i^{-s_i +\frac12}\
\end{align}
with
\begin{align}
  x_1= q^{u_2} e^{2\pi i (u_1+ \xi_1+\xi_2) }\ , & \qquad
  x_2 = q^{u_2} e^{ 2\pi i (u_1+\xi_1-\xi_2)}\ ,
  \nonumber \\
  y_1 = q^{u_2} e^{2\pi i (u_1+ \xi_1+z)}\ , & \qquad
  y_2 = q^{u_2} e^{2\pi i (u_1 + \xi_1 -z)} \ .
\end{align}
Integrating over the temporal holonomy $u_1$ provides the Gauss constraint
\begin{align}
  r_1+r_2 - s_1 - s_2 + 2  = p \ .
\end{align}
Then, the expression can be simplified into
\begin{align}
  \CE(\t;\xi_1,\xi_2,z) & =
  \sqrt{g^2\t_2} \int_0^1 du_2
  \sum_{p,\w,r_i,s_i}' \frac{(-1)^{s_1+s_2} }{\eta(q)^6}
  q^{-p\w} \left( q {\bar q}\right)^{\bar l_0}  q^{\frac{\sum_i (s_i-1/2)^2}{2}} S_{r_1}(q) S_{r_2}(q)
  \nonumber \\ & \qquad \times
  e^{2\pi i \xi_1p } e^{2\pi i \xi_2(r_1-r_2)} e^{-2\pi i z(s_1-s_2) }\ ,
\end{align}
where the symbol $\sum'$ denotes a constrained summation
\begin{align}
  \sum_{p,\w,r_i,s_i}' = \sum_{p,\w,r_i,s_i} \d_{r_1+r_2-s_1-s_2+2-p}\ .
\end{align}
In order to linearize the integral over the spatial holonomy $u_2$, we introduce
an additional continuous integration variable $u$
\begin{align}
  \left( q{\bar q} \right)^{\bar l_0} =
  \left(q{\bar q} \right)^{\frac{(p+g^2\w)^2 }{4g^2} + \frac{u_2(p+g^2\w)}{2} }
  \cdot \sqrt{\frac{4\t_2}{g^2}}\int_{-\infty}^\infty du \
  \left( q{\bar q} \right)^{\frac{u^2}{g^2} + i u u_2 }\ ,
\end{align}
and the elliptic genus can be expressed as
\begin{align}
  \CE & = 2\t_2 \int_{-\infty}^\infty  du\int_0^1 du_2 \
  \frac{1}{\eta(q)^6}\sum_{r_i,s_i,p,w}'
  \left(q{\bar q} \right)^{\frac{(p+g^2w)^2 }{4g^2} + \frac{u_2(p+g^2w)}{2}+\frac{u^2}{g^2} + i u u_2 }
  q^{-pw} q^{\frac{\sum_i(s_i-1/2)^2}{2}}
  \nonumber \\ & \qquad \times \Big.
  (-1)^{s_1+s_2} S_{r_1}(q) S_{r_2}(q)
  e^{2\pi i \xi_1 p} e^{2\pi i \xi_2 (r_1-r_2)} e^{-2\pi i z(s_1-s_2)}
   \ ,
  \nonumber \\ & =
  - \frac{1}{\pi \eta(q)^6}\int_{-\infty}^\infty du \ \sum'_{r_i,s_i,p,w}
  \frac{\left(q{\bar q} \right)^{\frac{(p+g^2w)^2 }{4g^2} +\frac{u^2}{g^2}}}{2iu + (p+g^2w)}
  \Big[ \left(q{\bar q} \right)^{iu+\frac{p+g^2w}{2}} - 1 \Big]
  q^{-pw} q^{\frac{\sum_i(s_i-1/2)^2}{2}}
  \nonumber \\ & \qquad \times \Big.
  (-1)^{s_1+s_2} S_{r_1}(q) S_{r_2}(q)
  e^{2\pi i \xi_1 p} e^{2\pi i \xi_2 (r_1-r_2)} e^{-2\pi i z(s_1-s_2)}\  .
  \label{Ham1}
\end{align}
This new variable $u$ is interpreted \cite{Ashok:2013zka} as a momentum conjugate to the non-compact
radial direction. Due to the imaginary exponent of the modular parameter $q$, the Hamiltonian
interpretation of the elliptic genus of the Taub-NUT space is not obvious yet.

Let us now separate the expression (\ref{Ham1}) into two pieces,
\begin{align}
  \CE = \CE_A + \CE_B\ ,
  \label{Ham2}
\end{align}
where the former contains the imaginary exponent of $q$ while the latter
has the real exponent
\begin{align}
  \CE_A & =   - \frac{1}{\pi \eta(q)^6}\int_{\mathbb{R}+i\e} du \ \sum'_{r_i,s_i,p,w}
  \frac{\left(q{\bar q} \right)^{\frac{(p+g^2w)^2 }{4g^2} +\frac{u^2}{g^2}}}{2iu + (p+g^2w)}
  \left(q{\bar q} \right)^{iu+\frac{p+g^2w}{2}}
  q^{-pw} q^{\frac{\sum_i(s_i-1/2)^2}{2}}
  \nonumber \\ & \qquad \times \Big.
  (-1)^{s_1+s_2} S_{r_1}(q) S_{r_2}(q)
  e^{2\pi i \xi_1 p} e^{2\pi i \xi_2 (r_1-r_2)} e^{-2\pi i z(s_1-s_2)}\  ,
  \nonumber \\
  \CE_B & =   + \frac{1}{\pi \eta(q)^6}\int_{\mathbb{R}+i\e} du \ \sum'_{r_i,s_i,p,w}
  \frac{\left(q{\bar q} \right)^{\frac{(p+g^2w)^2 }{4g^2} +\frac{u^2}{g^2}}}{2iu + (p+g^2w)}
  q^{-pw} q^{\frac{\sum_i(s_i-1/2)^2}{2}}
  \nonumber \\ & \qquad \times \Big.
  (-1)^{s_1+s_2} S_{r_1}(q) S_{r_2}(q)
  e^{2\pi i \xi_1 p} e^{2\pi i \xi_2 (r_1-r_2)} e^{-2\pi i z(s_1-s_2)}\  .
\end{align}
Here we assume that the parameter $g^2$ is generic so that $(p,w)=(0,0)$
is the only solution to an equation
\begin{align}
  p + g^2 w = 0 \ .
\end{align}
Note that the integration contour has to be chosen slightly above the real axis
to regularize the pole at $u=0$ when $p=w=0$.
Let us then shift both integration and summation variables as follows
\begin{align}
  \int_{\mathbb{R}+i\e}du \ q^{-pw}
  \frac{\left(q{\bar q} \right)^{\frac{(p+g^2w)^2 }{4g^2} +\frac{u^2}{g^2}}}{2iu + (p+g^2w)}
  \xrightarrow[u\to u+i\frac{g^2}{2}]{w\to w+1}
  \int_{\mathbb{R}+i\e-i\frac{g^2}{2}}du \ q^{-pw - p}
  \frac{\left(q{\bar q} \right)^{\frac{(p+g^2w)^2 }{4g^2} +\frac{u^2}{g^2}}}{2iu + (p+g^2w)}
  \left( q{\bar q} \right)^{i u + \frac{p+g^2w}{2}}\ .
  \nonumber
\end{align}
Using the identities below
\begin{align}
  S_{r-1}(q) = 1- S_{-r}(q) \ , \qquad
  q^r S_r(q) = S_{-r}(q)\ ,
\end{align}
one can manage to rewrite $\CE_B$ as follows
\begin{align}
  \CE_{B}  & = \frac{1}{\pi \eta(q)^6}\int_{\mathbb{R}+i\e-i\frac{g^2}{2}}du \
  \sum_{r_i,s_i,p,w}'
  \frac{\left(q{\bar q} \right)^{\frac{(p+g^2w)^2 }{4g^2} +\frac{u^2}{g^2}}}{2iu + (p+g^2w)}
  \left(q{\bar q} \right)^{iu+\frac{p+g^2w}{2}} q^{-pw} q^{\frac{\sum_i(s_i+1/2)^2}{2}}
  \nonumber \\ & \qquad \times(-1)^{s_1+s_2}
  \prod_{i=1}^2 \left( q^{-r_i-1} - S_{r_i+1}(q)\right) \times
  e^{2\pi i \xi_1 p} e^{2\pi i \xi_2 (r_1-r_2)} e^{-2\pi i z(s_1-s_2)}\ ,
  \nonumber \\ & =
  \frac{1}{\pi \eta(q)^6}\int_{\mathbb{R}+i\e-i\frac{g^2}{2}}du \
  \sum_{{\hat r}_i,{\hat s}_i,p,w}'
  \frac{\left(q{\bar q} \right)^{\frac{(p+g^2w)^2 }{4g^2} +\frac{u^2}{g^2}}}{2iu + (p+g^2w)}
  \left(q{\bar q} \right)^{iu+\frac{p+g^2w}{2}} q^{-pw} q^{\frac{\sum_i({\hat s}_i-1/2)^2}{2}}
  \nonumber \\ & \qquad \bigg. \times(-1)^{{\hat s}_1+{\hat s}_2}
  S_{{\hat r}_1}(q)S_{{\hat r}_2}(q)
  e^{2\pi i \xi_1 p} e^{2\pi i \xi_2 ({\hat r}_1-{\hat r}_2)} e^{-2\pi i z({\hat s}_1-{\hat s}_2)}
  + \big( \text{the rests} \big)\ ,
\end{align}
where we used for the last equality a change of variables
\begin{align}
  \hat r_i = r_i + 1\ , \qquad \hat s_i = s_i +1\ .
\end{align}
Plugging the above result back into (\ref{Ham2}), one can separate the elliptic genus
into a discrete and continuum part
\begin{align}
  \CE = \CE_\text{disc} + \CE_\text{rest}\ ,
\end{align}
as in~\eqref{Edecomp} of the main text.

\paragraph{$\CN=(0,4)$ elliptic genus}

Let us rewrite (\ref{(0,4)TN}) into the following form
\begin{align}
  \CE_{N=1}^{(0,4)}(\t;\xi_1,\xi_2) & = g^2 \int_0^1 du_1 \int_0^1 du_2 \ \frac{\eta(q)^2}
  {\myth1(\t, u_1+\t u_2 + \xi_1 + \xi_2) \myth1(\t, u_1+\t u_2 + \xi_1 -\xi_2) }
  \nonumber \\ & \times
  \sum_{(p,w)\in \mathbb{Z}^2} q^{w^2} q^{2 w u_2} e^{4\pi i w (u_1+\xi_1)} e^{-\frac{g^2}{\t_2}
  |u_1+\t u_2 + p + \t w|^2}
\end{align}
Using (\ref{identity1}), one can expand the elliptic genus as follows
\begin{align}
  \CE_{N=1}^{(0,4)}(\t;\xi_1,\xi_2) & = -\frac{\sqrt{g^2\t}}{\eta(q)^4} \int_0^1 du_1 \int_0^1 du_2
  \sum_{p,w,r_1,r_2} q^{w^2 + (r_1+r_2 + 1 + 2w) u_2} q^{l_0} {\bar q}^{\bar l_0}
  e^{2\pi i u_1 ( r_1+r_2 + 1+ 2w -p)}
  \nonumber \\ & \times
  e^{2\pi i \xi_1 ( r_1+r_2+1 + 2w)} e^{2\pi i \xi_2 (r_1-r_2)} S_{r_1}(q) S_{r_2}(q) \ .
\end{align}
Performing the integral over $u_1$ gives us the Gauss constraint
\begin{align}
  r_1 + r_2 + 1 + 2w = p \ ,
\end{align}
and the elliptic genus becomes
\begin{align}
  \CE_{N=1}^{(0,4)}(\t;\xi_1,\xi_2) & = - \frac{\sqrt{g^2\t}}{\eta(q)^4} \int_0^1 du_2
  \sum_{p,w,r_1,r_2}' q^{w^2-pw} \left( q \bar q \right)^{\bar l_0}
  e^{2\pi i p \xi_1 } e^{2\pi i \xi_2 (r_1-r_2)} S_{r_1}(q) S_{r_2}(q) \ ,
\end{align}
where the symbol $\sum'$ denotes a constrained sum
\begin{align}
  \sum_{p,w,r_1,r_2}' =  \sum_{p,w,r_1,r_2} \d_{r_1+r_2+1+2w -p}\ .
\end{align}
As in section 5.1, one can rewrite the elliptic genus as an integral
over a momentum $u$ conjugate to the non-compact ``radial'' variable,
\begin{align}
  \CE_{N=1}^{(0,4)}(\t;\xi_1,\xi_2) & = \frac{1}{\pi \eta(q)^4} \int_{-\infty}^{\infty} du
  \sum_{p,w,r_1,r_2}'  \frac{\left(q \bar q\right)^{\frac{(p+g^2 w)^2}{4g^2}+ \frac{u^2}{g^2}}}
  {2i u + (p + g^2 w)} \left[ (q\bar q)^{i u  + \frac12 (p+g^2 w)} - 1 \right]
  \nonumber \\ & \times
  q^{w^2-pw}
  e^{2\pi i p \xi_1 } e^{2\pi i \xi_2 (r_1-r_2)} S_{r_1}(q) S_{r_2}(q) \ ,
  \label{(0,4)TN2}
\end{align}
Following the procedure explained in section 5.1, one can then read off
from the above expression (\ref{(0,4)TN2}) the discrete part of the elliptic genus
\begin{align}
  \CE_{N=1,\text{disc}}^{(0,4)}(\t;\xi_1,\xi_2) = - \frac{1}{\eta(q)^4} \sum_{p,r_1,r_2} \sum_{w\in I_p}
  \d_{r_1+r_2+1+2w-p} \
  q^{w^2-pw} e^{2\pi i p \xi_1 } e^{2\pi i \xi_2 (r_1-r_2)} S_{r_1}(q) S_{r_2}(q)\ ,
  \label{(0,4)TN3}
\end{align}
where $I_p$ is defined as in (\ref{range}).

\newpage

\bibliographystyle{JHEP}
\bibliography{ALEF}

\def\cprime{$'$}
\providecommand{\href}[2]{#2}\begingroup\raggedright\begin{thebibliography}{10}

\bibitem{Eguchi:2004yi}
T.~Eguchi and Y.~Sugawara, {\it {$SL(2,R) / U(1)$ supercoset and elliptic
  genera of noncompact Calabi-Yau manifolds}},  {\em JHEP} {\bf 0405} (2004)
  014, [\href{http://xxx.lanl.gov/abs/hep-th/0403193}{{\tt hep-th/0403193}}].

\bibitem{Troost:2010ud}
J.~Troost, {\it {The non-compact elliptic genus: mock or modular}},  {\em JHEP}
  {\bf 1006} (2010) 104, [\href{http://xxx.lanl.gov/abs/1004.3649}{{\tt
  arXiv:1004.3649}}].

\bibitem{Eguchi:2010cb}
T.~Eguchi and Y.~Sugawara, {\it {Non-holomorphic Modular Forms and
  $SL(2,R)/U(1)$ Superconformal Field Theory}},  {\em JHEP} {\bf 1103} (2011)
  107, [\href{http://xxx.lanl.gov/abs/1012.5721}{{\tt arXiv:1012.5721}}].

\bibitem{Ashok:2011cy}
S.~K. Ashok and J.~Troost, {\it {A Twisted Non-compact Elliptic Genus}},  {\em
  JHEP} {\bf 1103} (2011) 067, [\href{http://xxx.lanl.gov/abs/1101.1059}{{\tt
  arXiv:1101.1059}}].

\bibitem{Ashok:2013zka}
S.~K. Ashok and J.~Troost, {\it {Elliptic genera and real Jacobi forms}},  {\em
  JHEP} {\bf 1401} (2014) 082, [\href{http://xxx.lanl.gov/abs/1310.2124}{{\tt
  arXiv:1310.2124}}].

\bibitem{Ashok:2013pya}
S.~K. Ashok, N.~Doroud, and J.~Troost, {\it {Localization and real Jacobi
  forms}},  \href{http://xxx.lanl.gov/abs/1311.1110}{{\tt arXiv:1311.1110}}.

\bibitem{Murthy:2013mya}
S.~Murthy, {\it {A holomorphic anomaly in the elliptic genus}},
  \href{http://xxx.lanl.gov/abs/1311.0918}{{\tt arXiv:1311.0918}}.

\bibitem{Pope:1978zx}
C.~Pope, {\it {Axial Vector Anomalies and the Index Theorem in Charged
  Schwarzschild and Taub-Nut Spaces}},  {\em Nucl.Phys.} {\bf B141} (1978) 432.

\bibitem{Pope:1981jx}
C.~Pope, {\it {The $\eta$ Invariant for Charged Spinors in Taub-Nut}},  {\em
  J.Phys.} {\bf A14} (1981) L133--L137.

\bibitem{Zwegers}
S.~{Zwegers}, {\it {Mock Theta Functions}},  {\em Ph.D thesis} (2008)
  [\href{http://xxx.lanl.gov/abs/0807.4834}{{\tt arXiv:0807.4834}}].

\bibitem{Dabholkar:2012nd}
A.~Dabholkar, S.~Murthy, and D.~Zagier, {\it {Quantum Black Holes, Wall
  Crossing, and Mock Modular Forms}},
  \href{http://xxx.lanl.gov/abs/1208.4074}{{\tt arXiv:1208.4074}}.

\bibitem{Gibbons:1979zt}
G.~Gibbons and S.~Hawking, {\it {Gravitational Multi-Instantons}},  {\em
  Phys.Lett.} {\bf B78} (1978) 430.

\bibitem{kronheimer1989}
P.~B. Kronheimer, {\it The construction of ale spaces as hyper-kahler
  quotients},  {\em Journal of Differential Geometry} {\bf 29} (1989), no.~3
  665--683.

\bibitem{Cherkis:1998xca}
S.~A. Cherkis and A.~Kapustin, {\it {$D(k)$ gravitational instantons and Nahm
  equations}},  {\em Adv.Theor.Math.Phys.} {\bf 2} (1999) 1287--1306,
  [\href{http://xxx.lanl.gov/abs/hep-th/9803112}{{\tt hep-th/9803112}}].

\bibitem{Cherkis:2003wk}
S.~A. Cherkis and N.~J. Hitchin, {\it {Gravitational instantons of type
  $D(k)$}},  {\em Commun.Math.Phys.} {\bf 260} (2005) 299--317,
  [\href{http://xxx.lanl.gov/abs/hep-th/0310084}{{\tt hep-th/0310084}}].

\bibitem{Witten:1993yc}
E.~Witten, {\it {Phases of $N=2$ theories in two-dimensions}},  {\em
  Nucl.Phys.} {\bf B403} (1993) 159--222,
  [\href{http://xxx.lanl.gov/abs/hep-th/9301042}{{\tt hep-th/9301042}}].

\bibitem{Benini:2012ui}
F.~Benini and S.~Cremonesi, {\it {Partition functions of $N=(2,2)$ gauge
  theories on S$^2$ and vortices}},
  \href{http://xxx.lanl.gov/abs/1206.2356}{{\tt arXiv:1206.2356}}.

\bibitem{Doroud:2012xw}
N.~Doroud, J.~Gomis, B.~Le~Floch, and S.~Lee, {\it {Exact Results in $D=2$
  Supersymmetric Gauge Theories}},  {\em JHEP} {\bf 1305} (2013) 093,
  [\href{http://xxx.lanl.gov/abs/1206.2606}{{\tt arXiv:1206.2606}}].

\bibitem{Sugishita:2013jca}
S.~Sugishita and S.~Terashima, {\it {Exact Results in Supersymmetric Field
  Theories on Manifolds with Boundaries}},  {\em JHEP} {\bf 1311} (2013) 021,
  [\href{http://xxx.lanl.gov/abs/1308.1973}{{\tt arXiv:1308.1973}}].

\bibitem{Honda:2013uca}
D.~Honda and T.~Okuda, {\it {Exact results for boundaries and domain walls in
  2d supersymmetric theories}},  \href{http://xxx.lanl.gov/abs/1308.2217}{{\tt
  arXiv:1308.2217}}.

\bibitem{Hori:2013ika}
K.~Hori and M.~Romo, {\it {Exact Results In Two-Dimensional $(2,2)$
  Supersymmetric Gauge Theories With Boundary}},
  \href{http://xxx.lanl.gov/abs/1308.2438}{{\tt arXiv:1308.2438}}.

\bibitem{Kim:2013ola}
H.~Kim, S.~Lee, and P.~Yi, {\it {Exact partition functions on $\mathbb{RP}^2$
  and orientifolds}},  {\em JHEP} {\bf 1402} (2014) 103,
  [\href{http://xxx.lanl.gov/abs/1310.4505}{{\tt arXiv:1310.4505}}].

\bibitem{Jockers:2012dk}
H.~Jockers, V.~Kumar, J.~M. Lapan, D.~R. Morrison, and M.~Romo, {\it
  {Two-Sphere Partition Functions and Gromov-Witten Invariants}},  {\em
  Commun.Math.Phys.} {\bf 325} (2014) 1139--1170,
  [\href{http://xxx.lanl.gov/abs/1208.6244}{{\tt arXiv:1208.6244}}].

\bibitem{Gomis:2012wy}
J.~Gomis and S.~Lee, {\it {Exact Kahler Potential from Gauge Theory and Mirror
  Symmetry}},  {\em JHEP} {\bf 1304} (2013) 019,
  [\href{http://xxx.lanl.gov/abs/1210.6022}{{\tt arXiv:1210.6022}}].

\bibitem{Gerchkovitz:2014gta}
E.~Gerchkovitz, J.~Gomis, and Z.~Komargodski, {\it {Sphere Partition Functions
  and the Zamolodchikov Metric}},
  \href{http://xxx.lanl.gov/abs/1405.7271}{{\tt arXiv:1405.7271}}.

\bibitem{Tong:2002rq}
D.~Tong, {\it {NS5-branes, T duality and world sheet instantons}},  {\em JHEP}
  {\bf 0207} (2002) 013, [\href{http://xxx.lanl.gov/abs/hep-th/0204186}{{\tt
  hep-th/0204186}}].

\bibitem{Harvey:2005ab}
J.~A. Harvey and S.~Jensen, {\it {Worldsheet instanton corrections to the
  Kaluza-Klein monopole}},  {\em JHEP} {\bf 0510} (2005) 028,
  [\href{http://xxx.lanl.gov/abs/hep-th/0507204}{{\tt hep-th/0507204}}].

\bibitem{Hori:2001ax}
K.~Hori and A.~Kapustin, {\it {Duality of the fermionic 2-D black hole and N=2
  liouville theory as mirror symmetry}},  {\em JHEP} {\bf 0108} (2001) 045,
  [\href{http://xxx.lanl.gov/abs/hep-th/0104202}{{\tt hep-th/0104202}}].

\bibitem{Hori:2002cd}
K.~Hori and A.~Kapustin, {\it {World sheet descriptions of wrapped NS
  five-branes}},  {\em JHEP} {\bf 0211} (2002) 038,
  [\href{http://xxx.lanl.gov/abs/hep-th/0203147}{{\tt hep-th/0203147}}].

\bibitem{Johnson:1996py}
C.~V. Johnson and R.~C. Myers, {\it {Aspects of type IIB theory on ALE
  spaces}},  {\em Phys.Rev.} {\bf D55} (1997) 6382--6393,
  [\href{http://xxx.lanl.gov/abs/hep-th/9610140}{{\tt hep-th/9610140}}].

\bibitem{Harvey:2014zra}
J.~A. Harvey, D.~Kutasov, and S.~Lee, {\it {Comments on Quantum Higgs Vacua}},
  \href{http://xxx.lanl.gov/abs/1406.6000}{{\tt arXiv:1406.6000}}.

\bibitem{Benini:2013nda}
F.~Benini, R.~Eager, K.~Hori, and Y.~Tachikawa, {\it {Elliptic genera of
  two-dimensional $N=2$ gauge theories with rank-one gauge groups}},  {\em
  Lett.Math.Phys.} {\bf 104} (2014) 465--493,
  [\href{http://xxx.lanl.gov/abs/1305.0533}{{\tt arXiv:1305.0533}}].

\bibitem{Benini:2013xpa}
F.~Benini, R.~Eager, K.~Hori, and Y.~Tachikawa, {\it {Elliptic genera of 2d
  $N=2$ gauge theories}},  \href{http://xxx.lanl.gov/abs/1308.4896}{{\tt
  arXiv:1308.4896}}.

\bibitem{Gadde:2013dda}
A.~Gadde and S.~Gukov, {\it {2d Index and Surface operators}},  {\em JHEP} {\bf
  1403} (2014) 080, [\href{http://xxx.lanl.gov/abs/1305.0266}{{\tt
  arXiv:1305.0266}}].

\bibitem{Harvey:2013mda}
J.~A. Harvey and S.~Murthy, {\it {Moonshine in Fivebrane Spacetimes}},  {\em
  JHEP} {\bf 1401} (2014) 146, [\href{http://xxx.lanl.gov/abs/1307.7717}{{\tt
  arXiv:1307.7717}}].

\bibitem{Hohenegger:2013ala}
S.~Hohenegger and A.~Iqbal, {\it {M-strings, elliptic genera and $\mathcal{N} =
  4$ string amplitudes}},  {\em Fortsch.Phys.} {\bf 62} (2014) 155--206,
  [\href{http://xxx.lanl.gov/abs/1310.1325}{{\tt arXiv:1310.1325}}].

\bibitem{David:2006yn}
J.~R. David and A.~Sen, {\it {CHL Dyons and Statistical Entropy Function from
  D$1$-D$5$ System}},  {\em JHEP} {\bf 0611} (2006) 072,
  [\href{http://xxx.lanl.gov/abs/hep-th/0605210}{{\tt hep-th/0605210}}].

\bibitem{Gauntlett:1999vc}
J.~P. Gauntlett, N.~Kim, J.~Park, and P.~Yi, {\it {Monopole dynamics and BPS
  dyons $N=2$ superYang-Mills theories}},  {\em Phys.Rev.} {\bf D61} (2000)
  125012, [\href{http://xxx.lanl.gov/abs/hep-th/9912082}{{\tt
  hep-th/9912082}}].

\bibitem{Gauntlett:1996cw}
J.~P. Gauntlett and D.~A. Lowe, {\it {Dyons and S duality in $N=4$
  supersymmetric gauge theory}},  {\em Nucl.Phys.} {\bf B472} (1996) 194--206,
  [\href{http://xxx.lanl.gov/abs/hep-th/9601085}{{\tt hep-th/9601085}}].

\bibitem{Lee:1996if}
K.-M. Lee, E.~J. Weinberg, and P.~Yi, {\it {Electromagnetic duality and $SU(3)$
  monopoles}},  {\em Phys.Lett.} {\bf B376} (1996) 97--102,
  [\href{http://xxx.lanl.gov/abs/hep-th/9601097}{{\tt hep-th/9601097}}].

\bibitem{Lee:1998nv}
K.-M. Lee and P.~Yi, {\it {Dyons in $N=4$ supersymmetric theories and three
  pronged strings}},  {\em Phys.Rev.} {\bf D58} (1998) 066005,
  [\href{http://xxx.lanl.gov/abs/hep-th/9804174}{{\tt hep-th/9804174}}].

\bibitem{Bak:1999da}
D.~Bak, C.-k. Lee, K.-M. Lee, and P.~Yi, {\it {Low-energy dynamics for 1/4 BPS
  dyons}},  {\em Phys.Rev.} {\bf D61} (2000) 025001,
  [\href{http://xxx.lanl.gov/abs/hep-th/9906119}{{\tt hep-th/9906119}}].

\bibitem{Bak:1999ip}
D.~Bak, K.-M. Lee, and P.~Yi, {\it {Quantum 1/4 BPS dyons}},  {\em Phys.Rev.}
  {\bf D61} (2000) 045003, [\href{http://xxx.lanl.gov/abs/hep-th/9907090}{{\tt
  hep-th/9907090}}].

\bibitem{Eguchi:1980jx}
T.~Eguchi, P.~B. Gilkey, and A.~J. Hanson, {\it {Gravitation, Gauge Theories
  and Differential Geometry}},  {\em Phys.Rept.} {\bf 66} (1980) 213.

\bibitem{Kac:1994kn}
V.~G. Kac and M.~Wakimoto, {\it {Integrable highest weight modules over affine
  superalgebras and number theory}},
  \href{http://xxx.lanl.gov/abs/hep-th/9407057}{{\tt hep-th/9407057}}.

\bibitem{Maldacena:2000kv}
J.~M. Maldacena, H.~Ooguri, and J.~Son, {\it {Strings in AdS$_3$ and the
  $SL(2,R)$ WZW model. Part 2. Euclidean black hole}},  {\em J.Math.Phys.} {\bf
  42} (2001) 2961--2977, [\href{http://xxx.lanl.gov/abs/hep-th/0005183}{{\tt
  hep-th/0005183}}].

\bibitem{Kim:2011mv}
H.-C. Kim, S.~Kim, E.~Koh, K.~Lee, and S.~Lee, {\it {On instantons as
  Kaluza-Klein modes of M5-branes}},  {\em JHEP} {\bf 1112} (2011) 031,
  [\href{http://xxx.lanl.gov/abs/1110.2175}{{\tt arXiv:1110.2175}}].

\bibitem{Kallen:2012va}
J.~Kallen, J.~Qiu, and M.~Zabzine, {\it {The perturbative partition function of
  supersymmetric 5D Yang-Mills theory with matter on the five-sphere}},  {\em
  JHEP} {\bf 1208} (2012) 157, [\href{http://xxx.lanl.gov/abs/1206.6008}{{\tt
  arXiv:1206.6008}}].

\bibitem{Kim:2012ava}
H.-C. Kim and S.~Kim, {\it {M5-branes from gauge theories on the 5-sphere}},
  {\em JHEP} {\bf 1305} (2013) 144,
  [\href{http://xxx.lanl.gov/abs/1206.6339}{{\tt arXiv:1206.6339}}].

\bibitem{Kim:2012qf}
H.-C. Kim, J.~Kim, and S.~Kim, {\it {Instantons on the 5-sphere and
  M5-branes}},  \href{http://xxx.lanl.gov/abs/1211.0144}{{\tt
  arXiv:1211.0144}}.

\bibitem{Haghighat:2013gba}
B.~Haghighat, A.~Iqbal, C.~Kozcaz, G.~Lockhart, and C.~Vafa, {\it {M-Strings}},
   \href{http://xxx.lanl.gov/abs/1305.6322}{{\tt arXiv:1305.6322}}.

\bibitem{Haghighat:2013tka}
B.~Haghighat, C.~Kozcaz, G.~Lockhart, and C.~Vafa, {\it {On orbifolds of
  M-Strings}},  \href{http://xxx.lanl.gov/abs/1310.1185}{{\tt
  arXiv:1310.1185}}.

\bibitem{Hosomichi:2014rqa}
K.~Hosomichi and S.~Lee, {\it {Self-dual Strings and 2D SYM}},
  \href{http://xxx.lanl.gov/abs/1406.1802}{{\tt arXiv:1406.1802}}.

\bibitem{Bak:2014xwa}
D.~Bak and A.~Gustavsson, {\it {Elliptic genera of monopole strings}},
  \href{http://xxx.lanl.gov/abs/1403.4297}{{\tt arXiv:1403.4297}}.

\bibitem{Gaiotto:2005gf}
D.~Gaiotto, A.~Strominger, and X.~Yin, {\it {New connections between 4-D and
  5-D black holes}},  {\em JHEP} {\bf 0602} (2006) 024,
  [\href{http://xxx.lanl.gov/abs/hep-th/0503217}{{\tt hep-th/0503217}}].

\bibitem{Dabholkar:2010rm}
A.~Dabholkar, J.~Gomes, S.~Murthy, and A.~Sen, {\it {Supersymmetric Index from
  Black Hole Entropy}},  {\em JHEP} {\bf 1104} (2011) 034,
  [\href{http://xxx.lanl.gov/abs/1009.3226}{{\tt arXiv:1009.3226}}].

\bibitem{Dabholkar:2008zy}
A.~Dabholkar, J.~Gomes, and S.~Murthy, {\it {Counting all dyons in N =4 string
  theory}},  {\em JHEP} {\bf 1105} (2011) 059,
  [\href{http://xxx.lanl.gov/abs/0803.2692}{{\tt arXiv:0803.2692}}].

\bibitem{Bringmann:2014nba}
K.~Bringmann, T.~Creutzig, and L.~Rolen, {\it {Negative index Jacobi forms and
  quantum modular forms}},  \href{http://xxx.lanl.gov/abs/1401.7189}{{\tt
  arXiv:1401.7189}}.

\bibitem{Dabholkar:2007vk}
A.~Dabholkar, D.~Gaiotto, and S.~Nampuri, {\it {Comments on the spectrum of CHL
  dyons}},  {\em JHEP} {\bf 0801} (2008) 023,
  [\href{http://xxx.lanl.gov/abs/hep-th/0702150}{{\tt hep-th/0702150}}].

\bibitem{Cheng:2007ch}
M.~C. Cheng and E.~Verlinde, {\it {Dying Dyons Don't Count}},  {\em JHEP} {\bf
  0709} (2007) 070, [\href{http://xxx.lanl.gov/abs/0706.2363}{{\tt
  arXiv:0706.2363}}].

\bibitem{Banerjee:2008yu}
S.~Banerjee, A.~Sen, and Y.~K. Srivastava, {\it {Genus Two Surface and Quarter
  BPS Dyons: The Contour Prescription}},  {\em JHEP} {\bf 0903} (2009) 151,
  [\href{http://xxx.lanl.gov/abs/0808.1746}{{\tt arXiv:0808.1746}}].

\bibitem{Cherkis:1997aa}
S.~A. Cherkis and A.~Kapustin, {\it {Singular monopoles and supersymmetric
  gauge theories in three-dimensions}},  {\em Nucl.Phys.} {\bf B525} (1998)
  215--234, [\href{http://xxx.lanl.gov/abs/hep-th/9711145}{{\tt
  hep-th/9711145}}].

\bibitem{Moore:2014gua}
G.~W. Moore, A.~B. Royston, and D.~Van~den Bleeken, {\it {Brane bending and
  monopole moduli}},  \href{http://xxx.lanl.gov/abs/1404.7158}{{\tt
  arXiv:1404.7158}}.

\bibitem{Ooguri:1995wj}
H.~Ooguri and C.~Vafa, {\it {Two-dimensional black hole and singularities of CY
  manifolds}},  {\em Nucl.Phys.} {\bf B463} (1996) 55--72,
  [\href{http://xxx.lanl.gov/abs/hep-th/9511164}{{\tt hep-th/9511164}}].

\bibitem{Kutasov:1995te}
D.~Kutasov, {\it {Orbifolds and solitons}},  {\em Phys.Lett.} {\bf B383} (1996)
  48--53, [\href{http://xxx.lanl.gov/abs/hep-th/9512145}{{\tt
  hep-th/9512145}}].

\bibitem{Ooguri:1997ih}
H.~Ooguri and C.~Vafa, {\it {Geometry of N=1 dualities in four-dimensions}},
  {\em Nucl.Phys.} {\bf B500} (1997) 62--74,
  [\href{http://xxx.lanl.gov/abs/hep-th/9702180}{{\tt hep-th/9702180}}].

\bibitem{Anselmi:1993sm}
D.~Anselmi, M.~Billo, P.~Fre, L.~Girardello, and A.~Zaffaroni, {\it {ALE
  manifolds and conformal field theories}},  {\em Int.J.Mod.Phys.} {\bf A9}
  (1994) 3007--3058, [\href{http://xxx.lanl.gov/abs/hep-th/9304135}{{\tt
  hep-th/9304135}}].

\bibitem{Gregory:1997te}
R.~Gregory, J.~A. Harvey, and G.~W. Moore, {\it {Unwinding strings and t
  duality of Kaluza-Klein and h monopoles}},  {\em Adv.Theor.Math.Phys.} {\bf
  1} (1997) 283--297, [\href{http://xxx.lanl.gov/abs/hep-th/9708086}{{\tt
  hep-th/9708086}}].

\bibitem{Cheng:2013wca}
M.~C.~N. Cheng, J.~F.~R. Duncan, and J.~A. Harvey, {\it {Umbral Moonshine and
  the Niemeier Lattices}},  \href{http://xxx.lanl.gov/abs/1307.5793}{{\tt
  arXiv:1307.5793}}.

\bibitem{Cheng:2014zpa}
M.~C.~N. Cheng and S.~Harrison, {\it {Umbral Moonshine and K3 Surfaces}},
  \href{http://xxx.lanl.gov/abs/1406.0619}{{\tt arXiv:1406.0619}}.

\end{thebibliography}\endgroup

\end{document}